\documentclass[aps,prd,showkeys,showpacs,amssymb,
amsfonts,epsf,preprintnumbers,nofootinbib,superscriptaddress]{revtex4}

\usepackage[dvips]{graphicx}
\usepackage{bm,latexsym,amsmath,amssymb,amsfonts,color}

\newcommand*{\cL}{{\cal L}}
\newcommand\beq{\begin{eqnarray}}
\newcommand\eeq{\end{eqnarray}}
\def\beqa{\begin{eqnarray}}
\def\eeqa{\end{eqnarray}}

\begin{document}

\title{Thick brane solutions}

\author{Vladimir Dzhunushaliev,$^{1,2}$
\footnote{
Email: vdzhunus@krsu.edu.kg}
Vladimir Folomeev,$^{2}$
\footnote{Email: vfolomeev@mail.ru}
and
Masato Minamitsuji$^3$}
\affiliation{$^1$Department of Physics and Microelectronic
Engineering, Kyrgyz-Russian Slavic University, Bishkek, Kievskaya Str.
44, 720021, Kyrgyz Republic \\ 
$^2$Institute of Physicotechnical Problems and Material Science of the NAS
of the
Kyrgyz Republic, 265 a, Chui Street, Bishkek, 720071,  Kyrgyz Republic \\
$^3$Center for Quantum Spacetime, Sogang University,
Shinsu-dong 1, Mapo-gu, 121-742 Seoul, South Korea
}

\begin{abstract}
This article 
gives a comprehensive
review on thick brane solutions and related topics.
Such models have attracted much attention from many aspects
since the birth of the brane world scenario.
In many works,
it has been usually assumed that a brane is an infinitely thin object;
however in more general situations,
one can no longer assume this.
It is also widely considered that
more fundamental theories such as string theory
would have a minimal length scale.
Many multidimensional field theories coupled to gravitation
have exact solutions of gravitating topological defects,
which can 
represent
our brane world.
The inclusion of brane thickness
can realize a variety of possible brane world models.
Given our understanding,
the known solutions can be classified into
topologically non-trivial solutions and trivial ones.
The former class contains solutions of
a single scalar (domain walls),
multi-scalar,
gauge-Higgs (vortices),
Weyl gravity
and so on.
As an example of the latter class,
we consider solutions of two interacting scalar fields.
Approaches to obtain cosmological equations in the thick
brane world are reviewed.
Solutions with spatially extended branes (S-branes) and
those with an extra time-like direction are also discussed.
\end{abstract}

\pacs{}
\keywords{Higher-dimensional gravity, Cosmology}
\maketitle

\tableofcontents

\section{Introduction}
\label{review}

At present,
the presence of extra dimensions is an integral part of almost all physical theories
playing 
a fundamental role 
in
trying to explain physical interactions
based on common principles.
Also,
any realistic candidate
for a grand unified theory, 
such as superstring/M-theory,
should be multidimensional 
by necessity,
otherwise, it will contain undesirable physical consequences.
Traditionally,
it has been considered that
in such theories the observable 4-dimensional spacetime
appears as a result of compactification of extra dimensions, where
the
characteristic size of extra dimensions becomes much less than that of
4-dimensional spacetime.
But, recently, a new idea, called {\it brane world},  has come up, where
we live on a thin leaf (brane) embedded into some multidimensional space
(bulk).
Because of the lack of experimental and observational data,
there is no reason
to prefer a particular class of multidimensional models of gravity.
Actually,
various kinds of multidimensional theory have been studied
in modern theoretical physics.

In this article, we will review brane world models where
our brane universe has a finite thickness\footnote{In this review, we will
basically follow the notations, sign conventions and definitions of fields which were adopted in the original papers.}.
The reason is mainly two-fold:
Firstly,
such solutions appear in various multidimensional field theories
coupled to gravity,
leading to a variety of possibilities of brane world,
which are interesting by itself.
Secondly, brane thickness should be an essential ingredient
to extend the idea of brane world to multidimensional spacetimes.

Before coming to our main subject,
it is important to keep the main history of multidimensional gravity in mind.
Thus,
we briefly 
discuss the history of
multidimensional gravity and especially
why the idea of brane world arose.

\subsection{Prebrane epoch of multidimensional theory}

\subsubsection{Why is the visible universe 4-dimensional?}

After the birth of Einstein's general relativity,
the natural question about the dimensionality of the world which we live in
appeared.
Within the framework of Einstein's gravitational theory, space and time are
unified and
it allows us to realize that the surrounding world is a
4-dimensional one. This statement is a consequence of a combination of the principle of Occam's razor and Einstein's idea about the
geometrical
nature of the gravitational field. Hereupon, the gravitational field is not
an external field on the background of
a specified spacetime,
but a geometrical characteristic,
i.e., metric, self-controlled on any sufficiently smooth manifold.

One of the questions inevitably arising in such a case is
about the dimensionality of our world. Naturally, it is necessary to consider
two questions concerning the dimensionality of the physical world:
\begin{itemize}
  \item
Why the physically observed dimensions of our Universe $= 3+1$   (space~+~time)?
  \item
If the real dimensions are still more than 4, why extra dimensions are not observable, and what about any consequences of 
multidimensionality
for a 4-dimensional observer?
\end{itemize}
Strictly speaking, the answer to the second question could serve as an answer to the first question. Indeed, if extra dimensions are
hidden from direct observation then our world looks 4-dimensional effectively.

Although at the present time, it is not possible to give an exact answer to the first question about reasons of the 4-dimensionality of our
world, investigations in this direction show
that the 4-dimensional spacetime is marked out somehow among spacetimes with a different
number of dimensions. Let us enumerate these properties:
\begin{itemize}
  \item
Einstein's equations 
in Lorentzian manifolds with dimensions $d<4$ lead to a flat metric.
  \item
Maxwell's equations are conformally invariant only 
in a four-dimensional spacetime.
  \item
In Newtonian gravity, circular orbits of point-like bodies
in {cental-symmetric fields} are stable only 
{on} manifolds with dimensions
$d<4$.
  \item
In spacetime with dimensions $d>4$ the Schr\"{o}dinger equation either
{does not have} any energy level,
or an energy spectrum is unbounded below.
  \item
Huygens principle is valid only for spaces with odd dimensions.
  \item
Quantum electrodynamics is renormalizable only in a spacetime with dimensions
$d<4$.
\end{itemize}

\subsubsection{Kaluza-Klein theory}

The present stage of the development of multidimensional theories of gravity began with
Kaluza's paper
\cite{kaluza} where the foundations of modern multidimensional gravitational theories have been laid.
The essence of 
Kaluza's idea
consists in the proposal that the 5-dimensional Kaluza-Klein gravitation is equivalent to
4-dimensional Einstein gravitation
coupled to Maxwell's electromagnetism.
In the primordial paper there was the proposal that
the
$G_{55}$ metric component is a constant, and it is not 
varied when obtaining
the
gravitational equations. Later on, this proposal was
excluded.
Let us show the basic advantages and disadvantages of the standard 5-dimensional Kaluza-Klein theory. The advantages are:
\begin{itemize}
  \item
The 5-dimensional Einstein equations could be reduced to
a form identical to the equations of
4-dimensional gravitation interacting with
electromagnetic and scalar fields.
  \item
The geodesics equations in a 5-dimensional space
lead to the equations of motion for charged particles in the electromagnetic and gravitational fields.
  \item
If one supposes that all quantities in the Kaluza-Klein theory do not depend on the 5-th coordinate $x^5$
then it transforms
as a transformation of coordinates in the following way:
  \begin{equation}
    {x'}^5 = x^5 + f(x^\mu), \qquad \mu = 0,1,2,3
  \label{secintr-10}
  \end{equation}
This coordinate transformation leads to the usual gradient transformations in
Maxwell's electromagnetism.
\end{itemize}
The disadvantages are the following:
\begin{itemize}
  \item
Why is the 5-th coordinate unobservable in our world?
  \item
Einstein noted \cite{einstein1} that it is not possible to give
a natural physical meaning to the metric component
 $G_{55}$.
  \item
The
equation for a scalar field $\mathcal R_{55} - G_{55} \mathcal R = 0$ gives
the stiff and unusual
connection between the 4-dimensional scalar curvature $R$ and the invariant
$F_{\mu \nu} F^{\mu \nu}$ of the electromagnetic field.
\end{itemize}

\subsubsection{Multidimensional theories}

Next,
multidimensional Kaluza-Klein theories 
were considered.
One of the most important questions is the question about
the
unobservability
and
independence of 4-dimensional quantities on extra dimensions.
The most general approach for a
solution of this problem consists
in obtaining the 
solutions of the multidimensional Einstein equations,
which yield characteristic linear 4-dimensional spacetime sizes of
many orders of magnitude bigger than linear sizes in extra coordinates.
It is also necessary that the
4-dimensional metric of the
obtained equations does not depend on extra coordinates.
This could be done by refusing
vacuum equations, 
and introducing 
torsion \cite{Thi72,Cha76,Dom78,Orz81} or quadratic terms of curvature
 \cite{Wet82},
or by adding 
multidimensional matter.
The last variant 
called ``spontaneous compactification'' of the extra dimensions 
was suggested by Cremmer \& Scherk \cite{Cre76,Cre77}.
Unfortunately, this approach
gave up 
Einstein's original idea about the geometrization
 of physics.
Following \cite{wesson},
multidimensional Kaluza-Klein theories
can be classified into 
compactified theories
\cite{Klein26a,Klein26b,Cre76,Cre77,Thi72,Cha76,Dom78,Orz81,Wet82}
and projective theories \cite{Veb31}-\cite{Sch95}.

According to
the definition,
compactified theories should satisfy the following properties:
\begin{itemize}
  \item
some periodicity condition
with respect to extra coordinates is fulfilled;
  \item
extra dimensions form some compact manifold $M$;
  \item
the linear sizes of this manifold $M$ are small in comparison 
with the sizes of spacetime
\cite{Klein26a,Klein26b}.
\end{itemize}
It is necessary to note that the metric
which satisfies the above properties
 is a solution of the vacuum (4+d)-dimensional Einstein equations
only
if $M$ is a flat manifold \cite{ACF87}.
Manifold $M$ with constant curvature can be a solution of the Einstein equations only
in the existence of
multidimensional matter \cite{Cre76,Cre77}, torsion \cite{Thi72,Cha76,Dom78,Orz81}
or quadratic terms
of curvature \cite{Wet82}.
In the projective geometry,  a point is a set of projective points for which all relations of $(n+1)$-homogeneous coordinates are the same.
In such a case, each projective line is a point in a real space. Such an approach leads to the conclusion that the metric does not depend on
an extra coordinate.
Projective field theories do not agree with experiment. At the present time, they are not considered as realistic candidates
for an unified field theory.

\subsubsection{Problem of unobservability of extra dimensions}

In the last few decades,
the idea that our space, indeed, has a number of dimensions more than four became quite popular again. It is supposed that many
problems of elementary particles physics could be solved by introducing
strings, supersymmetry, and by increasing 
the number of spacetime dimensions.
In increasing
the number of dimensions,
a legitimate question arises: why physical properties of extra dimensions differ strongly from
those of the observed 4-dimensional spacetime.
Two 
solutions to this question were suggested, namely a
mechanism of
spontaneous compactification and brane world models.

In the case of compactified extra dimensions,
it is supposed that these dimensions are rather small
and unobservable.
In the papers \cite{Cre76a}-\cite{scherk75} the modern approach to the problem of the unobservability of compactified extra dimensions
was introduced.
In a $d-$dimensional spacetime, it is possible to consider
gravity coupled to matter. For the purpose of compactification,
a special solution
of the block form $M^d = M^4 \times M^{d-4}$ is
searched for, where
$M^4$ is the 4-dimensional spacetime and
$M^{d-4}$ is a compact space
of extra dimensions.
$M^4$ and $M^{d-4}$ are spaces of constant curvature.
The block representation of the metric
is consistent with the gravitational equations only if one supposes that
the energy momentum tensor takes the block form
 $   T_{\mu\nu} = -\gamma_1 g_{\mu\nu}$ in $M^4$,
and
 $   T_{mn} = \gamma_2 g_{mn} $ in $M^{d-4}$,
where $g_{\mu\nu}$ and $g_{mn}$ are metrics of
$M^4$ and $M^{d-4}$.
There are 
various
mechanisms of spontaneous compactification:
Freund-Rubin compactification \cite{freund}
with a special ansatz for antisymmetric tensors;
Englert compactification \cite{englert}
setting a gauge field in an internal space equal to the spin connection;
suitable embedding in a gauge group \cite{volkov};
monopole or instanton mechanism \cite{randjbar};
compactification using scalar chiral fields \cite{omero},
and compactification using radiative corrections \cite{witten}.

\subsection{Brane world scenario}

In this subsection, the new approach to the problem
of unobservability of extra dimensions, called brane world scenario,
is explained.
This approach is quite different from the traditional compactification approach
and allows even 
noncompact extra dimensions.
According to the new approach, particles corresponding to electromagnetic, weak and strong interactions are
confined on some hypersurface (called a brane) which, in turn, is embedded in some multidimensional space (called a bulk). Only
gravitation and some exotic matter (e.g., the dilaton field) could propagate in the bulk. It is supposed that our Universe is such a brane-like object.
This idea for the first time
was formulated phenomenologically in \cite{akama}--\cite{gibbons},
and got confirmation within
the framework of string theory. Within the brane world scenario,
restriction on sizes of extra dimensions becomes weaker. It happens because
particles belonging to the Standard Model
propagate only in three space dimensions. But
Newton's law of gravitation is sensible to
existence of extra dimensions.

\vspace{0.3cm}

The prototype idea of brane world
appeared a rather long time ago.
Now the term ``brane model'' means different ways for solving some fundamental problems in high-energy physics.
The pioneering works in this direction
were
done in the papers  \cite{akama} and \cite{rubakov}.
These prototype models have been constructed phenomenologically and
considered as topological defects, i.e.,  thick branes
in the modern terminology.

In \cite{akama},
an idea that extra dimensions could be noncompactified was suggested.
It was supposed that we live in  a vortex enveloping multidimensional space. Akama showed that
"\ldots The Einstein equation is induced just as in Sakharov's pregeometry \ldots ". The main idea of this paper consists in using 
the Higgs lagrangian in a 6-dimensional flat spacetime
\begin{equation}
	\mathcal L = - \frac{1}{4} F_{MN} F^{M} + D_M \phi^\dag D^M \phi +
    a |\phi|^2 - b |\phi|^4 + c,
\label{Prebrane-10}
\end{equation}
where $F_{MN} = \partial_M A_N - \partial_N A_M$ and $D_M \phi = \partial_M + i e A_M$.
It allows one
to obtain the induced Einstein gravitation on
a
vortex. The corresponding field equations have the vortex solution
\begin{equation}
	A_M = \epsilon_{0123MN} A(r) \frac{X^N}{r},
    \phi = \varphi(r) e^{in\theta},
    r^2 = \left( x^5 \right)^2 + \left( x^6 \right)^2,
\label{Prebrane-20}
\end{equation}
where $A(r)$ and $\varphi(r)$ are the solutions of the Nielsen-Olsen
differential equations describing the vortex. Our Universe (the vortex)
is localized inside the region
 $\mathcal O(\epsilon) (\epsilon = 1/\sqrt{a})$. Further, introducing
 curvilinear coordinates
\begin{equation}
	X_M = Y^M \left( x^\mu \right) + n^M_m x^m,\quad
    M=0,1,2,3,5,6,\quad \mu=0,1,2,3,\quad m=5,6,
\label{Prebrane-30}
\end{equation}
where $X^M$ are the Cartesian coordinates and $n^M_m$ are the normal vectors of the vortex, one can show that the Einstein action is induced
{through} vacuum polarizations.

In the paper \cite{rubakov}
the idea that we live on a
domain wall (brane in modern language)
and " \ldots ordinary particles are confined inside a potential wall \ldots ",
was suggested.
The quantum toy model with 
Lagrangian 
\begin{equation}
	\mathcal L = \frac{1}{2} \partial_A \varphi \partial^A \varphi
-	\frac{1}{2} m^2 \varphi^2 -
	\frac{1}{4} \varphi^4, \quad A=0,1, \ldots , 4
\label{Prebrane-40}
\end{equation}
is under consideration, which
describes a scalar field in the (4+1)-dimensional spacetime $M^{(4,1)}$ with metric $g_{AB}=\mathrm{diag}(1,-1,-1,-1,-1)$. The classical field equations for the scalar field $\varphi$ admit a domain wall solution (which is
a (1+1)-dimensional kink)
\begin{equation}
	\varphi^{cl}(x) = \frac{m}{\sqrt \lambda} \tanh
	\left(
		\frac{mx^4}{\sqrt 2}
	\right).
\label{Prebrane-50}
\end{equation}
This kink solution can be considered as the domain wall. In the paper the possibility of trapping particles with spins $0$ and $1/2$ on the domain wall 
was also considered.

For particles with spin 0, the linearized equation of motion for the field $\varphi'=\varphi-\varphi^{cl}$ is under consideration
\begin{equation}
    - \partial_A \partial^A \varphi' - m^2 \varphi' -
    3 \lambda \left( \varphi^{cl} \right)^2 \varphi' = 0.
\label{Prebrane-60}
\end{equation}
It was shown that there exist three types of perturbations:
\begin{enumerate}
  \item the first one
is of particles confined inside the wall and with the energy $E={\vec k}\,^2$;
  \item the second one is
of particles confined inside the wall, 
but
with the energy $E=\vec k\,^2 + \frac{3}{2} m^2$;
  \item the third one is not confined inside the wall and at large $x^4$ the energy is $E=\vec k\,^2 + \left( k^4 \right)^2 + 2m^2$.
\end{enumerate}
The possibility of interaction of the first type of particles with creation of the third type of particles and with subsequent escape of
them beyond the domain wall
was
discussed.

Furthermore,
the question about trapping 
fermions on the domain wall was 
considered. The Lagrangian of massless fermions is
\begin{equation}
    \mathcal L_\psi = i \bar \Psi \Gamma^A \partial_A \Psi +
    \hbar \bar \Psi \Psi \varphi .
\label{Prebrane-70}
\end{equation}
It was shown that for such an interaction between
a fermion $\Psi$ and
a domain scalar field $\varphi$, the Dirac equation has a solution
\begin{equation}
    \Psi^{(0)}\left( x^0, \vec x, x^4 \right) = \psi(x^0, \vec x)
    \exp \left(
        -\hbar \int \limits_0^{x^4} \varphi^{cl}\left( {x'}^4 \right) d {x'}^4
    \right) ,
\label{Prebrane-80}
\end{equation}
where $\psi(x^0, \vec x)$ is a left-handed massless (3+1)-dimensional spinor.
It was also
shown that there exist two types of fermion perturbations:
\begin{enumerate}
  \item the first one
is of fermions \eqref{Prebrane-80}
confined inside the wall;
  \item the second one is
not
confined inside the wall and the energy
exceeds
  $\hbar m/\sqrt{\lambda}$.
It can be created only in high energy collisions.

\end{enumerate}
Note that in both papers
\cite{akama} and \cite{rubakov} the idea about noncompactified extra dimensions 
was suggested. The difference consists in that in
\cite{akama} 
our
4-dimensional world is {the}
interior of {a}
vortex, 
{while}
in \cite{rubakov}
{our}
4-dimensional world 
is {a} domain wall.

We should also mention interesting earlier attempts
in the context of supergravity and superstring theory.
In~\cite{Cvetic}, models of domain walls in $N=1$ supergravity theories were considered.
Using this model, it was shown that there exists a class of domain wall solutions which need not to be $\mathbb{Z}_2$ symmetric.
{Such a solution describes}
a stable domain wall that divides two isolated but non-degenerate supersymmetric
vacua, and at least one of them is
{an} anti-de Sitter one.
Moreover, it was shown that in supergravity theories gravity
plays  a non-trivial and crucial role for topological defects (in particular, for domain walls).

Ref. \cite{Anton} examined
the possibility that
some of the internal dimensions,
at relatively low energy scales,
may be of the order of a few TeV,
within the framework of
a
perturbative string theory
which could relate
the supersymmetry breaking scale to the size of
the internal
dimensions.

\vspace{0.3cm}

Although several ideas have been presented during the
80s and the early 90s,
one of the most striking facts which activated
the studies on brane models
was the development in superstring theory and M-theory
since the mid of
90s, especially the discovery of D-brane solutions \cite{db,db2}.
A valuable contribution to development of this model was made in the papers
\cite{arkanihamed1} and \cite{arkanihamed2}, following an earlier idea from~\cite{Anton}.
The authors considered a flat bulk geometry 
of $(4+d)$ dimensions,
when $d$ extra dimensions are compact with radius
$R$.
{
All the Standard Model particles are assumed to be localized
on the branes at the boundary of the $d$-dimensional compact space.
But the gravitational interaction feels the presence of extra dimensions
on the distance scales less than $R$
and differs from Newton's law there.}
{
By integrating over the internal space,
it turns out that}
the 4-dimensional Planck mass
$M_{\rm Pl}$ and the $(4+d)$-dimensional Planck mass $M_{\rm fund}$ are connected by the following relation:
\begin{equation}
    M_{\rm Pl}^2 = M_{\rm fund}^{2+d} R^d.
\label{review-270}
\end{equation}
Since Newton's law
has been checked up to the distances of the order of 0.1 mm,
$R$ can be of order 0.1 mm or less.

In the papers \cite{Gogberashvili:1998vx,Gogberashvili:1998iu,Gogberashvili:1999tb} the model of the Universe
expanding as a  3-shell in a in 5-dimensional spacetime was considered. It was shown that the model can solve
 the hierarchy problem.
Also the relation  between the size of the Universe and its thickness was found.

In the papers \cite{Randall:1999ee} and \cite{Randall:1999vf},  Randall and Sundrum 
considered a non-flat bulk space. The metric has the form
\begin{equation}
    ds^2 = e^{-2kr_c|y|}\eta_{\mu\nu}dx^\mu dx^\nu-r_c^2dy^2\,,
\label{RS-metric}
\end{equation}
 where $k$ is a parameter of order of the fundamental Planck  mass;
$r_c$ is 
the
radius of a 5-th coordinate. This metric is a solution of the 5-dimensional gravitational equations following from the Lagrangian
\begin{equation}
	S=\frac{M_\star^3}{16\pi}\int
	d^4x dy\sqrt{-g}\left(-R-\lambda\right)+\sum_{i=1,2}\int d^4x
	\sqrt{-h^{(i)}}\left({\cal L}^{(i)}-V^{(i)}\right)\,,
\end{equation}
 where $h_{\mu\nu}^{(i)}$, ${\cal L}^{(i)}$ and $V^{(i)}$ are the 4-dimensional metric, the Lagrangian and the vacuum energy on $(i)$th-brane,
 respectively. One can show that the relation between the 4-dimensional $M_{Pl}$ and the fundamental $M_*$ Planck masses is:
\begin{equation}
	M_{\rm Pl}^2=\frac{M^3_\star}{k}\left[1-e^{-2kr_c\pi}\right]
	\sim \frac{M^3_\star}{k}\,.
\label{rs-mass}
\end{equation}
Namely, the scale of physical phenomena on the brane is fixed by the value of the warp factor. On the brane near  $y = \pi$ (observed brane),
the conformal factor for the 4-dimensional metric is $\rho^2=e^{-2kr_c\pi}$,
and physical masses should be calibrated
{by taking
this factor into account}. For example, if $kr_c\sim 12$ then masses of order
TeV could be generated from the fundamental Planck mass $\sim 10^{16}$ TeV.

Various physical problems have
been considered in five-dimensional (and even multidimensional)
brane models, especially in the Randall-Sundrum models and their extensions.
See e.g.,\cite{tbr,tbr2,tbr3,tbr4,Mannheim} for reviews.
In many works, it was assumed that
the brane is infinitely thin.
Although in the thin brane approximation many interesting
results have been obtained,
in some situations the effects of the brane thickness cannot
be neglected.

\subsection{Thick branes}

From a realistic point of view,
a brane should have a thickness.
It is also widely considered that the most
fundamental theory
would have a minimal length scale.
In some cases the effects of brane thickness can be important.
The inclusion of 
brane thickness gives us
new possibilities and new problems.
In many multidimensional field theories coupled to gravity
there are solutions of topological defects.
They have lead to a richer variety of brane worlds
(see sections \ref{section-2} and \ref{section-3}).
At the same time,
brane thickness also brings ambiguity
in the definition of the effective four-dimensional quantities
(see section \ref{section-4}).

\vspace{0.3cm}

\begin{figure}[h]
\begin{minipage}[t]{.49\linewidth}
\begin{center}
 \includegraphics[width=8cm]{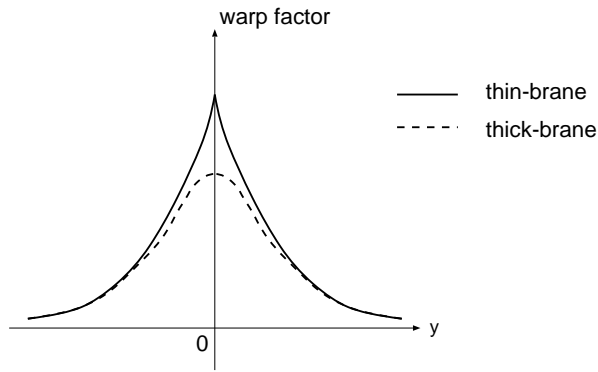}
	\caption{
The typical behavior of the warp factor $a(y)$
in the thin and the thick brane solutions are shown.
The dashed line denotes the warp factor of the thick brane, while
the solid one does that of the thin brane.}
\label{warp_factor of flat brane}
 \end{center}
\end{minipage}
\end{figure}

We now give our precise
definition of thick branes,
to avoid possible problems related to possible
differences in terminology.
Our definition is based on the following form of a multidimensional metric:
For five-dimensional problems,
the solutions whose metric are
\beq
	ds^2= a^2(y) g_{\mu\nu}dx^{\mu}dx^{\nu} - dy^2,
\label{intro_five}
\eeq
where $-\infty<y <\infty$ is the coordinate of the extra dimension,
are considered.
The four-dimensional $g_{\mu\nu}$ {is}
Minkowski or de Sitter (or anti-de Sitter) spacetime.
$a(y)$ is the warp function, which is regular,
has a peak at the brane
and falls off rapidly away from the brane.
A typical behavior of the warp function
in thin and thick brane solutions is shown in
Fig.~\ref{warp_factor of flat brane}.
When the $Z_2$-symmetry
is assumed, $a(y)=a(-y)$.
The normalizability of the graviton zero mode gives the
condition that $\int_{-\infty}^{\infty} dy a(y)^{4}$ is nonvanishing and finite.
In five-dimensional problems,
the thin brane approximation is valid
as long as the brane thickness cannot be resolved,
in other words,
if the energy scale of the brane thickness is much higher
than those in the bulk and on the brane.
In contrast,
when thickness becomes as large as the
scale of interest,
its effect is no longer negligible.

With
more than six-dimensional spacetimes, $n>1$,
it is assumed that the metric function has the form
\begin{equation}
	ds^2= a^2(r) g_{\mu\nu}dx^{\mu}dx^{\nu} - b^2(r) (dr^2 +  r^2 d \Omega ^2 _{n-1}).
\label{intro-10}
\end{equation}
The four-dimensional metric $g_{\mu\nu}$ also represents
Minkowski or de Sitter spacetime.
$d \Omega ^2 _{n-1}$ is the line element of an
 $(n-1)$-dimensional
compact manifold ({the} unit $(n-1)$-sphere in most cases).
 The extra coordinates are composed of the radial one $r$
($0<r<\infty$), and
those of the $(n-1)$-dimensional ones.
It is possible to put the brane at the center $r=0$
by an appropriate coordinate choice.
Here $a(r)$ is the warp factor,
which has a peak at the brane.
The functions $a(r)$ and $b(r)$ are regular everywhere and the integral
$
\int_0^{\infty} dr a^4 b^{n} r^{n-1}
$
is finite,
in order to obtain a
localized graviton zero mode.
In such higher-codimensional cases,
the situation is quite different from that in the five-dimensional one.
In the context of the thin brane approximation,
in approaching the brane, usually
self-interactions of bulk matter, e.g., gravitation, are divergent.
Then, it is impossible to describe the matter interactions localized
on the brane.
To study such models,
brane thickness becomes an essential ingredient which
plays the role of an effective UV cut-off.

Note that it is obvious that such a definition is rather ambiguous: for example, one can redefine the coordinate
$r$ in the following way
\begin{equation}
	\sqrt{b(r)} dr = dz,
\label{intro-20}
\end{equation}
and then the brane metric can be rewritten as follows
\begin{equation}
	ds^2= a^2(z) g_{\mu\nu}dx^{\mu}dx^{\nu}
- \left[ dz^2 +  r^2(z) d \Omega ^2 _{n-1} \right] .
\label{intro-30}
\end{equation}
In sections II-IV,
exact solutions of static thick brane solutions
in various classes of field theories
will be considered.
In sections V-VI,
time-dependent situations are considered, and then
the bulk metric is no longer the form of Eq. (\ref{intro_five})
or Eq. (\ref{intro-10}).

\subsection{Purpose and construction}
\label{purpose_of_rev}

In this paper, we will give a comprehensive review of works devoted to thick brane solutions, and
related topics.
To our knowledge there has been
no review on this subject
although an
enormous number of works have already been done.
Thus, now is an appropriate time to collect these works.
We hope that this paper becomes a good starting point for new studies.

In our understanding,
 all known thick brane solutions can be
classified into two large groups: {\it static} solutions (presented in Sections II-IV) and solutions {\it depending on time} (presented in Sections V-VI). In turn, the static solutions could be classified as:
\begin{enumerate}
  \item
  Topologically nontrivial thick branes:
These solutions are based on
the
assumption of existence of
some topological defects in spacetime.
 Solutions in this class are composed of thick branes made
of either a single scalar field
or non-interacting multi-scalar fields
(and so on).

 \item
  Topologically trivial thick branes:
These solutions can exist for interacting scalar fields but they are absent in the case of one scalar field.

\end{enumerate}

Several representative approaches to formulate cosmological
equations on thick branes will be reviewed (section \ref{section-4}).
Finally, 
the regular S-brane solutions
will also be explained (section \ref{section-5}).

\vspace{0.3cm}

As a final remark,
we will not discuss applications of thick brane models
to high-energy physics
and the problem of localization of field with various spins.
In this review, we are going to focus on their geometrical
and topological properties only.


\section{Topologically nontrivial thick branes}
\label{section-2}

\subsection{Definition of topologically nontrivial solutions}

In this review, we use the classification of thick brane solutions given in section \ref{purpose_of_rev}. According to this
classification, one type of solution is topologically nontrivial.
In this section we give the definition of
topologically nontrivial solutions. This classification is based on definitions of kink-like and monopole-like solutions.

Kink-like, topologically nontrivial solutions are considered in two dimensions (one space and one time dimensions). Such solutions come up 
in considerations of problems with one scalar field $\phi$ with a potential energy
having two or more degenerate vacuum states. One example is the well-known Mexican hat potential \cite{rajaraman}
$$
	U(\phi)=\frac{1}{4}\lambda\left(\phi^2 -
	\frac{m^2}{\lambda} \right)^2,
$$
where $\lambda$ and $m^2$ are coupling and mass parameters respectively. In this case, there exist localized solutions which go to $\phi_{\pm\infty}=\pm m/\sqrt{\lambda}$ (the so-called kink) asymptotically. The solution with an asymptotic behavior $\phi_{\pm\infty}=\mp m/\sqrt{\lambda}$ is known as the anti-kink. In the papers reviewed below the asymptotic values of
the scalar field $\phi_{\pm \infty}$ can be shifted by $\phi_0$, i.e.
$\phi_{\pm\infty}=\pm m/\sqrt{\lambda} + \phi_0$.

We now consider the case when the dimensionality of
space is $n>1$, and when the number of scalar fields is $m>1$. We take the following mapping of a sphere $S_{n-1}$ on a sphere $S_{m-1}$: sphere  $S_{n-1}$ is an infinitely remote  sphere in a space of extra dimensions, i.e. it is a sphere at $r\rightarrow\infty$ in the metric
\begin{equation}
	dl^2_{ed} =  b(r) (dr^2 +  r^2 d \Omega ^2 _{n-1}).
\label{topology-10}
\end{equation}
Here
 $dl^2_{ed}$ is the
metric in the extra dimensions; the metric $b(r) r^2 d \Omega ^2 _{n-1}$ is the metric on a sphere $S_{n-1}$.
 The sphere $S_{m-1}$ is a sphere in the abstract space of the scalar fields $\phi^a, a=1,2, \cdots m$.

\begin{figure}[h]
\begin{center}
\fbox{
 \includegraphics[width=8cm]{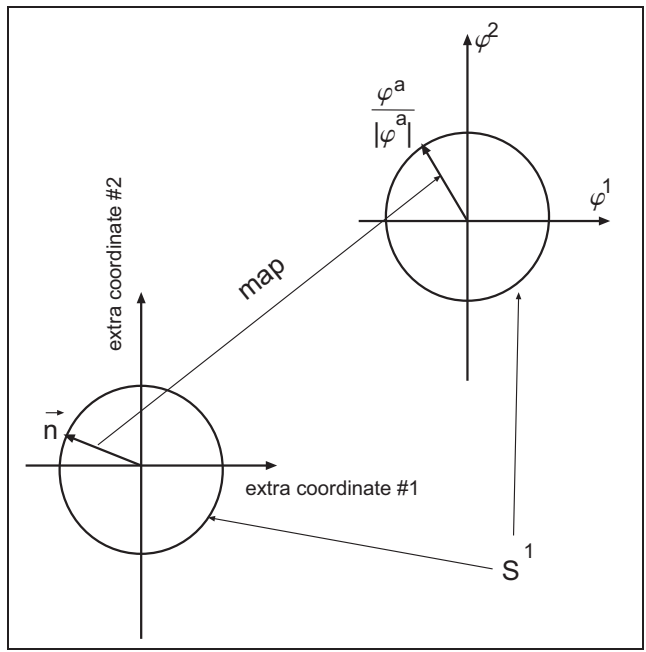}
}
	\caption{Map
	$\vec n = \frac{\vec r}{r} \rightarrow \left.
	\frac{\phi^a}{|\phi^a|}	\right|_{\vec r \rightarrow \infty}$
	for a case $n=m=2$}
 \label{map}
 \end{center}
\end{figure}

Using this set up the mapping is defined by
\begin{equation}
	\vec n = \frac{\vec r}{r} \rightarrow
	\left.
		\frac{\phi^a}{|\phi^a|}
	\right|_{\vec r \rightarrow \infty}.
\label{topology-20}
\end{equation}
A picture of this mapping is given schematically in Fig.~\ref{map}. This mapping maps the point
$\frac{\phi^a}{|\phi^a|} \in S_{m-1}$ in the space of the scalar fields to the vector $\vec n = \frac{\vec r}{r} \in S_{n-1}$ in a space of the extra dimensions.

From the topological point of view, we have the following map
\begin{equation}
	\pi_{n-1} \left( S_{m-1} \right) = S_{n-1} \rightarrow S_{m-1},
\label{topology-30}
\end{equation}
 where $\pi_{n-1} \left( S_{m-1} \right)$ is a homotopy group of the map
$S_{n-1} \rightarrow S_{m-1}$. It is known from topology that all nonsingular maps 
from one sphere $S_{n-1}$ into another one $S_{m-1}$
can be divided into homotopic sectors. Maps inside of one sector can pass continuously  each into other, but maps of two different sectors cannot pass continuously  each into other.

For example
\cite{steenrode}
\begin{eqnarray}
	\pi_n \left( S_n \right) &=& Z,
\label{topology-40}\\
	\pi_n \left( S_m \right) &=& 0 \quad \text{ by } \quad n<m,
\label{topology-50}\\
	\pi_n \left( S_1 \right) &=& 0 \quad \text{ by } \quad n>1,
\label{topology-60}
\end{eqnarray}
 where $Z$ is a group of integer numbers, i.e. the maps $S_n \rightarrow S_n$ are divided in to a discrete set of homotopic classes which are characterized by integers. In  \eqref{topology-50} and \eqref{topology-60}, zero on the right hand side means that the group is trivial, i.e. in this case all maps can be deformed each into other.

\emph{ We will call a thick brane solution topologically nontrivial if the corresponding homotopy group $\pi_{n-1} \left( S_{m-1} \right)$ is nontrivial}. Note that the case $n=1$ and $m=1$ is topologically nontrivial.

\subsection{Thick brane solutions with a single scalar field}

In this subsection, as the simplest example of a topologically
nontrivial solution, we review the five-dimensional
thick brane solutions with a single scalar field.

\subsubsection{General properties}

First, we review the general properties of thick brane solutions
of a single scalar field which have
the four-dimensional Poincare symmetry,
following Ref. \cite{bronnikov1,bronnikov2}.
We consider a five-dimensional Einstein-scalar theory
\begin{eqnarray}
 S =\frac{1}{2}	\int d^5 x \sqrt{-g}\left[ R
		-(\partial\phi)^2 - 2V(\phi)
  \right].\label{1}
\end{eqnarray}
In general,
the metric of a static five-dimensional spacetime
with the
four-dimensional Poincare symmetry
can be written as
\beq
ds^2=e^{2F(z)}\eta_{\mu\nu}dx^{\mu}dx^{\nu}+e^{8F(z)}dz^2.
\eeq
For simplicity,
it is assumed that the bulk geometry is $Z_2$-symmetric
across the center of the brane,
hence $F(-z)=F(z)$, $\phi(-z)=-\phi(z)$, and
the imposed boundary conditions are
\beq
F(0)=F'(0)=0,\quad \phi(0)=0.
\eeq
In Ref. \cite{bronnikov1,bronnikov2}, the properties of thick brane solutions
were
investigated
and
it 
was shown that
the only possible asymptotic geometry of the thick brane spacetime
is asymptotically AdS,
and is only possible if
the potential $V(\phi)$ has an alternating sign
and
satisfies the fine-tuning condition
\beq
\bar V(\infty)=0,\quad
\bar V(z)
:=\int_0^z \sqrt{-g} V(\phi(z)) dz
=\int_0^z e^{8F(z)} V(\phi(z)) dz.
\label{cond}
\eeq
The zero thickness limit is well-defined and
the solution reduces to the Randall-Sundrum model
if the asymptotic value of $V(\phi)$ approaches
a constant, which is the bulk cosmological constant.
Note that if the brane has the four-dimensional
de Sitter or anti-de Sitter
symmetry, the fine-tuning condition \eqref{cond}
may not be satisfied.


\subsubsection{Single scalar thick brane solutions}

In this subsection,
we review the various thick brane solutions
whose geometry can be Minkowski, de Sitter
or anti- de Sitter, although in terms of
the cosmological applications we focus on the first two cases.
Many of previously known solutions are discussed in
the context of the $Z_2$-symmetric bulk spacetime
but non-$Z_2$ symmetric solutions have also been reported.
The thick brane solutions of
more than two scalar fields or of other fields
will be reported, separately.

\paragraph{Set-up}

We focus on the Einstein-scalar theory given in
Eq. (\ref{1}).
Most of solutions reviewed here have the
following form of the metric
\beq
ds^2=dy^2+a(y)^2 \gamma_{\mu\nu}dx^{\mu}dx^{\nu},
\quad
\phi=\phi(y)\label{2}
\eeq
with the four-dimensional Ricci curvature tensor being
$R\big[\gamma_{\mu\nu}\big]=12 K$.
The constant
$K=0,1,-1$ represents the four-dimensional Minkowski,
de Sitter and AdS sections, respectively.
It is also assumed that
the scalar field
{depends}
only on the extra space coordinate $y$: $\phi=\phi(y)$.
If the $Z_2$-symmetry is imposed, namely $a(y)=a(-y)$ and $\phi(y)=-\phi(-y)$
(the center of the brane is placed at $y=0$),
the boundary conditions for metric and scalar field are given by
\beq
a'(0)=0 \quad
\phi(0)=0,
\eeq
where ``prime" denotes a derivative with respect to $y$.
The warp factor is normalized so that $a(0)=1$.
At the infinity $y\to \pm\infty$, the scalar field approaches
some constant value $\phi=\pm\phi_{\infty}$, where
for a  kink $\phi_{\infty}>0$ and for an anti-kink $\phi_{\infty}<0$.
By minimizing the action Eq. (\ref{1}),
the Einstein and scalar equations are explicitly given by
\beq
&&
6\frac{a'{}^2}{a^2}
-6\frac{K}{a^2}
=\frac{1}{2}\phi'{}^2-V,
\nonumber\\
&&3\frac{a''}{a}+3\frac{a'{}^2}{a^2}
-3\frac{K}{a^2}
=-\frac{1}{2}\phi'{}^2-V,
\nonumber\\
&&\phi''+\frac{4a'}{a}\phi'-\frac{dV}{d\phi}=0\,.
\eeq
Sometimes, it is useful to reduce the above Einstein-scalar equations into
first order equations.
This can be achieved by introducing the auxiliary
superpotential $W(\phi)$
\cite{DeWolfe:1999cp,grem,cehs,super3,sasakura1, sasakura2},
in terms of which the scalar field potential is given as
\beq
	V=-6W(\phi)^2
  +\frac{9}{2\gamma^2}
   \left( \frac{dW}{d\phi} \right)^2,\quad
\gamma:=\sqrt{1+\frac{K}{a^2 W^2}}.
\label{sup1}
\eeq
Using this,
the above equations become
\beq
\phi'=\frac{3}{\gamma}\frac{dW}{d\phi},\quad
\frac{a'}{a}=-\gamma W(\phi).
\label{sup2}
\eeq
For a given superpotential $W$ the scalar field profile
is obtained by integrating the first equation.
Then, the metric function is determined by integrating the
second equation.

More directly,
for a given metric function, one can determine the
scalar field profile and potential, using the procedure of \cite{kks}.
But sometimes, it is impossible to give the
potential as an analytical function of the scalar field,
rather than as a function of the bulk coordinate.
Another method \cite{kks,volkas} is
that first one gives the profile of the scalar field,
and then the metric function and the potential are determined
through the field equations.

\paragraph{Minkowski brane solutions}

Here the simplest cases are solutions which contain
the four-dimensional section which has the Poincare symmetry $K=0$.
The solutions which are obtained from the simple superpotential
\beq
	W=c\sin(b\phi)
\eeq
were discussed in \cite{grem,cehs,gio,flat4,super3,flat5}
(See also the textbook \cite{Mannheim}).
The corresponding scalar potential has the sine-Gordon form
\beq
	V(\phi)	=\frac{3}{2}c^2
	\left[ 3b^2\cos^2 \left( b\phi \right)
	   -4\sin^2\left( b\phi \right)
	\right]\,.
\eeq
The solution is given by
\beq
	a(y)=\left[ \frac{1}{\cosh(cb^2y)}\right]^{1/3b^2},\quad
	\phi(y)=\frac{2}{b}\arctan
	\left[ \tanh\left( \frac{3cb^2y}{2} \right) \right].
\eeq
In the asymptotic limit $y\to \infty$, $\phi(\pm\infty)= \pm  \pi/(2b)$
the potential approaches the negative values
$V(\pm\infty)=-6c^2$.
The bulk is asymptotically AdS.
This class of solutions also contains the solution derived in
\cite{kks},
\beq
	a(y) &=& \left[ \frac{1}{e^{-2n(y+y_0)}+e^{2n(y+y_0)}}
	\right]^{1/2n},
\nonumber\\
	\phi(y) &=& \pm \sqrt{\frac{6}{n}} \; \arctan (e^{2ny}),
\label{koyama}
\eeq
whose potential is given by
\beq
	W(\phi)=\cos\left( \frac{\sqrt{6n}}{3}\phi \right),
\quad
 V(\phi) = -6\cos^2\left( \frac{\sqrt{6n}}{3}\phi \right)
 +3n\sin^2\left( \frac{\sqrt{6n}}{3}\phi \right).
\eeq
The dimensionless parameter $n$ represents the brane thickness.
The solution is stable against perturbations.

In Ref. \cite{flat2}, another type of thick brane solution was discussed,
which was
obtained from the superpotential
\beq
	W(\phi)=\frac{av}{3}\phi\left( 1-\frac{\phi^2}{3v^2}\right),
\eeq
leading to the potential
\beq
	V(\phi)	=\frac{\lambda}{4}\left( \phi^2-v^2 \right)^2
	-\frac{\lambda}{27M^2} \phi^2\left( \phi^2-\frac{v^2}{3} \right)^2,
\eeq
where $\lambda:=2a^2/v^2$ is scalar self-coupling constant and the gravitational scale $M$ is shown explicitly.
In the absence of gravity $M\to \infty$, the theory reduces to the
$\lambda\phi^4$ scalar field model.
The scalar and metric functions are given by
\beq
	\phi(y) &=& v\tanh(ay),
\nonumber\\
	M^3\ln a(y) &=&  C_0-\frac{8}{27}v^2\ln \left[\cosh(ay)\right]
	-\frac{v^2}{54}\tanh^2(av),
\eeq
where $C_0$ is an integration constant which represents the
overall scaling of our four-dimensional world
(See also the textbook \cite{Mannheim}).
It is trivial to check that the geometry reduces to the
five-dimensional Minkowski geometry
in the decoupling limit $M\to \infty$.
In Ref. \cite{ringeval} the potential
$$
V(\phi)	=-V_0+\frac{\lambda}{4}\left( \phi^2-v^2 \right)^2
$$
was considered. Here $V_0>0$. This potential gave rise to
thick Minkowski brane solution with the asymptotically AdS bulk.

\paragraph{de Sitter brane solutions}

Here, we review some known thick de Sitter brane solutions
with $K=+1$.

In Ref. \cite{kks}, the stable de Sitter brane version of the solution
Eq. (\ref{koyama}) was discussed, whose metric function
is given by
\beq
	a(y)=\left[ \frac{1}{\sinh^{-2n}(y+y_0)+\sinh^{2n}(y+y_0)}
	\right]^{1/2n}.
\eeq
But there is no analytic form for
$\phi(y)$ and $V(\phi)$ \cite{kks}.

In Ref. \cite{sasakura1, sasakura2},
for the scalar field potential
\beq
	V(\phi)=\frac{9H^2(\beta^2-1)}{2\beta^2}
	+\frac{15H^2(\beta^2+1)}{2\beta^2}
	\cos\left(\sqrt{\frac{2}{3}}\phi\right)\,,
\eeq
the stable solutions
\beq
	a(y) &=& -{\rm sn} (Hy, i\beta^{-1}),
\nonumber\\
	\phi(y) &=& \sqrt{6} \; \arctan
	\left[ \frac{{\rm cn} (Hy,i\beta^{-1})}
	            {\beta{\rm dn}(Hy, i\beta^{-1}) }
	\right]
\eeq
were derived,
where the elliptic functions are defined by
\beq
	{\rm sn}^{-1}(z,k)
	=\int_0^z
	\frac{dx}{\sqrt{(1-x^2)(1-k^2x^2)}},
\eeq
${\rm cn}(z)=(1-{\rm sn}^2(u,k))^{1/2}$,
and
${\rm dn}(z)=(1-k^2 {\rm sn}^2(u,k))^{1/2}$.


Other interesting solutions were given for an axionic scalar field potential \cite{wang,mns1,mns2,irreg1,irreg2,irreg3}
\beq
	V(\phi)=V_0\cos^{2(1-\sigma)}\left(\frac{\phi}{\phi_0}\right),
	\quad
	\phi_0:=\sqrt{3\sigma(1-\sigma)},
\label{wang-10}
\eeq
where $0<\sigma <1$.
The $Z_2$-symmetric solution \cite{wang} is given by
\beq
	a(z)=\left[ \cosh\left(\frac{Hz}{\sigma}\right)\right]^{-\sigma},
	\quad
	\phi=\phi_0\arcsin\left[\tanh\left( \frac{ Hz}{\sigma} \right)
	\right],
\label{wang-20}
\eeq
where $dy=a(z) dz$ and the brane expansion rate $H$ is related to the
thickness and potential parameter through
\beq
H^2=\frac{2\sigma V_0}{3\big(1+3\sigma\big)}.
\eeq
The physical thickness of the brane is given by $\sigma/H$. In the limit $\sigma\to 0$, the solution smoothly approaches a thin de Sitter brane embedded in a five-dimensional Minkowski bulk.

Extensions of Eq. \eqref{wang-20} to the non-$Z_2$ symmetric bulk were considered in Ref. \cite{irreg1,irreg2,irreg3}.
Other thick de Sitter brane solutions with more complicated metric functions
were reported in Ref. \cite{volkas}.

\subsection{Thick brane solutions in generalized scalar field theories}

In this subsection,
we review the thick brane solutions in
more general theories of a single scalar field,
i.e.,
a scalar field with a non-standard kinetic term
\cite{bglm,agsw1,agsw2,ygfr},
a scalar field non-minimally coupled to gravity \cite{flat1},
a phantom scalar field \cite{flat6,Gogberashvili:2009yp},
and a tachyon field \cite{tac}.

\subsubsection{Scalar field with a non-standard kinetic term}

The paper \cite{bglm} discussed the case of a scalar field
with non-standard kinetic term.
Such a model was initially considered
as the
kinetic generalization of the single field inflation models \cite{adm,gm}
and extensively studied for modeling inflation and dark energy.
Here it is applied to a five-dimensional space-time
to find new thick brane solutions.
The starting point is the action
\beq
S=\int d^5 x \sqrt{g} \Big(-\frac{1}{4}R+{\cal L}(\phi,X)\Big),
\eeq
where $g_{ab}$ is the five-dimensional metric
with the signature $(+,-,-,-,-)$, and the five-dimensional
gravitational constant is chosen $G^{(5)}=1/(4\pi)$.
The indices $M,N,\cdots$ and $\mu,\nu,\cdots$ run
over the five- and four-dimensional space-time, respectively.
We now define $X:=(1/2)g^{MN}\nabla_{M}\phi\nabla_{N}\phi$.
Then, the equations of motion are given by
\beq
 G^{MN}\partial_{M}\partial_{N}\phi
+2X{\cal L}_{X\phi}
-{\cal L}_{\phi}
=0\,,
\eeq
where
$
G^{MN}=\cL_{X}g^{MN}+\cL_{XX}\nabla^{M}\phi\nabla^{N}\phi$.
The Minkowski thick brane solutions whose metric is given by
\beq
ds^2=e^{2A(y)}\eta_{\mu\nu}dx^{\mu}dx^{\nu}-dy^2,
\eeq
are considered.
The equations of motion are
\beq
&&\Big(\cL_{X}+2X\cL_{XX}\Big)\phi''
-\Big(2X\cL_{X\phi}-\cL_{\phi}\Big)
=-4 \cL_{X}\phi' A'\,,
\nonumber\\
&& A''=\frac{4}{3}X\cL_X,
\nonumber\\
&&\Big( A'\Big)^2
=\frac{1}{3}\Big(\cL- 2X\cL_X\Big)\,,
\eeq
where the prime is derivative with respect to $y$.
For convenience,
the (superpotential) function $W(\phi)$, such that
$\cL_{X}\phi'= W_{\phi}$,
is introduced.
The scalar field Lagrangian is assumed to be
\beq
\cL= F(X)-V(\phi)\,,
\eeq
where $F$ can be an arbitrary function of $X$.
Note that $F(X)=X$ for the standard canonical scalar field.
The equations of motion now reduce to
a set of the  first order equations that
\beq
&&F'(X)\phi'= \frac{1}{2}W_{\phi},
\nonumber\\
&&
F-2F'(X) X-V(\phi)
=\frac{1}{3}W^2.
\eeq
In Ref. \cite{bglm},
two specific examples of the choice of
the function $F$
were considered: (I) $F_I=X+\alpha |X| X$,
and (II) $F_{II}=- X^2 $.

\paragraph{Case (I) }

In case (I), $F_I=X+\alpha |X| X$, it is difficult to derive solutions for general values of
$\alpha$.
Thus,
Ref. \cite{bglm} considered the case of small $\alpha$
and discussed perturbations around the case of a
canonical scalar field.
Then, the behaviors of scalar and metric
{functions} are given by
\beq
\phi(y)=\phi_0(y)
      -\frac{\alpha}{2}\phi_0'(y)W(\phi_0(y))\,,
\quad
A(y)=A_0(y)+\frac{\alpha}{12}W(\phi_0(y))^2,
\eeq
where $\phi_0(y)$ is the solution in the case of $\alpha=0$
(the canonical case).
For the specific choice that $W(\phi)=3a \sin(b\phi)$,
one finds the scalar potential
\beq
V=\frac{9}{8}a^2 b^2 \cos^2(b\phi)
-3a^2\sin^2(b\phi)
-\frac{81\alpha}{64}a^4 b^4 \cos^4(b\phi)
+O(\alpha^2)
\,.
\eeq
The scalar and metric functions were found as
\beq
&&
\phi(y)=\frac{1}{b}\arcsin\Big[\tanh\Big(\frac{3}{2}ab^2 y\Big)\Big]
-\frac{9\alpha a^2 b}{4}
\frac{\sinh \Big(\frac{3}{2}a b^2 y\Big)}
{\cosh^2 \Big(\frac{3}{2}a b^2 y\Big)}
+O(\alpha^2),
\nonumber\\
&& A(y)=-\frac{2}{3b^2}\ln
\left(
\cosh\Big(\frac{3}{2}ab^2y\Big)
\right)
+\frac{3a^2\alpha}{4}
\tanh^2\Big(\frac{3}{2}ab^2y\Big)
+O(\alpha^2)\,.
\eeq
For larger value of $\alpha$, the peak of the warp factor $A$ is spread out
and the energy density at the center of the brane decreases.

\paragraph{Case (II)}

In case (II),  $F_{II}=- X^2 $,
the choice of the superpotential,
\beq
W(\phi)=9 a^3 b^2 \sin (b\phi)
   \Big(2+\cos^2(b\phi)\Big),
\eeq
leads to
\beq
V(\phi)
=\frac{243}{8}a^4 b^4 \cos^4(b\phi)
-27 a^6b^4\sin^2(b\phi) \Big(2+\cos^2(b\phi)\Big)^2.
\eeq
An exact solution was found
\beq
&&\phi(y)
=\frac{1}{b}\arcsin
\big(\tanh (3 a b^2 y)\big)\,,
\quad
A(y)=
-\frac{a^2}{6}\tanh^2\big(3ab^2 y\big)
-\frac{2}{3}a^2
\ln\Big(\cosh (3a b^2 y)\Big)\,.
\eeq
The effective brane thickness is of order
$(ab^2)^{-1}$, which can become larger for smaller $a$ and $b$.

The work of \cite{agsw1,agsw2} considered
the model composed of the non-linear kinetic term $F(X)=X|X|$ and
the self-interacting potential $V(\phi)=\lambda^4 (\phi^2-\phi_0^2)^2$,
where $\phi=\pm \phi_0$ corresponds to the global minima and
$\lambda$ is the self-coupling of the scalar field.
A solitonic solution coupled to gravity,
which can be interpreted as a thick brane,
was obtained
by employing a
series expansion method and a
numerical one in \cite{agsw2},
while an exact solution without gravity was originally derived in \cite{agsw1}.
The solution is stable against the linear perturbations.
In addition,
this model has an appealing property that the propagation of perturbations of the scalar outside the thick brane is completely suppressed and
the scalar field is automatically restricted to the brane.

\vspace{0.3cm}

The recent work of \cite{ygfr}
 discussed the de Sitter or AdS thick brane solutions
in more general class of models
that
\begin{eqnarray}
\cL= F(X,\phi)-V(\phi)\,.
\end{eqnarray}
The line elements of the de Sitter and AdS brane solutions are assumed to be
\begin{equation}
ds^2=e^{2A(y)}\big(dt^2-e^{2\sqrt{\Lambda}t}(dx_1^2+dx_2^2+dx_3^2)\big)-dy^2,
\end{equation}
and
\begin{equation}
ds^2=e^{2A(y)}\big(e^{-2\sqrt{-\Lambda}x_3}(dt^2-dx_1^2-dx_2^2)+dx_3^2\big)-dy^2,
\end{equation}
respectively.
The choice of $F(X,\phi)= X \phi^m$, where $m=0,1,2,\cdots$ is considered.
The solutions are given in assuming that $A=\ln \big(\cos(by)\big)$.
Note that in this ansatz there are naked singularities at
$y=\pm y^{\ast}:=\pm \pi/(2b)$.

The de Sitter brane solutions are obtained as follows:
For $m=2n$ where $n=0,1,2,\cdots$,
\begin{eqnarray}
\phi(y)=
\big[(n+1)\beta {\rm arcsinh}(\tan (by)) \big]^{1/(n+1)},
\end{eqnarray}
where $\beta=\frac{1}{b}\sqrt{(3/2)(b^2-\Lambda)}$, for the potential
\begin{equation}
V(\phi)=3b^2
    -\frac{9}{4}\big(b^2-\Lambda\big)\cosh^2
\left[\frac{\phi^{n+1}}{(n+1)\beta}\right].
\end{equation}
Note that for an odd $n$ the solution of $\phi(y)$ is invalid for
a negative $y$,
while for an odd $n$ the solution of $\phi(y)$ is valid everywhere
in $-y^{\ast}<y<y^{\ast}$.
For $m=2n+1$ where $n=0,1,2,\cdots$, the de Sitter brane solutions are given by
\begin{eqnarray}
\phi(y)=
\big[(n+\frac{1}{2})\beta {\rm arcsinh}(\tan (by)) \big]^{2/(2n+1)},
\end{eqnarray}
where $\beta=\frac{1}{b}\sqrt{(3/2)(b^2-\Lambda)}$, for the potential
\begin{equation}
V(\phi)=3b^2
    -\frac{9}{4}\big(b^2-\Lambda\big)
\cosh^2
\left[\frac{2\phi^{(2n+1)/2}}{(2n+1)\beta}\right].
\end{equation}
The solution is invalid for a negative $\phi$.
The de Sitter brane solutions exist only for $0<\Lambda<b^2$.
The AdS brane solutions are obtained by reversing the sign of $\Lambda$.
Then, $\Lambda$ must take a value in the range of $-b^2<\Lambda<0$.

The case of $m=0$ recovers the solutions
{with naked singularities in the bulk}
obtained in the case of a
canonical scalar field
\cite{grem2}.

\subsubsection{Non-minimally coupled scalar field}

Ref. \cite{flat1} considered thick brane solutions of
a scalar field non-minimally coupled to the scalar curvature:
\beq
S=\int d^5 x\sqrt{-G}
\Big[f(\phi) R-\frac{1}{2}\Big(\partial \phi\Big)^2
-V(\phi)
\Big]\,,
\eeq
where $f(\phi)$ denotes the direct coupling of the scalar field
to gravity
and the five-dimensional metric $G_{ab}$ has the signature $(-,+,+,+,+)$.
The above action is conformally related to the Einstein frame
action with the Ricci scalar term ($2M^3 R$)
via the conformal transformation $G_{ab}\to {\tilde G}_{ab} f(\phi)/(2M^3)$,
where $M$ is the five-dimensional gravitational scale.
The metric ansatz was taken to be
\beq
ds^2= e^{2A}\eta_{\mu\nu}dx^{\mu}dx^{\nu}+dy^2 .
\eeq
The coupling function $f(\phi)$ was chosen to be
$f(\phi)= 2M^3-\frac{1}{2}\xi \phi^2$,
where the cases of $\xi=0$ and $\xi=3/16$
represent the minimal and
conformal couplings, respectively.
The Einstein equations are given by
\beq
&&V(\phi)
=-\frac{3}{2}\Big(2M^3-\frac{\xi}{2}\phi^2\Big)
    \Big(2\dot{A}^2+\ddot{A}\Big)
+\frac{7}{2}\xi \dot{A}\dot{\phi}\phi
+\xi \dot{\xi}^2\,,
\nonumber\\
&&
\frac{1}{2}\dot{\phi}^2
=-\frac{3}{2}\Big(2M^3-\frac{\xi}{2}\dot{\phi}^2\Big)\ddot{A}
 -\frac{\xi}{2}\dot{A}\dot{\phi}\phi
 +\xi \ddot{\phi}\phi
 +\xi \dot{\phi}^2\,.
\eeq
In the examples below, the scalar field potential is obtained
once the scalar and metric profiles are found.

In the simplest case, $\xi=0$,
the solution was given by
\beq
&&\phi(y)=\phi_0 \tanh (ay),
\quad
e^{A}
=\big(\cosh (ay)\big)^{-\gamma}
e^{-\gamma \tanh^2(ay)/4},
\eeq
where $\gamma=\phi_0^2/(9M^3)$.
For a non-zero coupling constant $\xi\neq 0$, by choosing
\beq
e^{A(y)}=\big(\cosh (ay)\big)^{-\gamma}\label{mia}
\eeq
the solution $\phi(y)=\phi_0 \tanh (ay)$ was obtained
with
\beq
\gamma=2\Big(\frac{1}{\xi}-6\Big),
\quad
\phi_0
=a^{-1}\dot{\phi}(0)
=\big(2M^3\big)^{1/2}
\sqrt{\frac{6(1-6\xi)}{\xi(1-2\xi)}}.
\eeq
The above solution exists for $0<\xi<1/6$.
The scalar curvature of this solution
approaches $R\to -20 a^2\big(\xi^{-1}-6\big)^2$ for $y\to \infty$,
namely the bulk is asymptotically AdS.
Eq. (\ref{mia}) is also the metric function of the solution whose
scalar field profile is given by
$
\phi(y)=\phi(0)\big(\cosh (ay)\big)^{-1},
$
with $\phi^2(0)=12M^3(\xi^{-1}-6)/(3-16\xi)$.

Another solution with
$\phi(y)=\phi_0 \tanh (ay)$,
where $\phi_0=2\sqrt{\xi^{-1}M^3}$,
and
\beq
A(y)=-4\ln \Big(\cosh (ay)\Big)
+\frac{1}{3}\big(8-\xi^{-1}\big)
\tanh^2(ay)
F_{PFQ}\Big(\Big\{1,1,7/6\Big\},\Big\{3/2,2\Big\},\tanh^2(ay)\Big)\,,
\eeq
where $F_{PFQ}$ represents the generalized hypergeometric function,
was also found.
Note that in all the above nontrivial solutions
 there is no smooth limit to the minimally coupled case
$\xi= 0$.

\subsubsection{Phantom scalar field}

Ref. \cite{flat6} considered a model of a bulk phantom scalar field:
\beq
S=\int d^5 x\sqrt{-g}
\Big[
\frac{1}{2\kappa_5^2}
\Big(R-2\Lambda\Big)
+\frac{1}{2}\big(\nabla \phi\big)^2
-V\big(\phi\big)
\Big],
\eeq
where $g_{ab}$ is the five dimensional metric with signature $(-,+,+,+,+)$
and $V(\phi)$ is the potential of the scalar field $\phi$.
Note that sign of the kinetic term is now negative, i.e, {\it phantom}.
The thick brane solutions for the phantom scalar field with the ansatz
\beq
ds^2=dy^2+e^{-2f(y)}\eta_{\mu\nu}dx^{\mu}dx^{\nu}
\eeq
and $\phi=\phi(y)$ are considered.
The equations of motion are given by
\beq
&&f''=-a \big(\phi'\big)^2,
\nonumber\\
&&
\big(f'\big)^2
=\frac{a}{4}\Big(-(\phi')^2-2V\Big)-\frac{\Lambda}{6}\,,
\nonumber\\
&&\phi''-\frac{4f'}{f}\phi'=-\frac{dV}{d\phi}\,,
\eeq
where $a:=\kappa_5^2/3$.
An explicit solution was obtained
for the sine-Gordon potential
\beq
V(\phi)=B\Big(1+\cos\big(\frac{2\phi}{A}\big)\Big)
\eeq
as
\beq
&&f(y)=-\frac{a}{\kappa_1}\sqrt{\frac{|\Lambda|}{6}}
\ln \cosh\Big(\frac{\kappa_1}{a}y\Big),
\nonumber\\
&&
\phi(y)
=2A \arctan \Big[{\rm exp} \Big(\frac{\kappa_1}{a}y\Big)\Big]
-\frac{\pi A}{2},
\eeq
where $\kappa_1:=\big(1/A^2\big)\sqrt{|\Lambda|/6}$.
Here the constant $B$ has to be chosen as
$
B=\big(|\Lambda|/6a^2\big)\Big(a-1/(4A^2)\Big)
$.
The solution has a growing warp factor for the larger value of $|y|$,
in contrast to the case of the normal kinetic term.

In \cite{Gogberashvili:2009yp},
by using a massless
scalar phantom/ghost field,
an analytic time-dependent warped solution
forming a standing wave in the bulk was found.
The nodes of the standing wave are interpreted as
the different four-dimensional Minkowski vacua, called {\it islands} in Ref. \cite{Gogberashvili:2009yp}, having different physical parameters
such as gravitational and cosmological constants.
This model possesses the following characteristic features:
1) The ordinary four-dimensional matter can be localized on the tensionless branes, i.e. on the nodes.
2) It is possible to realize a new gravitational localization mechanism when matter fields can reside only on Minkowski {\it islands}, where
the background space-time does not oscillate.

\subsubsection{Tachyon field}

As a final example of a generalized scalar field theory,
consider a model of a tachyon field \cite{tac}whose action is given by
\beq
S=
\int d^5 x\sqrt{-g}
\Big[\frac{1}{2\kappa_5^2}
\Big(R-2\Lambda\Big)
+V(T)\sqrt{1+g^{MN}\partial_M T\partial_N T}
\Big]
\,,
\eeq
where $\Lambda_5$
and $V(T)$
are the bulk cosmological constant 
and the tachyon potential, respectively.
$g$ is the determinant of the five-dimensional metric
$g_{ab}$ whose signature is $(-,+,+,+,+)$.
It is assumed that
the metric and tachyon field take the following form
\beq
ds^2=dy^2+e^{2f(y)}\eta_{\mu\nu}dx^{\mu}dx^{\nu}
\eeq
and $T=T(y)$.

A thick brane solution is obtained
for non-zero cosmological constant $\Lambda_5\neq 0$.
The potential is explicitly given as a function of $y$, as
\beq
V(y)
=\frac{1}{\kappa_5^2}
 \frac{1}{\cosh^2 \big(b y\big)}
\sqrt{\big(\Lambda_5+6b^2\big)\sinh^2(by)
+\Lambda_5+6(b^2-H^2)
}
\sqrt{\big(\Lambda_5+6b^2\big)\sinh^2(by)+(\Lambda_5-6H^2)}\,,
\eeq
where $b$ is an arbitrary constant.
The warp and tachyon functions are given by
\beq
&&f(y)=\ln \cosh (b y),
\nonumber\\
&&
T(y)=-\frac{i}{b}
\sqrt{\frac{3(b^2+H^2)}{\Lambda_5-6H^2}}
{\rm Elliptic}
F\Big[
iby,\frac{\Lambda_6+6b^2}{\Lambda-6H^2}
\Big]\,.
\eeq
The warp factor grows for large $y$.

\subsection{Other types of thick branes}

\subsubsection{Multi-scalar}

The brane models with two interacting scalar fields considered in section \ref{two_scalar_field}
refer to the so-called  non-topological solutions in the terminology of Ref. \cite{rajaraman}. At the same time, there
exist other types of solutions - topological ones. These solutions describe our 4-dimensional Universe as a
global topological defect in a multidimensional spacetime. Such a topological defect can be created by a
set of scalar fields $\phi^a$ describing some ``hedgehog'' configuration \cite{Olasagasti:2000gx,Olasagasti1,Ghergetta1}.

The main idea of the research in this direction is illustrated by the example of
the paper \cite{Olasagasti:2000gx} where the $(p-1)$-brane solutions (the number $p$ refers to the coordinates on the brane) are considered.
It is assumed that there is a global defect in $n$ extra dimensions
which is described by a multiplet of $n$ scalar fields, $\phi^a$,
whose Lagrangian is~\footnote{In the original paper \cite{Olasagasti:2000gx}, the sign of the kinetic term is different but it is wrong.}
\begin{equation}
\label{Lagr_mult_fields}
L=-\frac{1}{2}\partial_A\phi^a\partial^A\phi^a-V(\phi)~,
\end{equation}
where the potential $V(\phi)$ has its minimum on the $n$-sphere $\phi^a\phi^a=\eta^2$.  As an
example, the potential
\beq
\label{pot_Mex}
V(\phi)=\frac{\lambda}{4}\left(\phi^a\phi^a-\eta^2\right)^2
\eeq
was used. Using the following {\it ansatz} for the metric:
\beq
ds^2=A(\xi)^2 d\xi^2+\xi^2 d\Omega^2_{n-1}
+B(\xi)^2 {\hat g}_{\mu \nu} dx^{\mu} dx^{\nu},
\label{ansatz}
\eeq
where $d\Omega^2_{m}$ stands for the metric on the
 unit $m$-sphere, and
the spherical coordinates in the extra dimensions are defined by the
usual relations, $\xi^a=\{\xi \cos\theta_1, ~\xi \sin\theta_1\cos\theta_2,
... \}$.

Using a different metric {\it ansatz}, a few different types of solutions were found. The first type, without a
cosmological constant, exists for all $n\geq 3$, and is  very similar to the global monopole
solutions from \cite{Barriola}.  The brane worldsheet is flat, and there is
a solid angle deficit in the extra dimensions.

For a positive cosmological constant, $\Lambda > 0$, the solutions describe spherical branes in an
inflating multidimensional universe. In the limit $\eta\rightarrow 0$, when the gravitational effect of the
defect can be neglected, the universe can be pictured as an expanding $(p+n-1)$-dimensional
sphere with a brane
wrapping around it in the form of a sphere of lower dimensionality $(p-1)$.
Another class of solutions has curvature singularities even in the absence of a defect
($\eta = 0$), and these solutions were considered as unphysical. Finally, there are
solutions having the geometry of a $(p + 1)$-dimensional de Sitter space,
with the remaining $(n - 1)$ dimensions having the geometry of a cylinder.

For a negative cosmological constant, $\Lambda < 0$, three classes of solutions were found.
The first two are essentially analytic
continuations of the positive-$\Lambda$ solutions.  The third class is similar to the Randall-Sundrum
($n = 1$) and the Gregory ($n = 2$) solutions, and exhibit an exponential warp factor.

In \cite{Ghergetta1} a model of a global topological defect with the potential \eqref{pot_Mex}
was considered. The authors 
were looking for solutions decaying far from the brane as $B(\xi)\sim\exp(-c~\xi)$.
It was shown that there exist solutions with the spherical radius $R\rightarrow const$ at large $\xi$,
i.e., the extra dimensions form an $n$-dimensional cylinder $R_+\times S^{n-1}$. Such solutions exist only
for $n\geq 3$.

The branes considered above refer to Minkowski branes with some non-zero bulk cosmological constant.
In the paper \cite{Delsate} a model with some extra 4-dimensional $\Lambda$-term is considered
(i.e. a de Sitter brane). This work investigated global monopole solutions  in the Goldstone model
with $n$ scalar fields with the Lagrangian \eqref{Lagr_mult_fields}.
Within the framework of this model, it was shown that new types of solutions exist
(mirror symmetric solutions).  These solutions have periodic matter and metric functions, and require a
fine tuning of two cosmological constants.

\subsubsection{Vortex}

If one replaces the domain wall with some more complicated  topological defect, such as
string-vortices or monopoles (see above), new possibilities
for gravitational localization to the brane appear. Braneworld models using a 6-dimensional vortex
were proposed in Ref. \cite{Giovannini:2001hh}. In this model gravity was localized on an
Abrikosov-Nielsen-Olsen vortex in the context of the Abelian Higgs model in 6 dimensions.
In this model, the conservation of a topological charge guarantees the stability of the configuration.
In principle the presence of gravity can change the situation, but
one can hope that ``gravitating'' topological defects can provide potential
candidates for stable configurations.

For this purpose, the total action of the gravitating Abelian Higgs model in six dimensions
can be chosen as
\begin{equation}
S= S_{\rm brane} + S_{\rm grav}~,
\label{total}
\end{equation}
where $S_{\rm brane}$ is the gauge-Higgs action and $S_{\rm grav}$
is the six-dimensional generalization of the Einstein-Hilbert action.
More specifically,
\begin{equation}
S_{\rm brane}=\int
d^6x\sqrt{-G}{\cal L}_{\rm brane},~~
{\cal L}_{\rm brane}=
\frac{1}{2}({\cal D}_{A}\phi)^*{\cal D}^{A}\phi-\frac{1}{4}
F_{AB}F^{AB}
-\frac{\lambda}{4}\left(\phi^*\phi-v^2\right)^2~,
\label{a1}
\end{equation}
where ${\cal D}_{A}=\nabla_{A}-ieA_{A}$ is the gauge covariant derivative,
while $\nabla_{A}$ is the generally covariant derivative, and the bulk action is
\begin{equation}
	S_{\rm grav}=-\int d^6x \sqrt{-G}
	\left( \frac{R}{2\chi}+\Lambda\right),
\label{a2}
\end{equation}
where $\Lambda$ is a bulk cosmological constant, $\chi = 8\pi
G_6=8\pi/M_6^4$  and $M_{6}$ denotes the six-dimensional Planck
mass. In Eq. (\ref{a1}), $v$ is the vacuum expectation value
of the Higgs field which
determines the masses of the Higgs boson and of the gauge boson
\begin{equation}
m_{H} = \sqrt{2 \lambda}\, v,\,\,\,m_{V} = e\, v.
\end{equation}
Choosing the six-dimensional metric as
\begin{eqnarray}
ds^2=G_{AB} dx^{A} dx^{B} =
M^2(\rho)g_{\mu\nu}dx^\mu dx^\nu-d\rho^2-L(\rho)^2d\theta^2,
\label{metric_vortex}
\end{eqnarray}
where $\rho$ and $\theta$ are, respectively,  the bulk radius and the
bulk angle, $g_{\mu\nu}$ is the four-dimensional  metric and
$M(\rho)$, $L(\rho)$ are the warp factors. The Abrikosov-Nielsen-Olsen ansatz
for the gauge-Higgs system reads:
\begin{eqnarray}
&& \phi(\rho,\theta) =vf(\rho)e^{i\,n\,\theta},
\nonumber\\
&&A_{\theta}(\rho,\theta)  =\frac{1}{e}[\,n\,-\,P(\rho)] ~,
\label{NO}
\end{eqnarray}
where $n$ is the winding number.

Choosing the corresponding regular boundary conditions, one can find that
outside the core of the string all source terms vanish and a general solution to
the equations can be easily found for the case $\Lambda \leq
0$ (solutions for the case of a
positive cosmological constant in the bulk were studied in
\cite{m2}) as follows
\begin{equation}
m(x)=- c~\frac{1-\epsilon\,e^{ 5 c x }}
{1+\epsilon e^{ 5 c x}}\,,
\label{mm}
\end{equation}
where $\epsilon$ is an integration constant; $c = \sqrt{- \mu/10} > 0$; $\mu$ is some combination
of the 6-dimensional Einstein gravitational constant, $\Lambda$-term and the mass of the Higgs;
$m(x)$ and $x$ are the rescaled variables $M(\rho)$ and $\rho$, respectively.
Once $m(x)$ is known, the function
$\ell(x)$ (the rescaled $L(\rho)$) can also be determined.

To clarify the question about  gravitational localization on the brane, one can
demand the finiteness of the four-dimensional Planck mass
\begin{equation}
M_{P}^2 = \frac{4 \pi M^4_{6}}{m_{H}^2} \int dx M^2(x) {\cal L}(x)
< \infty~.
\label{4dpl}
\end{equation}
It was shown in \cite{Giovannini:2001hh} that
if $\epsilon > 0$ or $\epsilon \leq -1$, then for $x \rightarrow
\infty$ one has $M\sim {\cal L} \sim e^{cx}$ so that the integral
in (\ref{4dpl}) diverges and gravity cannot be localized.

If $\epsilon=0$ the solution is simply
\begin{equation}
m(x) = \ell(x) = -c,
\label{eps=0}
\end{equation}
and the warp factors
will be exponentially  decreasing as a function of the bulk radius:
\begin{eqnarray}
&& M(x) = M_0 e^{-c x} ,\,\,\,
\nonumber\\
&& {\cal L}(x) = {\cal L}_0 e^{-c x}.
\label{LM}
\end{eqnarray}
The solution of Eqs. (\ref{eps=0})--(\ref{LM})
leads to the localization of gravity and to a
smooth AdS geometry far from the string core.

If $-1<\epsilon < 0$, $M(x_0)=0$ where $x_0=
\frac{1}{5c}\log{1/|\epsilon}|$. The geometry is singular at
$x=x_0$ thus  $x<x_0$ should be required. In spite of the fact
that ${\cal L}$ diverges at $x=x_0$, the integral (\ref{4dpl})
defining the four-dimensional Planck mass is finite. So, these geometries
can be potentially used for the localization of gravity, provided the
singularity at $x_0$ is resolved in some way, e.g. by string theory.
These types of singular solutions were discussed in \cite{m2} for the case
of positive cosmological constant.

Finally, if the  bulk cosmological constant is zero,  the solutions have a
power-law behavior,  namely
\begin{equation}
M(x) \sim x^{\gamma},\,\,\,{\cal L}(x) \sim x^{\delta},
\end{equation}
with
\begin{equation}
d \gamma + \delta=1, \,\,\,\, d \gamma^2 + \delta^2 =1 \, ,
\end{equation}
where $d=4$ is the number of dimensions of the metric $g_{\mu\nu}$.
These solutions belong to the  Kasner class. The
Kasner conditions leave open only two possibilities: either $\delta =
1$ and $\gamma = 0$ or $\gamma= 2/5$ and $\delta = -3/5$. None of them
lead to the localization of gravity.

Summarizing the results of the paper \cite{Giovannini:2001hh},
it was shown that the localization of gravity is possible on a ``thick''
string, and a fine-tuning condition was found which led to a set of
physically interesting solutions. Since the described geometries  are regular,
gravity can be described in classical terms both in the  bulk and on
the vortex.

Similar considerations allowed the authors of \cite{RandjbarDaemi:2003qd} to conclude
that within the framework of  an anomaly-free Abelian Higgs model coupled to gravity in a
6-dimensional spacetime, it is possible to construct an effective $D = 4$ electrodynamics of charged particles
interacting with photons and gravitons. In this work both  gauge field and gravity are localized
near the core of an Abrikosov-Nielsen-Olsen vortex configuration.

Further considerations of questions concerning localization of  gauge fields on a gravitating  vortex in a 6-dimensional spacetime can be found in the papers \cite{Giovannini:2002,RandjbarDaemi:2002}. Note that most of the calculations are performed numerically or asymptotically, and there are no general analytic solutions for 6-dimensional vortices. However, in order to address some questions (for example, stability analysis) the availability of exact  solutions is desirable. In Ref. \cite{TorrealbaS.:2008zi} some exact solutions for 6-dimensional vortices were found. In particular, using the special boundary conditions, the several kink solutions were found.
The first solution was
\begin{equation}
	f=f_{0}\arctan \left( \sinh \frac{\beta r}{\delta}
	\right) ,\qquad f_{0}=2\sqrt{%
	\delta },
\label{ffo}
\end{equation}%
with the corresponding metric ``warp'' factor
\begin{equation}
	M(r)=\cosh ^{-\delta }\left( \frac{\beta r}{\delta} \right),
	\qquad \delta >0,\qquad \beta >0,
\label{Mgrande}
\end{equation}
 where $\delta $ could be seen as the
wall thickness and $\beta $ as a parameter depending on the cosmological
constant. The effective potential for this case was
\begin{equation}
V(f)=\frac{2\beta ^{2}}{\delta }[(1+5\delta )\cos ^{2}(f/f_{o})-5\delta ].
\label{VdeF}
\end{equation}

The second and third solutions were, respectively
\begin{eqnarray}
	f &=&f_{0}\tanh \left( a r\right) ,\qquad \text{and \ }
\label{ftanh1} \\
	\text{\ }f &=&f_{0}\arctan \left( a r\right) ,
\label{ftanh2}
\end{eqnarray}%
with the metric ``warp" factors, respectively
\begin{eqnarray}
	M(r) &=&\exp \left[\frac{1}{24}f_{0}^{2} \;
	\mathrm{sech} (a r) \right]\cosh^{-f_{o}^{2}/6}(a r),
	\qquad \text{and  }
\label{Mtan1} \\
	M(r) &=&\exp \left[-\frac{a}{2}f_{0}^{2} r\arctan (a r) \right],
\label{Mtan2}
\end{eqnarray}
and with a  polynomial potential
identical to the Mexican hat potential, or a potential containing polynomial and
non-polynomial terms
\begin{eqnarray}
	V(f) &=&\frac{a^{2}}{2} \left[ \left( f^{2}-f_{o}^{2} \right)^{2}-
	\frac{5}{9}	f^{2} \left(f^{2}-3f_{o}^{2} \right) \right],\text{and}
\label{VdF2} \\
	V(f) &=&\frac{a^{2}f_{o}^{2}}{2} \left\{
	\cos ^{4} \left(\frac{f}{f_{o}} \right)-\frac{5}{16}
		f_{o}^{2} \left[\frac{f}{f_{o}} +
		\frac{1}{2}\sin \left(\frac{f}{f_{o}} \right) \right]
	\right\}^{2}.
\label{Vdf3}
\end{eqnarray}
Using these exact solutions, it was shown that the zero mode of the linearized gravity spectrum is
localized on the 3-brane and 
the massive modes are not bounded at all.

\subsubsection{Weyl gravity model}

In Refs.~\cite{ariasetal,Barbosa-Cendejas:2005kn,Barbosa-Cendejas:2006bh}, thick brane solutions in a
pure geometric Weyl 5D spacetime, which constitutes a non-Riemannian generalization of Kaluza-Klein gravity, are considered.
A Weyl geometry is an affine manifold specified by $(g_{MN},\omega_M)$ with $M,N=0,1,2,3,5$,
where $g_{MN}$ is the metric tensor, and $\omega_M$ is a ``gauge" vector involved in the definitions of the affine connections of the manifold.

Besides the scalar curvature $R$, the action used in this approach also contains terms describing
a Weyl scalar $\omega$
\begin{equation}
	\label{action_pg} S_5^W =\int\limits_{M_5^W}\frac{d^5x\sqrt{|g|}}{16\pi
	G_5}e^{\frac{3}{2}\omega}[R+3\tilde{\xi}(\nabla\omega)^2+6U(\omega)],
\end{equation}
where $M_5^W$ is a Weyl manifold specified by the pair $(g_{MN},\omega)$. The Weylian Ricci tensor reads
\begin{equation}
R_{MN}=\Gamma_{MN,A}^A-\Gamma_{AM,N}^A+\Gamma_{MN}^P\Gamma_{PQ}^Q-\Gamma_{MQ}^P\Gamma_{NP}^Q,
\end{equation}
where
\begin{equation}
	\Gamma_{MN}^C=\{_{MN}^{\;C}\}-\frac{1}{2}\left(
	\omega_{,M}\delta_N^C+\omega_{,N}\delta_M^C-g_{MN}\omega^{,C}\right)
\end{equation}
are the affine connections on $M_5^W$, $\{_{MN}^{\;C}\}$ are the Christoffel symbols and $M,N=0,1,2,3,5$; the constant $\tilde{\xi}$ is an arbitrary coupling parameter, and $U(\omega)$ is a self--interaction potential for the scalar field $\omega$.

The 5-dimensional metric describing a thick brane is assumed to have the form
\begin{equation}
	ds_5^2=e^{2A(y)}\eta_{mn}dx^m dx^n+dy^2,
\label{line}
\end{equation}
where $e^{2A(y)}$ is the warp factor depending on the extra coordinate $y$, and $m,n=0,1,2,3$, $\eta_{mn}$ is the 4D Minkowski metric.

Further, the conformal transformation $\widehat{g}_{MN}=e^{\omega}g_{MN}$, mapping the Weylian action
(\ref{action_pg}) into the Riemannian one is carried out via
\begin{equation}
	S_5^R=\int\limits_{M_5^R}\frac{d^5x\sqrt{|\widehat
	g|}}{16\pi G_5} \left[
		\widehat R+3{\xi}(\widehat\nabla\omega)^2+6\widehat	U(\omega)
	\right],
\label{confaction}
\end{equation}
where $\xi=\tilde{\xi}-1$, $\ \widehat U(\omega)=e^{-\omega}
U(\omega)$ and all hatted magnitudes and operators are defined in the Riemann frame.
Now the metric \eqref{line} takes the form
\begin{equation}
	\widehat{ds}_5^2=e^{2\sigma(y)}\eta_{nm}dx^n
	dx^m+e^{\omega(y)}dy^2,
\label{conflinee} \end{equation}
where $2\sigma=2A+\omega$.
If one introduces new functions
$X=\omega'$ and $Y=2A'$ \cite{ariasetal} then the Einstein equations read
\begin{eqnarray}
	X'+2YX+\frac{3}{2}X^2=\frac{1}{\xi}\frac{d\widehat
	U}{d\omega}e^{\omega},
\label{fielde1} \\
	Y'+2Y^2+\frac{3}{2}XY=\left(-\frac{1}{\xi}\frac{d\widehat
	U}{d\omega}+4\widehat U\right)e^{\omega}.
\label{fielde2}
\end{eqnarray}
As pointed out in \cite{ariasetal}, this system of equations can be easily solved if one uses the condition $X=kY$, where $k$ is an arbitrary constant parameter, but excluding the value $k=1$.

In this case both field equations in \eqref{fielde1} \eqref{fielde2} reduce to a
single differential equation
\begin{equation}
\label{finale}
	Y'+\frac{4+3k}{2}Y^2=\frac{4\lambda}{1+k}e^{(\frac{4k\xi}{1+k}+1)\omega},
\end{equation}
where $\lambda$ is some constant parameter.
The authors  of \cite{ariasetal} first considered the problem for a case of $Z_2$-symmetric manifolds.
Here, the two following cases are considered:

\begin{description}

\item
(i) $\xi=-(1+k)/(4k)$ and leaving $k$ arbitrary except for $k=-4/3$,

\item
(ii) $k=-4/3$ and leaving $\xi$ arbitrary except for $\xi=-1/16$.

\end{description}

\paragraph*{The case (i)}
For this case,
a solution was found for the
simplified \eqref{finale} in the form
\begin{eqnarray}
	\omega(y)&=&b k \ln[\cosh(ay)],\\
	e^{2A(y)} &=&[\cosh(ay)]^b,
\label{sol1}
\end{eqnarray}
where
$$
a=\sqrt{\frac{4+3k}{1+k}2\lambda}, \quad b=\frac{2}{4+3k}.
$$

\paragraph*{The case (ii)}
The authors of \cite{Barbosa-Cendejas:2005kn,Barbosa-Cendejas:2006bh} considered another simplified case with
$k=-4/3$ and $\xi$ an arbitrary parameter.
Then Eq.~\eqref{finale} takes the form
\begin{equation}
	Y'+12\lambda e^{p\omega}=0 \qquad \mbox{\rm or}\qquad
	\omega''-16\lambda e^{p\omega}=0,
\label{diffeqw}
\end{equation}
where $p=1+16\xi$.
Note that the choice $\xi=-1/16$ leads to a constant potential and
thus the case $\xi\neq -1/16$ is assumed.
For such a choice of the parameter,
one can find the following solution for $\omega$ and $A$
\begin{eqnarray}
	\omega &=& -\frac{2}{p}\ln\left\{\frac{\sqrt{-8\lambda
	p}}{c_1}\cosh\left[c_1(y-c_2)\right]\right\},
\label{pairsolut1} \\
	e^{2A} &=& \left\{\frac{\sqrt{-8\lambda
	p}}{c_1}\cosh\left[c_1(y-c_2)\right]\right\}^{\frac{3}{2p}}.
\label{pairsolut2}
\end{eqnarray}
The following physically interesting problems were considered in detail:
\begin{itemize}
	\item $\lambda>0$, $p<0$, $c_1>0$. In this case the solution is a thick brane, and the fifth coordinate changes within the limits
$-\infty<y<\infty$.
	\item $\lambda>0$, $p>0$, $c_1=iq_1$. In this case the solution describes a compact manifold along the extra dimension with $-\pi\le q_1(y-c_2)\le\pi$ and consequently this solution is not a thick brane.
\end{itemize}

If one calculates the function
 $\omega'(y)$ for both the cases
(i) 
and
(ii), 
then one has
\begin{eqnarray}
	\omega'(y) &=& ab \tanh(ay), \qquad {\rm \,\, (i)}
\label{topl1} \\
	\omega'(y) &=& -\frac{2 c_1}{p} \tanh\left[ c_1 \left( y-c_2 \right) \right],\qquad {\rm \,\, (ii)}
\label{topl2}
\end{eqnarray}
respectively.
As one can see, both solutions are kink-like. Thus these solutions are topologically nontrivial solutions.

The de Sitter thick brane solution in Weyl gravity
was investigated in Ref. \cite{dS_Weyl}.
To show the analytic form of the solution,
it is convenient to transform the fifth coordinate
to the conformal coordinate $z$, defined by $dz=e^{-A(z)}dy$,
\begin{equation}
	ds_5^2=e^{2A(z)}\left(-dt^2+e^{2\beta t}d{\bf x}^2+dz^2\right),
\label{line_ds}
\end{equation}
where $\beta$ is the expansion rate of the de Sitter spacetime.
For the potential
\begin{equation}
U(\omega)=\frac{1+3\delta}{2\delta}\beta^2 e^{-\omega}
\Big(\cos\frac{\omega}{\omega_0}\Big)^{2(1-\delta)},
\end{equation}
where $\omega_0=\sqrt{3\delta(1-\delta)}$ ($0<\delta<1$),
the solutions are given by
\begin{eqnarray}
\omega=\omega_0\arctan \left(\sinh \frac{\beta z}{\delta}\right),\quad
e^{2A}=\cosh^{-2\delta}\Big(\frac{\beta z}{\delta}\Big)
e^{2\omega_0\arctan \big(\frac{\beta z}{\delta}\big)}.
\end{eqnarray}
After transforming back to the proper coordinate $y$,
the geometry is asymmetric across the $y=0$ surface.
But the solution still has a kink-like profile and connects two Minkowski vacua.

\subsubsection{Bloch brane}

It is possible to borrow ideas from condensed matter physics and apply them to
brane world models. Of interest is the situation which occurs in ferromagnetic systems,
in which one has an Ising or Bloch type wall.
An Ising wall is a simple interface with no internal structure
whereas a Bloch wall is an interface which has a nontrivial
internal structure. One can find thick brane solutions with internal structures
like that of the Bloch wall.

In the paper \cite{Bazeia:2004dh},
the thick brane solution with two scalar fields $(\phi, \chi)$,
called a Bloch brane, was considered.
The solution has an internal structure because of the dependence on $\chi$.
For this case, the action is taken to have the form
\begin{equation}
	S_c=\int d^4x\,dy\sqrt{|g|}\left[ -\frac14 R+
	\frac12\partial_a\phi\partial^a\phi+\frac12\partial_a\chi\partial^a\chi-
	V(\phi,\chi)
	\right].
\end {equation}
The thick brane metric is
\begin{equation}
	ds^2=g_{ab}dx^adx^b=e^{2A(y)}\eta_{\mu\nu}dx^{\mu}dx^{\nu}-dy^2,
\end {equation}
where $a,b=0,1,2,3,4,$ and $e^{2A}$ is the warp factor. $x^4=y$ is the extra dimension coordinate. The potential is given by
\begin{eqnarray}
	V_c(\phi,\chi) &=& \frac18 \left[\left(\frac{\partial
	W_c}{\partial\phi}\right)^2 + \left(\frac{\partial
	W_c}{\partial\chi}\right)^2 \right]-\frac13 W^2_c ,
\label{gpot}\\
	W_c &=& 2\left( \phi-\frac13\phi^3-r\phi\chi^2 \right),
\label{gpot2}
\end{eqnarray}
where
$r$ is a constant. It is readily seen that this potential is unbounded below
at $\chi=const$
\begin{equation}
	V_c(\phi,\chi)
	\stackrel{\left| \phi \right| \rightarrow \infty}{\longrightarrow}
	- \infty .
\end{equation}
The corresponding equations describing two gravitating scalar fields are
\begin{eqnarray}
	\phi^{\prime\prime}+4A^\prime\phi^\prime&=& \frac{\partial
	V(\phi,\chi)}{\partial\phi},
\\
	\chi^{\prime\prime}+4A^\prime\chi^\prime&=& \frac{\partial
	V(\phi,\chi)}{\partial\chi},
\\
	A^{\prime\prime}&=&-\frac23\,\left(\phi^{\prime2}+\chi^{\prime2}
\right),
\\
	A^{\prime2}&=&\frac16\left(\phi^{\prime2} +
	\chi^{\prime2}\right)-\frac13 V(\phi,\chi),
\end{eqnarray}
where prime stands for derivative with respect to $y$.
One can reduce these equations to the form
\begin{eqnarray}
	\phi^{\prime} &=& \frac12\,\frac{\partial W_c}{\partial\phi},
\\
	\chi^{\prime} &=& \frac12\,\frac{\partial W_c}{\partial\chi},
\\
	A^\prime &=& -\frac13\,W_c. \label{blocheq}
\end{eqnarray}
\paragraph{Bloch walls}
The solution of these equations is
\begin{eqnarray}
	\phi(y)&=&\tanh(2ry),
\label{phi}\\
	\chi(y)&=&\pm\sqrt{\frac1r-2\,}\;{\rm sech}(2ry).
\label{chi}\\
	A(y) &=& \frac1{9r}\left[(1-3r)\tanh^2(2ry)-2\ln\cosh(2ry)\right].
\label{A}
\end{eqnarray}
The brane has a typical width as large as $1/r$.
In the case $r<1/2$,
the two-field solution represents a Bloch-type brane,
where $(\phi,\chi)$ represents an elliptic trajectory in the field space.
For $r>1/2$, the two-field solution is changed to a one-field solution,
i.e., an Ising-type brane, connecting two minima $(\phi,\chi)=(\pm 1,0)$
straight along the line $\chi=0$ in the field space.
Equation \eqref{phi} shows that this solution is a topologically nontrivial one.

\paragraph{General method}
In Ref. \cite{dfh},
the method to obtain more general (often numerical) Bloch brane solutions
was given. The starting point was the general class of the potential
\beq
 W_c(\phi,\chi)
=2\phi\Big[\lambda\Big(\frac{\phi^2}{3}-a^2\Big)
   +\mu \chi^2
          \Big]\,,
\eeq
which includes the previous case, for the choice
that $a=1$, $\lambda=-1$ and $\mu=-r$.
From Eq. (\ref{blocheq}), $d\phi/d\chi=W_{c,\phi}/W_{c,\chi}$,
where $W_{c,\phi}:= \partial W_c/\partial \phi$,
one obtains
\beq
\frac{d\phi}{d\chi}
=\frac{\lambda(\phi^2-a^2)+\mu \chi^2}
      {2\mu\phi\chi}\,.\label{pc}
\eeq
Introducing the new variable $\rho=\phi^2-a^2$, one can replace
Eq. (\ref{pc}) with an ordinary differential equation for $\rho$
with respect to $\chi$, which can be analytically solved as
\beq
&&\rho(\chi)=\phi^2-a^2
=c_0 \chi^{\lambda/\mu}
          -\frac{\mu}{\lambda-2\mu}\chi^2,\quad (\lambda\neq 2\mu),
\nonumber\\
&&\rho(\chi)=\phi^2-a^2
=\chi^{2}
          \Big(\ln (\chi)+ c_1\Big),
\quad (\lambda= 2\mu),
\label{rho_r}
\eeq
where $c_0$ and $c_1$ are integration constants.
Substituting Eq. (\ref{rho_r}) into the differential equation for $\chi$,
$d\chi/dy=(1/2)W_{c,\chi}$,
\beq
&&\frac{d\chi}{dy}
=\pm
2\mu\chi
\sqrt{a^2+
c_0 \chi^{\lambda/\mu}
          -\frac{\mu}{\lambda-2\mu}\chi^2},\quad (\lambda\neq 2\mu),
\nonumber\\
&&\frac{d\chi}{dy}=
\pm 2\mu\chi
\sqrt{a^2+
\chi^{2}
\Big(\ln (\chi)+ c_1\Big)},
\quad (\lambda= 2\mu).
\label{cr}
\eeq

For the metric function $A$,
\beq
\frac{dA}{dy}=\frac{dA}{d\chi}\frac{d\chi}{dy}
=\frac{dA}{dy}W_{c,\chi}=-\frac{1}{3}W_c,
\eeq
which leads to
$dA/d\chi=-(1/3)(W_c/W_{c,\chi})$.
This can be solved as
\beq
&&A(\chi)
=\alpha_0
+\frac{2a^2}{9}\ln \chi
-\frac{1}{9}\frac{\lambda-3\mu}{\lambda-2\mu}\chi^2
-\frac{c_0}{9}\chi^{\lambda/\mu},\quad (\lambda\neq 2\mu)
\nonumber \\
&&
A(\chi)
=\alpha_1
+\frac{2a^2}{9}\ln \chi
-\frac{3\mu+\lambda c_1}{18\mu}\chi^2
-\frac{\chi^2}{6\mu}
 \chi^2
 \Big(\ln \chi-\frac{1}{2}\Big)
,\quad (\lambda= 2\mu),\label{A_r}
\eeq
where $\alpha_0$ and $\alpha_1$ are integration constants
which are chosen so that $A(y=0)=0$.

To summarize, the general method to obtain solutions is as follows:
First, for a given set of parameters $\lambda$, $a$ and $\mu$,
one solves Eq. (\ref{cr})
and determines the profile of $\chi$ as a function of $y$.
Then, from Eq. (\ref{rho_r}) and (\ref{A_r}),
one can determine the profile of $\phi$ and $A$ as the function of $y$,
through $\chi(y)$.
In general, Eq. (\ref{cr}) does not have exact solutions
except for the limited cases discussed below.
Note that the solutions discussed in Ref. \cite{Bazeia:2004dh} correspond to
$c_0=0$. It should be stressed that the model
admits a particular class
of solutions which cannot be obtained from the method given above.
A set of solutions such that $\chi=0$
(for which Eq. (\ref{pc}) does not make any sense)
do not have any internal structure for the brane, although
the system admits a solution given by $\phi=\pm a \tanh (\lambda a y)$.

According to the method shown above, in Ref. \cite{dfh},
several new analytic thick brane solutions were found for $c_0\neq 0$,
which were called {\it degenerate Bloch walls} and {\it critical Bloch walls}.
Here,  by ``degenerate" we mean that one cannot specify
the property of the solutions
only by fixing the potential parameters $\lambda$, $a$ and $\mu$,
but one also needs to fix the integration constant $c_0$.
In fact,  
the solutions still have remarkable variety for different values of $c_0$.

\paragraph{Degenerate Bloch walls}
For the case $\lambda=\mu$ and $c_0<-2$,
the following solution was obtained
\beq
&&
\chi(y)=\frac{2a}{\sqrt{c_0^2-4}\cosh (2\mu a y)-c_0},
\quad
\phi(y)=\frac{\sqrt{c_0^2-4}}{\sqrt{c_0^2-4}\cosh (2\mu a y)-c_0},
\eeq
and
\beq
e^{2A(y)}
=N
\left[
\frac{2a}{\sqrt{c_0^2-4}\cosh (2\mu a y)-c_0}
\right]^{(4a^2)/9}
\exp
\left[
\frac{2a \big(c_0^2-c_0\sqrt{c_0^2-4}\cosh(2\mu a y)-4a\big)}
{9\big(\sqrt{c_0^2-4}\cosh(2\mu a y)-c_0\big)^2}
\right],
\eeq
where $N$ is chosen so that $A(y=0)=0$.

For the choice $\lambda=4\mu$ and $c_0<1/16$,
a similar solution was obtained
\beq
&&\chi(y)=
-\frac{2a}{\sqrt{\sqrt{1-16 c_0}\cosh (4a\mu y)+1}},
\quad
\phi(y)=\sqrt{1-16c_0}a
\frac{\sinh (4\mu a y)}
     {\sqrt{\sqrt{1-16 c_0}\cosh (4a\mu y)+1}},
\eeq
and
\beq
e^{2A}
&=&N
\left[
\frac{-2a}{\sqrt{\sqrt{1-16 c_0}\cosh (4a\mu y)+1}}
\right]^{16a^2/9}
\exp
\left\{
-\frac{4a^2}{9}
\left[
\frac{1+32 a^2c_0+\sqrt{1-16c_0}\cosh (4\mu a y)}
     {\big(\sqrt{1-16c_0}\cosh(4\mu a y)+1\big)^2}
\right]
\right\}
\eeq
and again $N$ was chosen so that $A(y=0)=0$.
An interesting feature of these solutions is that,
for some value of $c_0$, $\phi$ exhibits a double kink profile.
This represents the formation of a double wall structure,
extended along the spatial dimension.

\paragraph{Critical Bloch walls}

For the case that $\lambda=\mu$ and $c_0=-2$,
the solution
\beq
&&\chi=\frac{a}{2}\Big[1\pm\tanh(\mu a y)\Big],
\quad
\phi=\frac{a}{2}\Big[\tanh(\mu a y)\mp 1\Big],
\eeq
and
\beq
e^{2A}
=N\left[\frac{a}{2}\left(1\pm \tanh (\mu a y)\right)\right]^{2a^2/9}
\exp
\left\{
-\frac{a^2}{9}
\left[
\left(1\pm\tanh(\mu a y)\right)^2
-\frac{2}{a}
\left(1\pm \tanh(\mu a y)\right)^2
\right]
\right\},
\eeq
was obtained.
Similarly, for the case that $c_0=1/16$ and $\lambda=4\mu$,
another solution was obtained of the form
\beq
\chi=-\sqrt{2}a
     \frac{\cosh(\mu a y)\pm \sinh (\mu a y)}
          {\sqrt{\cosh (2\mu a y)}},\quad
\phi
=\frac{a}{2}
\big(1\mp \tanh (2\mu a y)\big),
\eeq
and
\beq
e^{2A}
=N\left[\frac{-a e^{\pm \mu a y}}{\sqrt{\cosh (2\mu ay)}}\right]^{16a^2/9}
&\times&
\exp
\Big\{
-\frac{2a^2}{9}
\tanh (2a\mu y)
\big[
a^2 \tanh (2a\mu y)
\mp (1+2 a^2)
\big]
\Big\}\,.
\eeq
The novelty in these cases is the fact that both, $\phi$ and
$\chi$ fields have a kink-like profile and the warp factor has the remarkable
feature of having two Minkowski-type regions.


\section{Topologically trivial thick branes}
\label{section-3}

\subsection{Thick branes of two strongly interacting scalar fields}
\label{two_scalar_field}

Here, we review the topologically trivial solutions.
For example, one possibility is to consider two nonlinear gravitating
scalar fields which can create a 4-dimensional brane in a multidimensional spacetime.
Such problems have been investigated for the 5-dimensional case in~\cite{Dzhun,phantom}, for the 6-dimensional
case in~\cite{Dzhun1}, and for the 7- and the 8-dimensional cases in~\cite{Dzhun2}.
From the physical point of view, the situation is as follows:
an interaction potential of these fields has two local and two global minima.
Thus there are two different vacuums. The multidimensional space is filled with these
scalar fields which are located at the vacuum in which the scalar
fields are at the local minimum. Therefore there is a defect in the form of a
4-dimensional brane on the background of this vacuum.

The general approach to such solutions is as follows: the action
of $D=4+n$ dimensional gravity for all models
can be written as ~\cite{Singl}:
\begin{equation}
\label{Daction}
S = \int d^Dx\sqrt {^Dg} \left[ -\frac{M^{n+2}}{2}R + L_m
\right]~,
\end{equation}
where $M$ is the fundamental mass scale, and $n$ is a number of extra dimensions.
The Lagrangian, $L_m$, for the two interacting scalar fields $\varphi, \chi$
takes the form:
\begin{equation}
\label{lagrangian}
    L_m =\epsilon\left[\frac{1}{2}\partial_A\varphi\partial^A\varphi+
	\frac{1}{2}\partial_A\chi\partial^A\chi-V(\varphi,\chi)\right]~,
\end{equation}
where the potential energy $V(\varphi,\chi)$ is:
\begin{equation}
\label{pot2}
    V(\varphi,\chi)=\frac{\Lambda_1}{4}(\varphi^2-m_1^2)^2+
    \frac{\Lambda_2}{4}(\chi^2-m_2^2)^2+\varphi^2 \chi^2-V_0.
\end{equation}
(This potential was used in~\cite{Dzhunushaliev:2006di} as
an effective description of a condensate of gauge field in SU(3) Yang-Mills theory,
i.e. the scalar fields were taken as effective fields which described a condensate of Yang-Mills
fields.) Here the capital Latin indices run over $A, B =0, 1, 2, 3, ..., D$ and
the small Greek indices $\alpha, \beta =0, 1, 2, 3$ refer to four
dimensions; $\Lambda_1, \Lambda_2$ are the self-coupling constants, $m_1, m_2$
are the masses of the scalar fields $\varphi$ and $\chi$, respectively; $V_0$ is an arbitrary
normalization constant which can be chosen based of physical motivations;
$\epsilon=+1$ for usual scalar fields, and
 $\epsilon=-1$ for phantom scalar fields.

The use of two fields ensures the presence of two global minima of the potential \eqref{pot2} at
$\phi~=~0, \chi~=~\pm~m_2$ and two local minima at $\chi~=~0, \phi~=~\pm~m_1$ for the
values of the parameters $\Lambda_1, \Lambda_2$ used in the above papers. The conditions for existence of the
local minima are: $\Lambda_1>0, m_1^2>\Lambda_2 m_2^2/2$, and for the
global minima they are: $\Lambda_2>0, m_2^2>\Lambda_1 m_1^2/2$. Because of these minima
there were solutions localized on the brane for the 5-, 6-, 7- and 8-dimensional cases ~\cite{Dzhun}-\cite{Dzhun2}.
In these different cases the solutions go asymptotically to one of the local minima.

Variation of the action \eqref{Daction} with respect to the
$D$-dimensional metric tensor $g_{AB}$ leads to Einstein's
equations:
\begin{equation}
\label{EinsteinEquation}
R^{A}_B - \frac{1}{2}\delta^{A}_B R = \frac{1}{M^{n+2}} T^{A}_B,
\end{equation}
where $R^{A}_B$ and $T^{A}_B$ are the $D$-dimensional Ricci and the energy-momentum
tensors, respectively. The corresponding scalar field equations can be obtained from \eqref{Daction}
by variation with respect to the field variables $\varphi,\chi$. These equations are
\begin{equation}
\label{FieldEquations}
\frac{1}{\sqrt{^D \! g}}\frac{\partial}{\partial x^A}\left[\sqrt{^D \! g}\,\,
g^{AB}
\frac{\partial (\varphi,\chi)}{\partial x^B}\right]=-\frac{\partial V}{\partial (\varphi,\chi)}.
\end{equation}

\subsubsection{General equations}
Using the generalized $D$-dimensional metric~\cite{Singl}
\begin{equation}
\label{metric_n}
ds^2= \phi ^2(r) \eta_{\alpha \beta }(x^\nu)dx^\alpha dx^\beta -
\lambda (r) (dr^2 +  r^2 d \Omega ^2 _{n-1}) ~,
\end{equation}
where $d \Omega ^2 _{n-1}$ is the solid angle for the $(n-1)$ sphere,
one can rewrite the Einstein equations \cite{Singl} in the form
\begin{eqnarray}
\label{Einstein-na}
    3 \left( 2\frac{\phi ^{\prime \prime}}{\phi} -
    \frac{\phi ^{\prime}}{\phi} \frac{\lambda ^{\prime}}{\lambda }
    \right) + 6 \frac{(\phi ^{\prime})^2}{\phi ^2}+(n-1)&&
\nonumber \\
    \times \left[
    3\frac{\phi ^{\prime}}{\phi}\left(\frac{\lambda^\prime}{\lambda}+\frac{2}{r}\right)+
    \frac{\lambda^{\prime\prime}}{\lambda}-
    \frac{1}{2}\frac{\lambda ^{\prime}}{\lambda}\left(\frac{\lambda^\prime}{\lambda}-
    \frac{6}{r}\right)+ \frac{n-4}{4}\frac{\lambda^\prime}{\lambda}
    \left(\frac{\lambda^\prime}{\lambda}+
    \frac{4}{r}\right)  \right]&& \\ \nonumber
    =-\frac{2\lambda}{M^{n+2}}T^\alpha_\alpha,&&  \\
\label{Einstein-nb}
    12 \frac{(\phi ^{\prime})^2}{(\phi)^2 }+
    (n-1) \left[ 4 \frac{\phi ^{\prime}}{\phi}\left(\frac{\lambda^\prime}{\lambda}+
    \frac{2}{r}\right)+
    \frac{n-2}{4}\frac{\lambda^\prime}{\lambda} \left(\frac{\lambda^\prime}{\lambda}+
    \frac{4}{r}\right) \right] &&\\ \nonumber
    =    -\frac{2\lambda}{M^{n+2}}T^r_r~,&& \\
\label{Einstein-nc}
    4 \left( 2\frac{\phi ^{\prime \prime}}{\phi} -
    \frac{\phi ^{\prime}}{\phi} \frac{\lambda ^{\prime}}{\lambda }
    \right) + 12 \frac{(\phi ^{\prime})^2}{\phi ^2}+(n-2)&&
\nonumber \\
   \times \left[ 4\frac{\phi ^{\prime}}{\phi}\left(\frac{\lambda^\prime}{\lambda}+
    \frac{2}{r}\right)+
    \frac{\lambda^{\prime\prime}}{\lambda}-
    \frac{1}{2}\frac{\lambda^{\prime}}{\lambda}
    \left(\frac{\lambda^\prime}{\lambda}-\frac{6}{r}\right)+
    \frac{n-5}{4}\frac{\lambda^\prime}{\lambda}
    \left(\frac{\lambda^\prime}{\lambda}+\frac{4}{r}\right) \right]&& \\ \nonumber
     =  -\frac{2\lambda}{M^{n+2}}T^\theta_\theta~,&&
\end{eqnarray}
where $T^\alpha_\alpha, T^r_r$ and $T^\theta_\theta$ are  $\left(^\alpha_\alpha\right), \left(^r_r\right)$
and $\left(^\theta_\theta\right)$ components of the energy-momentum tensor, respectively.
The scalar field equations become
\begin{eqnarray}
  \label{Field_eqn_phi}
  \varphi'' + \left(
        \frac{n-1}{r} + 4 \frac{\phi'}{\phi} +\frac{n-2}{2} \frac{\lambda'}{\lambda}
    \right) \varphi &=& \lambda \varphi \left[
        2 \chi^2 + \Lambda_1 \left( \varphi^2 - m_1^2 \right)
    \right] ,\\
    \chi'' + \left(
        \frac{n-1}{r} + 4 \frac{\phi'}{\phi} +\frac{n-2}{2} \frac{\lambda'}{\lambda}
    \right) \chi &=& \lambda \chi \left[
        2 \varphi^2 + \Lambda_2 \left( \chi^2 - m_2^2 \right)
    \right].
  \label{Field_eqn_chi}
\end{eqnarray}


The set of equations \eqref{Einstein-na}-\eqref{Field_eqn_chi} is a set of nonlinear equations with solutions
whose behavior very much depends on values of the parameters $m_1, m_2, \Lambda_1, \Lambda_2$.
As shown in Refs.~\cite{Dzhun}-\cite{Dzhun2}, by specifying some values of the self-coupling constants
$\Lambda_1, \Lambda_2$, the problem of searching for {\it regular} solutions to systems similar to the system
given by \eqref{Einstein-na}-\eqref{Field_eqn_chi}
reduced to the evaluation of eigenvalues of the parameters $m_1, m_2$. Only for specific values of these parameters
one was able to find {\it regular} solutions with {\it finite} energy.

\subsubsection{5-dimensional problems}
\label{5-dim_problem}
\begin{figure}[h]
\begin{minipage}[t]{.49\linewidth}
\begin{center}
 \includegraphics[width=8cm]{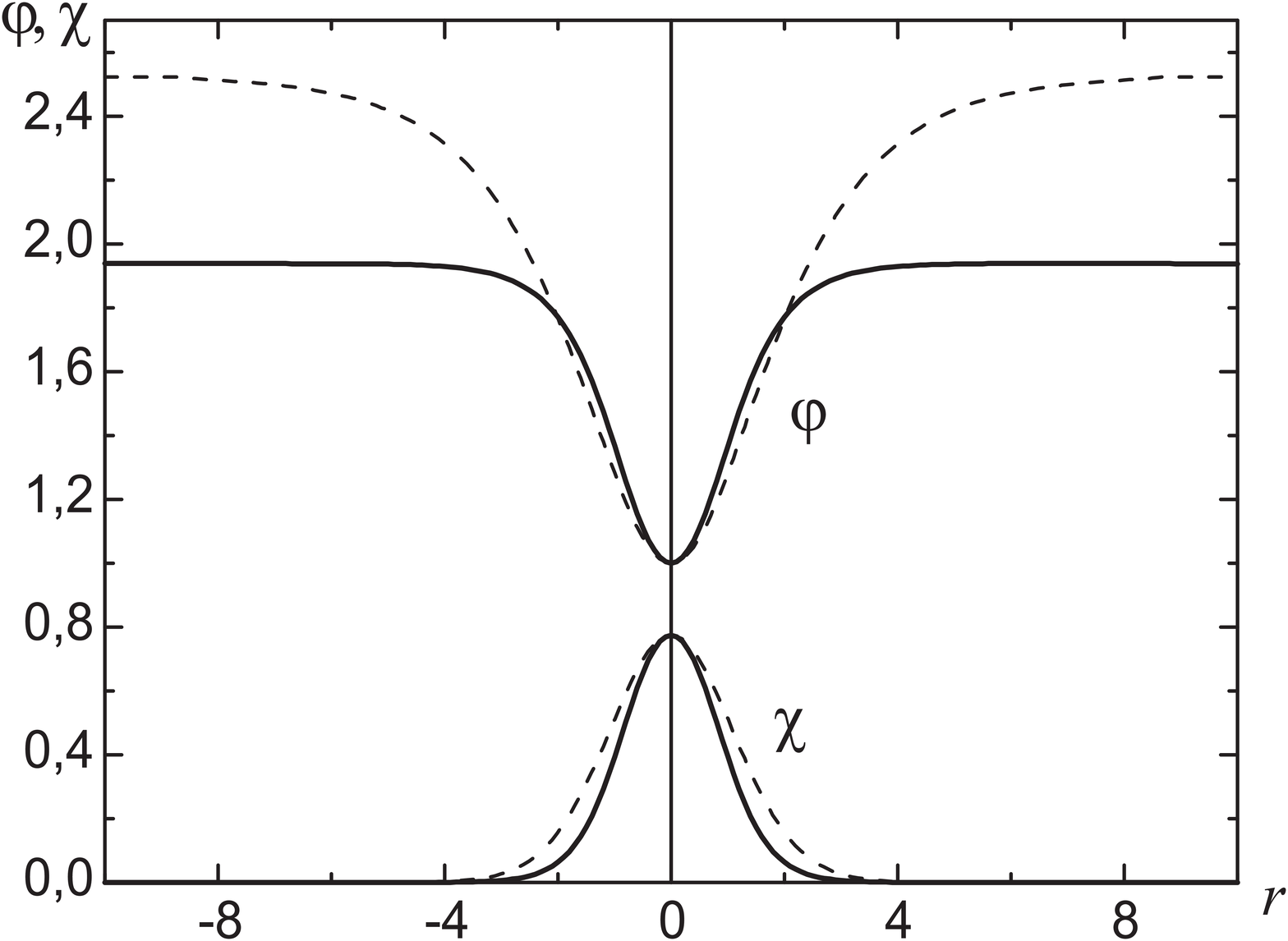}
	\caption{The scalar fields profiles for the cases
	$\epsilon=-1$ (solid lines) and $\epsilon=+1$ (dashed lines). 5D case.}
 \label{fields_phantom}
 \end{center}
\end{minipage}\hfill
\begin{minipage}[t]{.49\linewidth}
 \begin{center}
 \includegraphics[width=8cm]{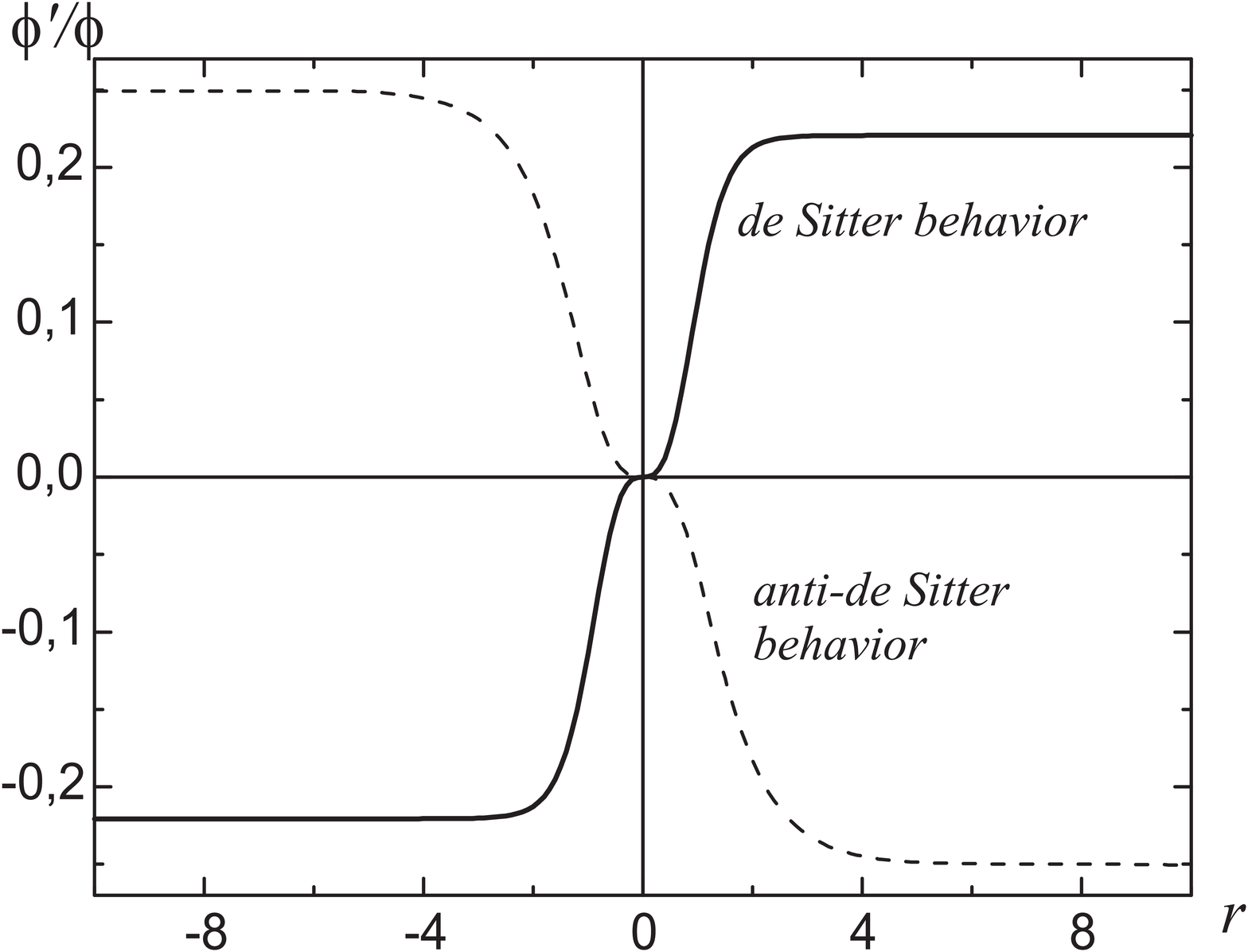}
	\caption{The profiles of the functions $\mathcal{H}=\phi^\prime/\phi$ for the cases
	$\epsilon=-1$ (solid line) and $\epsilon=+1$ (dashed line). 5D case.}
 \label{metric_phantom}
 \end{center}
\end{minipage}\hfill
\end{figure}

The usual scalar field case with $\epsilon=+1$ was considered in \cite{Dzhun}, and the phantom field case with $\epsilon=-1$ was
considered in ~\cite{phantom}.
These problems were solved by the numerical shooting method. The results are presented in Figs.~\ref{fields_phantom}-\ref{metric_phantom}. These solutions describe a thick brane in a 5-dimensional bulk. As pointed out in Ref.~\cite{phantom}, there are two distinctive features of the solutions for the phantom versus the usual scalar fields:
a) they have different asymptotic behavior; b) the solutions with phantom scalar fields are stable, and the solutions with the usual scalar fields are not (at least for the parameters $\Lambda_1, \Lambda_2, \varphi(0), \chi(0)$ used in \cite{phantom}).

\subsubsection{6,7,8-dimensional problems}

\begin{figure}[h]
\begin{minipage}[t]{.49\linewidth}
\begin{center}
 \includegraphics[width=8cm]{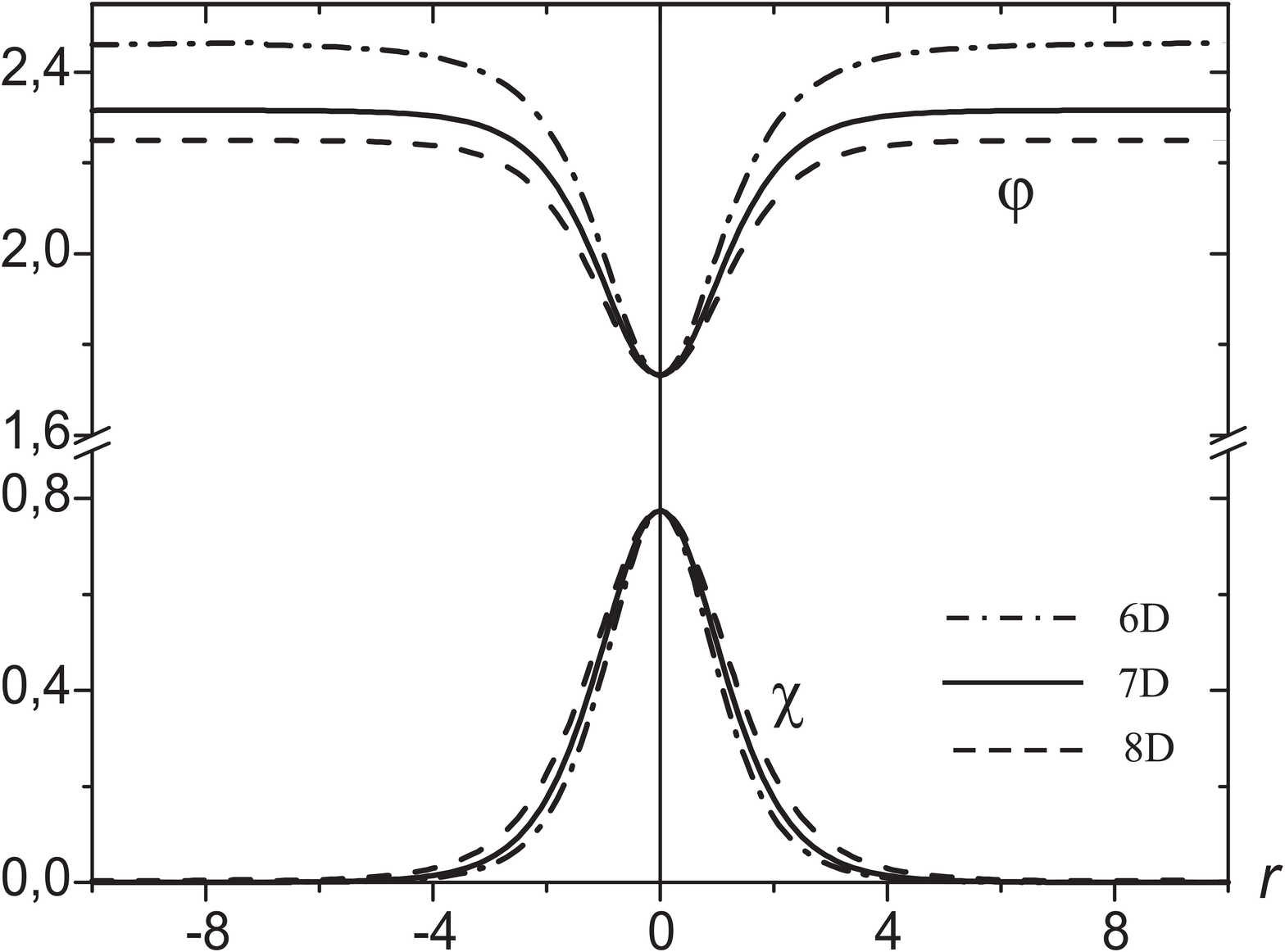}
	\caption{The scalar fields profiles for the 6,7,8-dimensional problems.}
 \label{scalar_fields_6_7_8}
 \end{center}
\end{minipage}\hfill
\begin{minipage}[t]{.49\linewidth}
 \begin{center}
 \includegraphics[width=8cm]{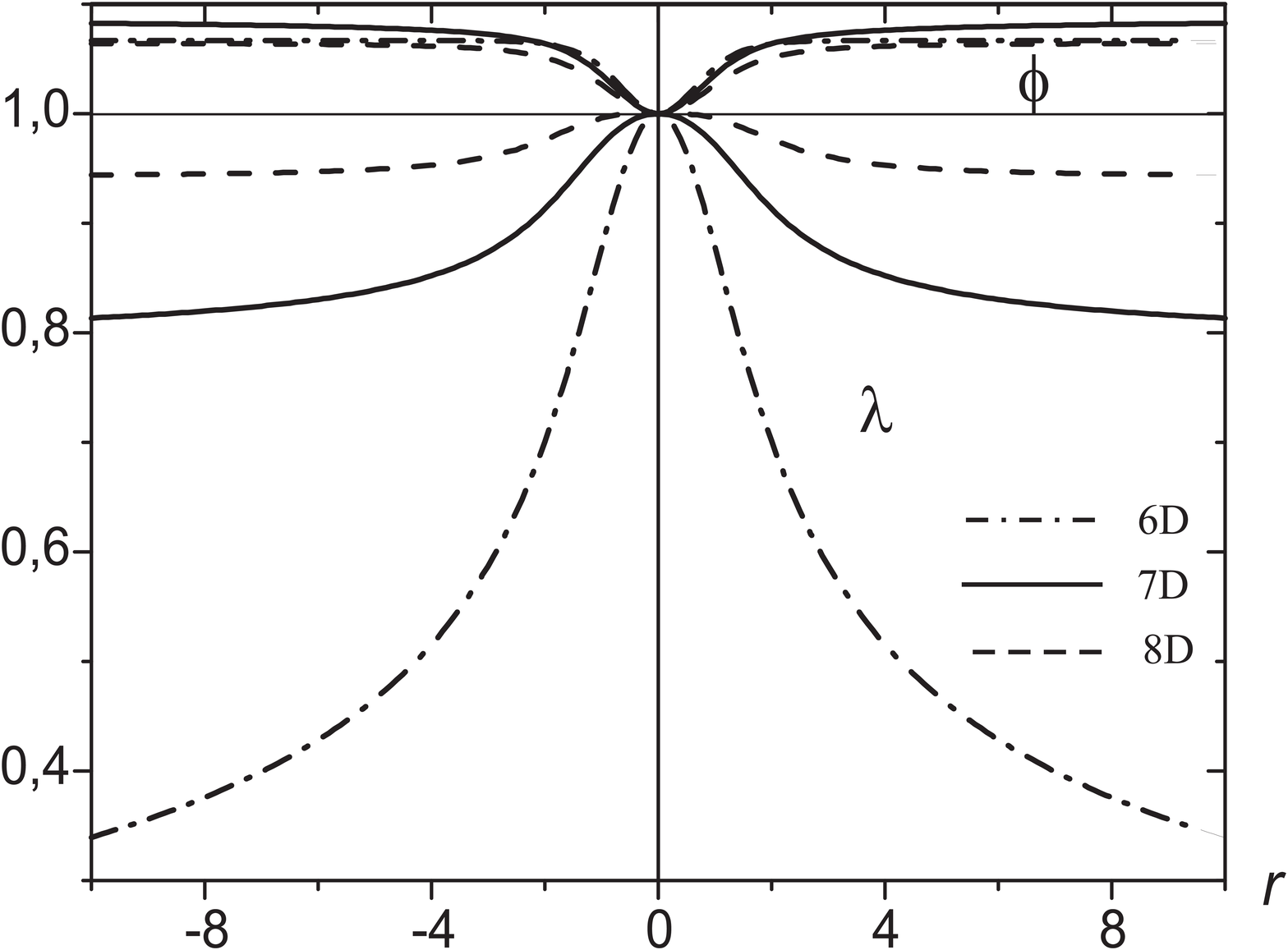}
	\caption{The profiles of the metric functions $\phi$ and $\lambda$
	for the  6,7,8-dimensional problems.}
 \label{metric_functions_6_7_8}
 \end{center}
\end{minipage}\hfill
\end{figure}

These cases were considered in the papers \cite{Dzhun1,Dzhun2}. The results are presented in Figs.~\ref{scalar_fields_6_7_8}-\ref{metric_functions_6_7_8}.
All the results obtained in \cite{Dzhun,phantom} for the 5-dimensional case and in \cite{Dzhun1} for the 6-dimensional case, and also in
\cite{Dzhun2} for the 7- and 8-dimensional cases show that it is in
principle possible to localize scalar fields with the potential \eqref{pot2}
to a brane in any dimensions. This lends support to the idea that there exist similar regular solutions is possible
in a larger number (any number) of extra dimensions (a discussion of this question is given in \cite{Dzhun2}).

\subsubsection{Phantom thick de Sitter brane solutions in multidimensions}

\begin{figure}[h]
\begin{minipage}[t]{.48\linewidth}
   \begin{center}
    \includegraphics[width=8cm]{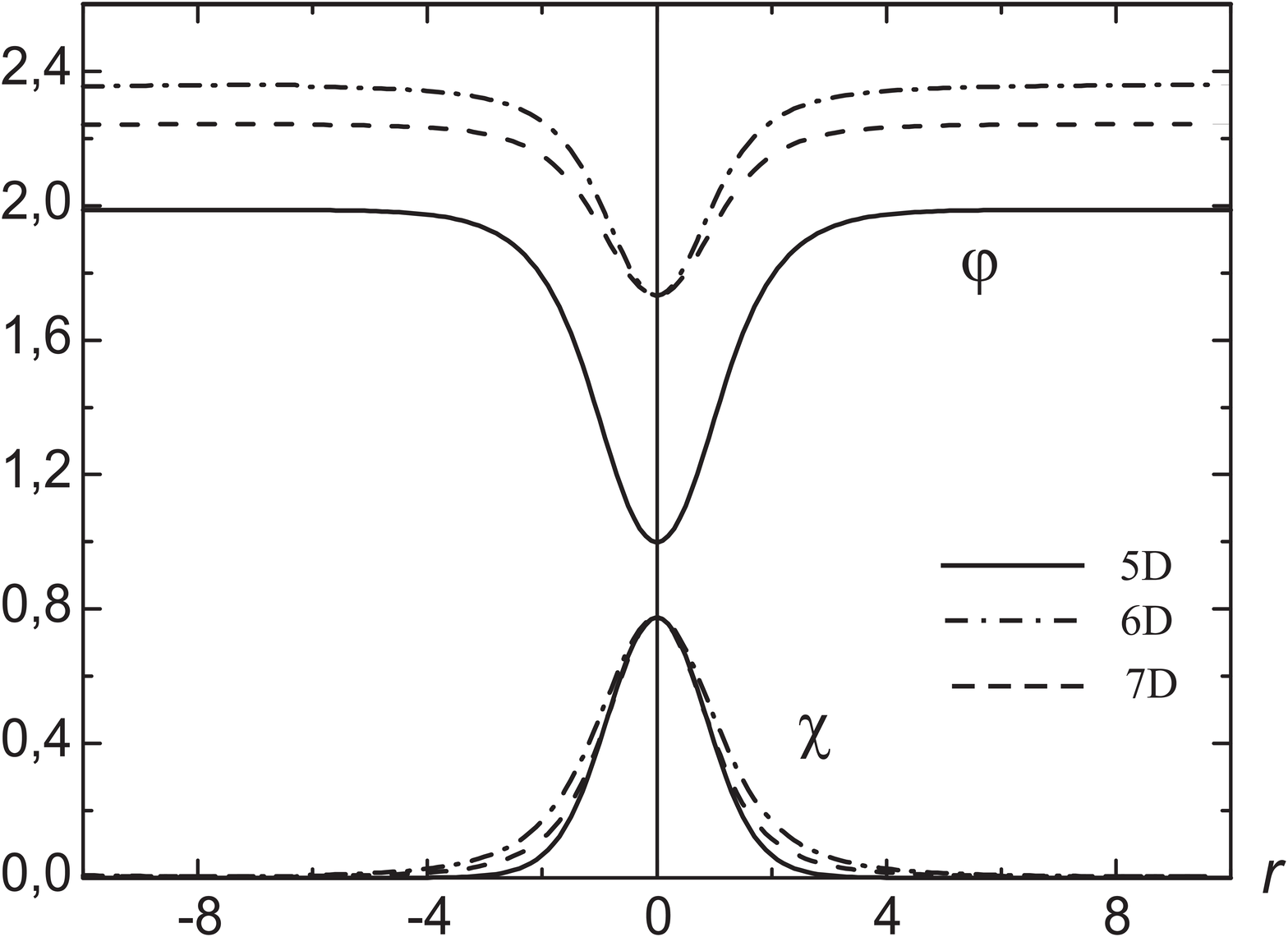}
        \caption{ The scalar field configurations $\varphi$ and $\chi$
are shown
as functions of the dimensionless $r$.
}
\label{scalar_fields_ds}
   \end{center}
 \end{minipage}\hfill
\hspace{0.4cm}
\begin{minipage}[t]{.48\linewidth}
   \begin{center}
    \includegraphics[width=8cm]{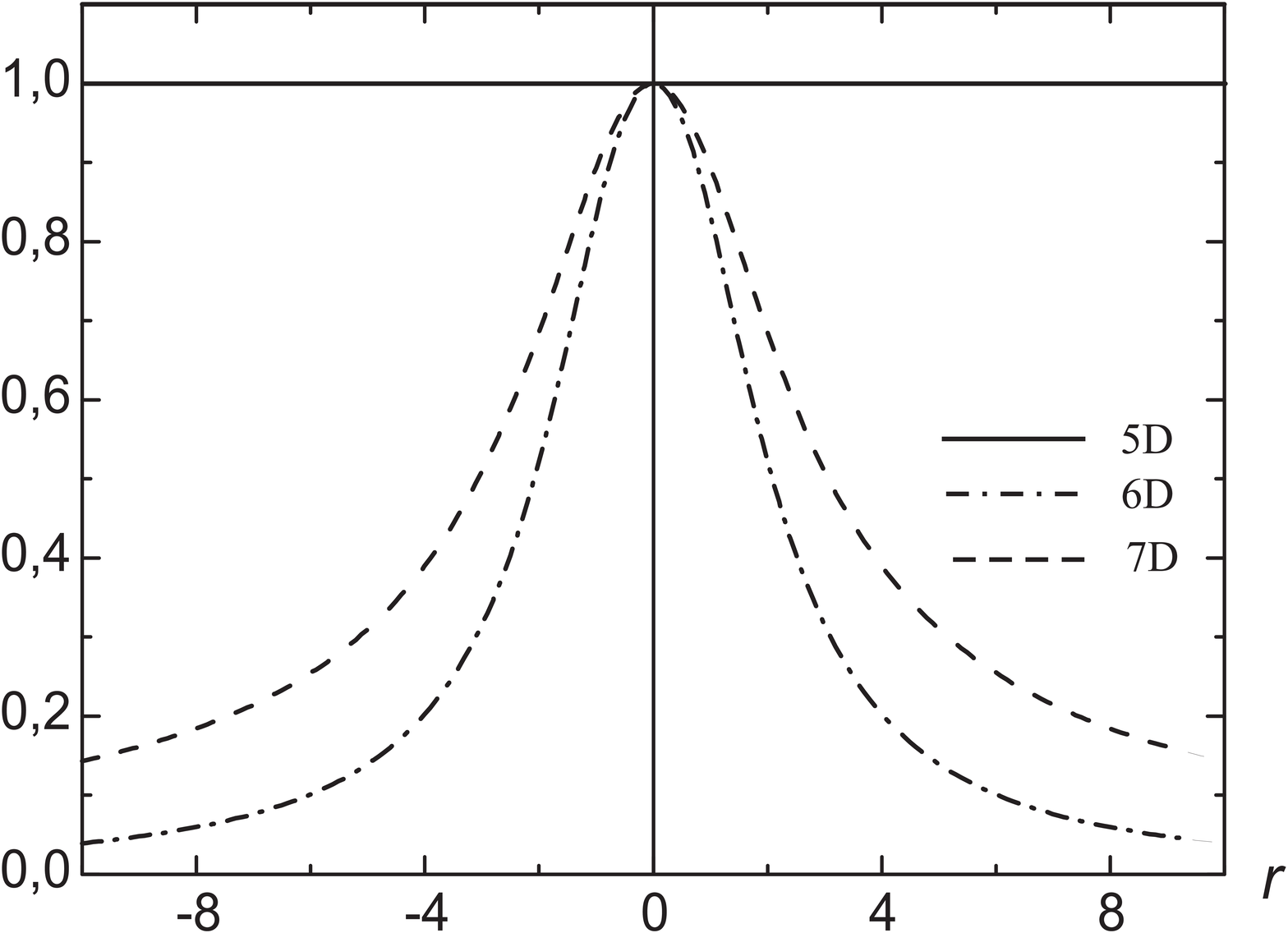}
        \caption{The metric function $\lambda(r)$ is shown
as a function of the dimensionless $r$.
}
\label{metric_lambda_ds}
   \end{center}
   \end{minipage}\hfill
\end{figure}

\begin{figure}[h]
\begin{minipage}[t]{.48\linewidth}
   \begin{center}
    \includegraphics[width=8cm]{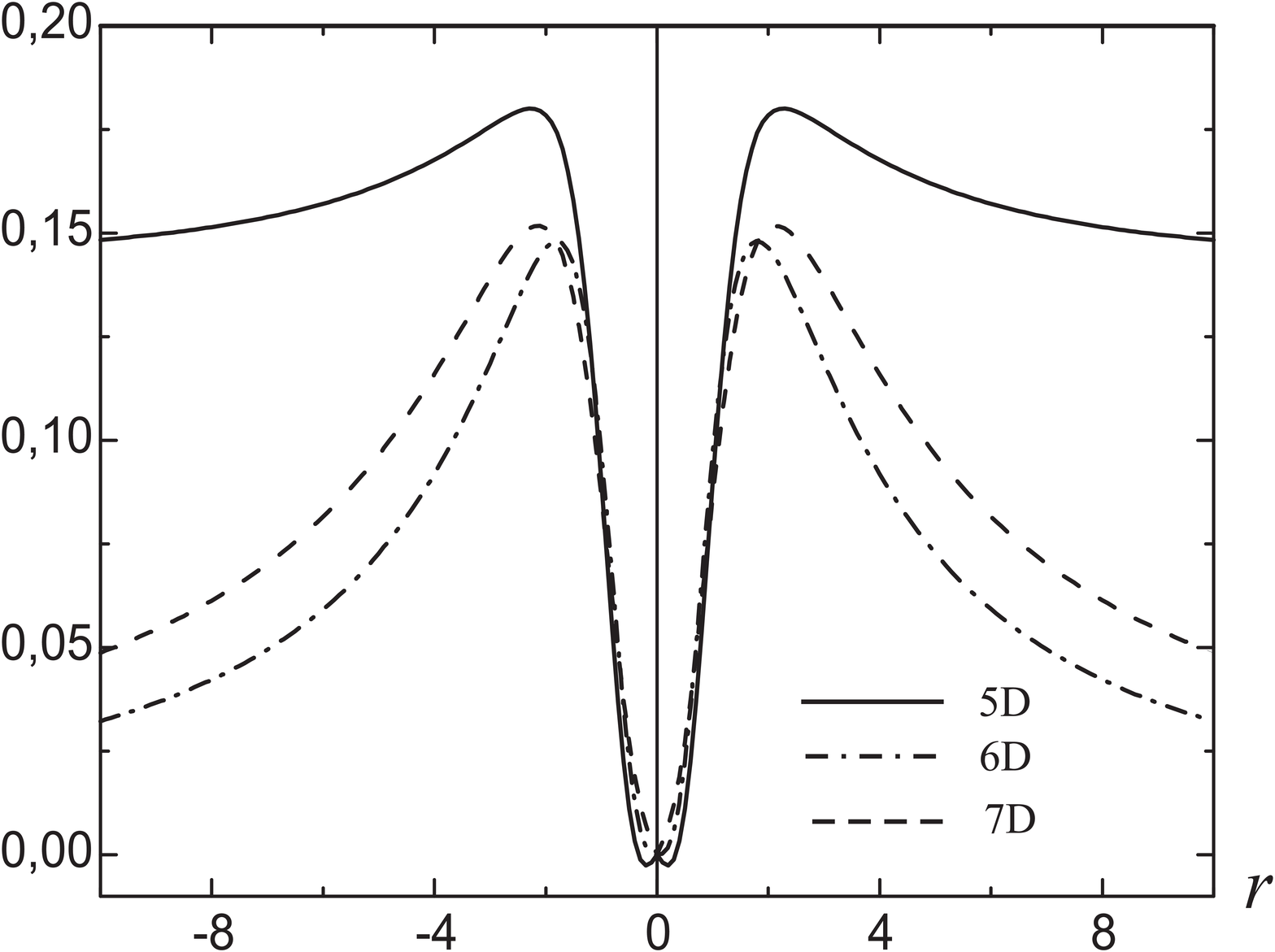}
        \caption{The function $\phi'/\phi$ is shown
a function of dimensionless $r$.
}
\label{hubble_ds}
   \end{center}
   \end{minipage}\hfill
\hspace{0.4cm}
\begin{minipage}[t]{.48\linewidth}
   \begin{center}
    \includegraphics[width=8cm]{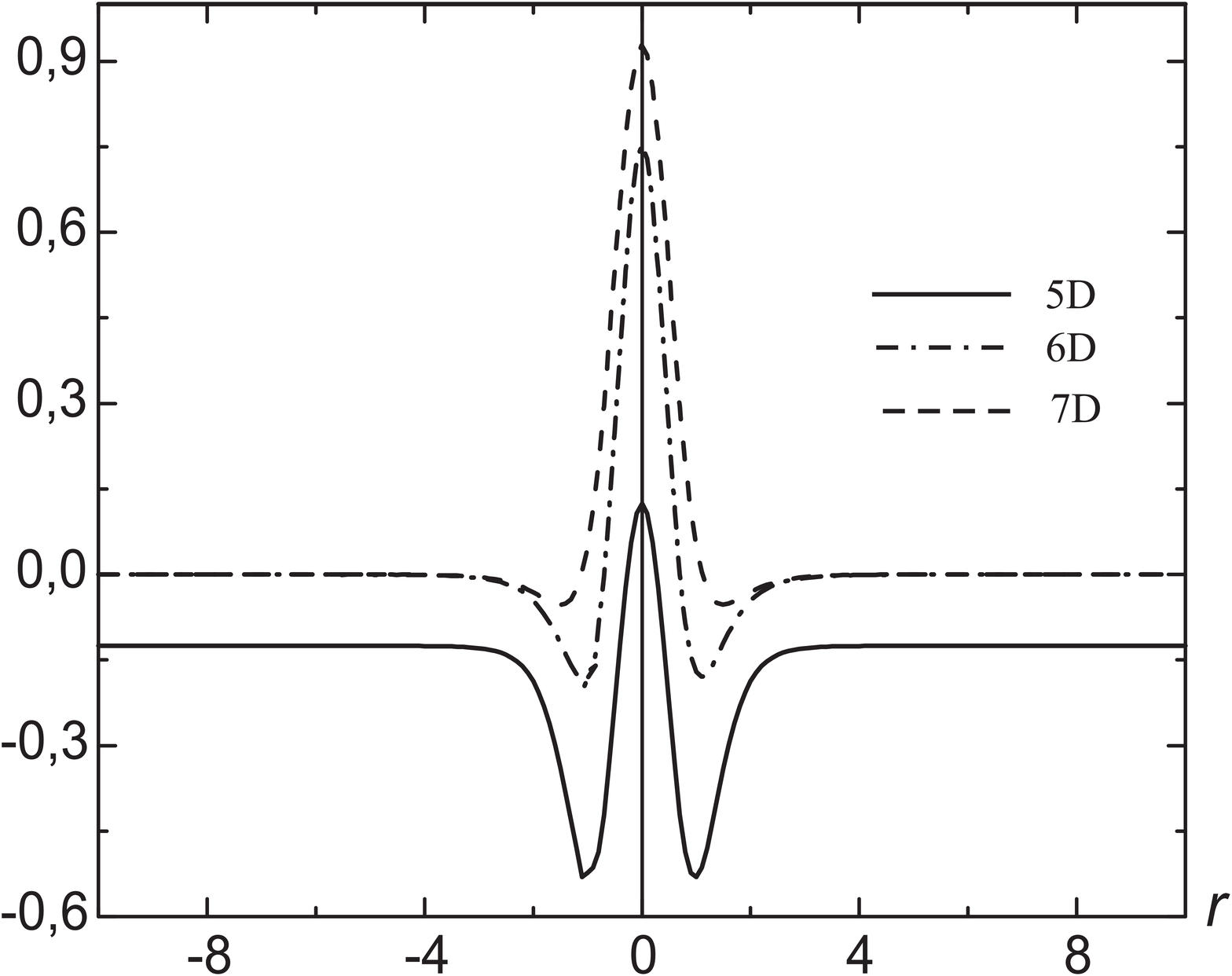}
        \caption{The energy density $T_0^0=\epsilon\left[\frac{1}{2\lambda}\left(\varphi^{\prime 2}+\chi^{\prime 2}\right)+V\right]$ is shown
a function of dimensionless $r$.
}
\label{energy_density_ds}
   \end{center}
   \end{minipage}\hfill
\end{figure}

Thick de Sitter brane solutions for phantom scalar fields
were considered in \cite{Dzhunushaliev:2008hq} for 5-, 6- and 7-dimensional cases. For this purpose the following metric was used
\begin{equation}
\label{metric_n_ds}
ds^2= a ^2(r) \gamma_{\alpha \beta }(x^\nu)dx^\alpha dx^\beta -
\lambda (r) (dr^2 +  r^2 d \Omega ^2 _{n-1})~,
\end{equation}
where $d \Omega ^2 _{n-1}$ is the solid angle of the $(n-1)$ sphere.
$\gamma_{\alpha \beta}$ is the metric of the four-dimensional de Sitter
space whose Ricci curvature tensor is given by
$R_{\mu\nu}[\gamma]=3H^2 \gamma_{\mu\nu}$. Then one has Einstein equations similar to
equations \eqref{Einstein-na}-\eqref{Einstein-nc} with the following extra terms on the left hand side:
$-6{H^2\lambda}/{\phi^2}$ in the $\left(^\alpha_\alpha\right)$ component, and
$-12{H^2\lambda}/{\phi^2}$ in the $\left(^r_r\right)$ and $\left(^\theta_\theta\right)$ components.
The scalar field equations remain the same.
The results of numerical calculations are presented in Figs.~\ref{scalar_fields_ds}-\ref{energy_density_ds}.

As one can see from these figures, there exist five-, six- and seven-dimensional thick de Sitter brane solutions of two interacting ({\it phantom}) scalar fields. It was also shown in \cite{Dzhunushaliev:2008hq} that asymptotically there exist anti-de Sitter spacetimes for the five-dimensional case and flat spacetimes for the six- and seven-dimensional cases. For the five-dimensional case, it was shown that the solutions are stable. It is also quite reasonable to expect that the thick brane solutions in multidimensions are also stable since the spacetime and vacuum structures are essentially the same as in the case of five spacetime dimensions.


\section{Branes with unusual source functions}

Here we present the
thick brane solutions created by special source functions \cite{gosi1, gosi2}.
It was shown  that by going from 5D to 6D it is possible to trap fields
of all spins (i.e. spins 0, $\frac{1}{2}$, 1, 2) to the
4D spacetime using only
gravity. The ``trapping" provided by the 6D solution of \cite{gosi2} is automatic since
 the proper distance 
from the brane at $r=0$ is
finite
and
any field can only be distributed within
some finite distance 
from the brane.

The 6D action was chosen as
\begin{equation}\label{6daction}
S = \int d^6x\sqrt {-g} \left[ \frac{M^4}{2}R +\Lambda + L_m
\right]~,
\end{equation}
where $M$, $R$, $\Lambda$ and $L_m$ are 
the fundamental
scale, the scalar curvature, the cosmological constant and the
matter field Lagrangian, respectively.
All 
these physical quantities
refer to the $6$-dimensional spacetime with the signature
{of} $(+ - - - - -)$.
The variation of the action \eqref{6daction} with respect to the
$6$-dimensional metric tensor $g_{AB}$ leads to
{the Einstein equations}
\begin{equation}\label{EinsteinEquation2}
R_{AB} - \frac{1}{2}g_{AB} R = \frac{1}{M^4}\left( g_{AB} \Lambda
+ T_{AB} \right)~,
\end{equation}
where $R_{AB}$ and $T_{AB}$ are the Ricci and the energy-momentum
tensors, respectively.
{The capital} Latin indices run 
$A, B,\cdots= 0, 1, 2, 3, 5, 6 $.

The 4-dimensional Einstein equations are 
$R_{\mu\nu }^{(4)} - \frac{1}{2} \eta_{\mu\nu} R^{(4)} = 0$,
i.e., the ordinary vacuum equations without no cosmological
constant.
The Greek indices $\alpha, \beta,... = 0, 1, 2, 3$ refer to
{the ordinary} four dimensions.
The ansatz functions for the source are
\begin{equation} \label{source}
T_{\mu\nu} =  - g_{\mu\nu} F(r), ~~~ T_{ij} = - g_{ij}K(r), ~~~
T_{i\mu} = 0 ~.
\end{equation}
The small Latin indices refer to the extra coordinates $i, j  = 5, 6$. Finally
the metric takes the form
\begin{equation}\label{ansatzA}
ds^2  = \phi ^2 (r)\eta _{\mu \nu } dx^\mu
dx^\nu   - \lambda (r)(dr^2  + r^2d\theta ^2)~ .
\end{equation}
The metric of the ordinary 4-space,
$\eta_{\alpha \beta }$, has the signature {of} $(+,-,-,-)$.
The 4D and 2D
``warp" factors, namely the ansatz functions $\phi (r)$ and $\lambda (r)$,
depend only
on the extra radial coordinate, $r$.
The Einstein field equations
\eqref{EinsteinEquation2} in terms of the ansatz functions are \cite{gosi2}
\begin{eqnarray}\label{Einstein6a}
3 \frac{\phi^{\prime \prime}}{\phi} + 3 \frac{\phi ^{\prime}}{r
\phi} + 3 \frac{(\phi^{\prime})^2}{\phi ^2} +
\frac{1}{2}\frac{\lambda ^{\prime \prime}}{\lambda }
-\frac{1}{2}\frac{(\lambda^{\prime})^2}{\lambda^2} +
\frac{1}{2}\frac{\lambda^{\prime}}{r\lambda }
= \frac{\lambda }{M^4}[F(r) - \Lambda ] ~,  \\
\label{Einstein6b}
\frac{\phi^{\prime} \lambda^{\prime}}{\phi \lambda } + 2
\frac{\phi^{\prime}}{r\phi} + 3 \frac{(\phi^{\prime})^2}{\phi
^2} = \frac{\lambda }{2 M^4}[K(r) - \Lambda ] ~, \\
\label{Einstein6c}
2\frac{\phi^{\prime \prime}}{\phi} - \frac{\phi^{\prime} \lambda
^{\prime}}{\phi \lambda } + 3\frac{(\phi^{\prime})^2}{\phi^2} =
\frac{\lambda }{2 M^4}[K(r) - \Lambda ] ~,
\end{eqnarray}
where the prime $=\partial / \partial r$. These equations are 
the $(\alpha, \alpha)$, $(r,r)$, and $(\theta, \theta)$ components,
respectively.
{Two of the three equations \eqref{Einstein6a} -- \eqref{Einstein6c}
are independent, and
they can be reduced to a set of
two equations for $\phi(r) , \lambda (r)$.}
An analytic solution to these equations is found
with the ansatz functions of the form
\begin{equation} \label{phi-lambda}
\phi (r) = \frac{c^2+ ar^2}{c^2+ r^2}~ \stackrel{r \rightarrow \infty}{\longrightarrow} a, \qquad
\lambda (r) = \frac{c^4}{(c^2 + r^2)^2} \stackrel{r \rightarrow \infty}{\longrightarrow} \frac{c^4}{r^4}~,
\end{equation}
where $c$ and $a$ are {constants}.
The source functions are
\begin{equation} \label{FK}
F(r) = \frac{f_1}{2 \phi (r) ^2} +\frac{3 f_2}{4 \phi (r)} \stackrel{r \rightarrow \infty}{\longrightarrow}
\frac{f_1}{2 a ^2} +\frac{3 f_2}{4 a}, ~~~~~ K(r) =
\frac{f_1}{\phi (r) ^2} +\frac{f_2}{\phi (r) }~ \stackrel{r \rightarrow \infty}{\longrightarrow}
\frac{f_1}{a ^2} +\frac{f_2}{a },
\end{equation}
where the constants $f_1 = -\frac{3\Lambda}{5} a$ and $f_2=\frac{4
\Lambda}{5}(a+1)$ are determined by the 6D cosmological constant
$\Lambda$, and the constant $a$, from the 4D warp function $\phi
(r)$.

A drawback of this solution is that the matter sources are put in by hand via the
ansatz functions $F(r)$ and $K(r)$ rather than
{given on some realistic foundation of the field theory.}
Also from $T_{\mu \nu}$ from \eqref{source} and $F (r)$ from \eqref{FK} (as well
as using the asymptotic values of $F(r)$ at $r=0$ and $r = \infty$)
one sees that the energy density is negative on the brane
and decreases to {a 
negative 
value at $r = \infty$.}

In the paper \cite{Singl_2} these solutions
were generalized to the case of
$n > 2$ extra dimensions.  As with the 6D solution, the obtained solutions
provide {the universal, gravitational trapping mechanism}
{of fields with various spins of from 0 to 2}.
Considering
the magnitude of the  ansatz functions for the stress-energy
which are necessary
to realize this trapping solution one finds physically
reasonable properties for a range of parameters $n$, $\Lambda$, $\epsilon$ (brane width) and $a$.
In addition, for $n = 2$ and $n = 3$, it is possible to have
the 
stress-energy tensor
which approaches zero in the limit of $r \to \infty$.
For $n = 4$,
either $F(r)$ or $K(r)$ goes to zero, but not both do.

\section{Cosmological thick branes}
\label{section-4}

In the previous sections, thick brane solutions which contain Minkowski
or de Sitter four-dimensional geometry {were} considered.
For such models, the bulk metric functions depend only
on the radial bulk coordinate
and solutions can be obtained, just by solving the
ordinary differential equations.
But if a thick brane which has {a} less symmetric,
Friedmann-Robertson-Walker (FRW) cosmological geometry,
{then}
in most of cases the metric functions also depend on time,
and to find a solution, one has to solve partial
differential equations.
The cosmological thick brane solutions
{were}
not explored much, except for works e.g., \cite{vol2}.

In this section,
several approaches to formulate the cosmological equations,
in the presence of arbitrary matter
inside the brane, are reviewed.
Because of the subtlety about how to define and
handle the effective four-dimensional quantities,
there {is} no unique formulation.
Thus, we review several representative approaches
both in five- and multidimensional spacetimes.

\subsection{Codimension-one model}

\subsubsection{Thin brane cosmology}

The interest in the localized matter
distributions in the context of gravity
has a long history.
Because of the difficulty to treat
{a thick wall},
in the early times it had been considered
to idealize a wall as an infinitesimally
thin object \cite{sen1,sen2,sen3}.
Then, {the}
interest {had} come again with the studies on
cosmological phase transitions and formation of topological defects.
Again, these defects were mainly assumed to be infinitesimally
thin \cite{vilenkin1,vilenkin2}.
The formulation of the equations
of motion of a singular domain wall
was summarized by Israel \cite{israel}.
In the context of the Lanczos-Darmois-Israel formalism,
a thin shell is regarded as
an idealized object with zero thickness.

Recent developments of string and M-theory have motivated us
to study cosmology in brane world.
The concept of a domain wall 
was applied
very frequently
to treat a self-gravitating brane mathematically in
the multidimensional general relativity.
Cosmological equations on {a} thin brane distribution
embedded into a five-dimensional bulk spacetime is easily derived
by integrating the five-dimensional Einstein equations
with a $\delta$-functional contribution into the
energy-momentum tensor \cite{bdl_a,bdl_b}
or
by employing the Israel junction conditions \cite{kraus1,kraus2}.
The thin brane cosmology
for a given bulk geometry is uniquely determined.
As the simplest example,
in an AdS-Schwarzschild bulk spacetime with the $Z_2$ symmetry
with respect to the brane,
the effective Friedmann equation on the brane is given by
\begin{eqnarray}
  \left( {\frac{\dot{a}}{2}} \right)^2+\frac{K}{ a^2}
= \frac{1}{3}\Lambda_4
+\frac{8\pi G_4}{3}\rho
\Big(1+\frac{\rho}{2\sigma}\Big)
+{M\over a^4},
\label{effFreidmann}
\end{eqnarray}
where $a$ and $\rho$ are the scale factor and energy density of the brane,
respectively.
$\sigma$ represents the brane tension.
The {AdS} curvature radius is given
by $\ell^2=-6/\Lambda$, where
$\Lambda$ is a (negative) {bulk} cosmological constant.
The constant $M$ is related to the mass of a black hole sitting in the bulk.
The four-dimensional Newton and cosmological constants are given by
\begin{eqnarray}
8\pi G_4=\frac{\kappa_5^2}{\ell},\quad
\Lambda_4= \frac{\kappa_{5}^4}{12}\sigma^2-\frac{3}{\ell^2}
         =\frac{\kappa_{5}^4}{12}\sigma^2+\frac{1}{2} \Lambda_5\,.\label{4dcos}
\end{eqnarray}
In the effective Friedmann equation~(\ref{effFreidmann}),
two corrections to the conventional
{one} appear.
The first correction is
the term which is proportional to the square of energy density on the brane,
which may be important in the early universe.
The second contribution is the so-called  dark radiation which is proportional to the mass of the black hole $M$ and a purely geometrical effect.

\subsubsection{Thickness of domain walls in general relativity}

In contrast to the case of a thin brane, {the}
thickness brings new subtlety.
Here, we see the brief history on the treatment of the thickness
of a domain wall.
Early attempts to formulate the thickness were
mainly motivated {in modeling}
the late time cosmological phase transition \cite{cosmo1}.
Silveira \cite{cosmo2} studied the dynamics of a spherical thick domain wall
by defining an average of radius $R$, weighted by the energy density.
It 
{was}
also unclear whether
the Israel conditions are applicable to the domain wall problem
and
the zero thickness limit of a thick domain wall is really equivalent
to a thin domain wall dominated by the Israel conditions.
There {were}
several works related to this issue \cite{ray,gt, cosmo3}.
The work of \cite{cosmo4} discussed a modification of the Israel thin shell equations in the Einstein-scalar theory
and concluded that the effect of thickness at first approximation was
effectively to reduce the energy density of the wall
in comparison with the thin case, leading to a faster
collapse of a spherical wall in vacuum.
The work of \cite{cosmo5} applied the expansion method of
the wall action in terms of
the small thickness parameter
and showed that the effective action for a thick domain consists of the usual Nambu-Goto term and
a contribution proportional to the induced Ricci scalar curvature.

\subsection{Thick brane cosmology}

Here,
several representative approaches
to discuss cosmology on a thick brane are reviewed.

\paragraph{Averaging approach}
In the approach employed in Ref. \cite{langl},
the effective brane quantities are obtained by integrating
the
five-dimensional ones
over the brane thickness.
The brane has a finite thickness between $y=-y_0/2$ and $y=y_0/2$,
where $y$ is the proper coordinate along the extra dimension.
For simplicity, it is assumed that the brane thickness is
time-independent.

There is ambiguity of the definition of
the effective four-dimensional quantities.
Here,
the averaging prescription that one integrates a
five-dimensional quantity $Q(t,y)$ over the brane thickness
\beq
 \langle Q(t)\rangle
=\frac{1}{y_0}\int_{-y_0/2}^{y_0/2} Q(t,y)dy,
\eeq
is adopted.
For convenience,
{the dimensionless quantities}
\beq
&&\epsilon:=\frac{\kappa_5^2}{6}y_0^2
         \langle \rho \rangle,\quad
\alpha:=\frac{a(y_0/2)}{\langle a \rangle},
\nonumber\\
&&\eta:=\frac{\langle \rho a^2\rangle}
             {\langle \rho \rangle\langle  a^2\rangle},\quad
\tilde\eta=\frac{\langle a^2\rangle}{\langle a\rangle^2}
\eeq
are introduced.
The parameter $\eta$ represents the homogeneity of the
matter distribution over the brane.
By imposing $p_r(\pm y_0/2)=0$, where $p_r$ is the pressure along the
extra direction, and
by integrating the Einstein equations,
the effective Friedmann equation is found to be
\beq
 H^2
=\frac{2}{y_0^2}
  \left(
   \alpha^2+\epsilon \eta-\frac{y_0^2}{\ell^2}\tilde{\eta}
  \right)
\left[1\pm
\sqrt{1
-\frac{(\epsilon\eta-\tilde \eta\frac{y_0^2}{\ell^2})^2
   +\frac{Cy_0^2}{\langle a\rangle^4}-\frac{\alpha^4y_0^2}{\ell^2}}
{(\alpha^2+\epsilon\eta-\tilde \eta\frac{y_0^2}{\ell^2})^2}}
\right],
\eeq
where $C$ is an integral constant.
The constant $C$ must satisfy the inequality that
\beq
C\frac{y_0^2}{\langle a\rangle^2}
\leq
2\alpha^2\Big(\eta\epsilon-\tilde \eta\frac{y_0^2}{\ell^2}\Big)
+\alpha^4\Big(1-\frac{y_0^2}{\ell^2}\Big).
\eeq
In the thin brane limit $y_0\to 0$, we find
\beq
H^2\approx
\frac{1}{\alpha^2}
\left(
\frac{\kappa_5^2}{36}\langle \rho \rangle^2\eta^2
+\frac{C}{a^4}
-\frac{\alpha^4}{\ell^2}
\right),
\eeq
which corresponds to the well-known thin brane cosmology
with the quadratic dependence on the energy density of a brane.
In the opposite limit  $y_0\to \infty$,
the generalized Friedmann equation reduces to
\beq
H^2\simeq
\frac{2\epsilon \eta }{y_0^2}
-\frac{2\tilde\eta}{\ell^2}
+\frac{2}{y_0}
\sqrt{\frac{\alpha^2\big(\alpha^2-\eta\big)}{\ell^2}
-\frac{C}{\langle a\rangle^4}},
\eeq
up to $O(y_0^{-1})$.
In the case of a Minkowski bulk $y_0\to 0$, the inequality yields
$C\to 0$ and
\beq
H^2\approx \frac{2\epsilon \eta}{y_0^2}
=\frac{\kappa_5^2\eta }{3y_0}\langle\rho \rangle,
\eeq
which corresponds to the standard four-dimensional cosmology
with  {effective Newton's}
constant $\kappa_4^2:=\kappa_5^2\eta/y_0$.
In the homogeneous bulk  $\eta=1$,
the result is in agreement with the Kaluza-Klein picture.
Thus, the above equations smoothly interpolate
{between the Kaluza-Klein and thin brane pictures of cosmology}.
{The extension}
of this method to the DGP model was discussed in Ref. \cite{thick_dgp}.

\vspace{0.3cm}

\paragraph{Two thin-shell approximation}
Another 
different approach based on the gluing of
the regular thickened brane with two different spacetime
geometries along two regular hypersurfaces
was considered in  \cite{cosmo7,cosmo8}.
The purpose was to understand the dynamics of a localized matter distribution of any kind, confined within two different spacetimes or matter phases.
It would be enable one to have any topology and any spacetime on each
side of {a} thick wall or brane.
By construction, {the matching}
is regular and there is no singular surface whatsoever
in this formulation.
Therefore, {the} junction conditions for the induced metric
extrinsic curvature tensors
on the thick wall boundaries with the two embedding spacetimes can be applied
as
\beq
\left[h_{ab}\right]_{\Sigma_i}=0,\quad
\left[K_{ab}\right]_{\Sigma_i}=0,
\eeq
where $\Sigma_i$ (i=1,2) represents each regular boundary located
between the core brane region and each bulk region.
By applying the formalism developed in \cite{cosmo10},
in Ref. \cite{cosmo11a}
cosmology on the brane in an AdS-Sch bulk
was investigated.
It {turns} out that the first modified Friedmann
equation becomes
\beq
H^2+\frac{K}{a^2}=\frac{8\pi G_4}{\rho}
+\frac{\kappa_5^4\rho^2}{36}
+\frac{\Lambda_4}{3}
+\frac{C}{a^4},
\eeq
where the effective four-dimensional cosmological
and {gravitational constants}
can be read off
\beq
\frac{\Lambda_4}{3}=\frac{\Lambda}{6}+\frac{w^2 \Lambda^2}{9},
\quad
8\pi G_4=\frac{\kappa_5^2w (-\Lambda)}{3}.\label{id}
\eeq
$2w$ is the proper separation between two regular walls.
In contrast to the thin brane case,
the linear dependence on the brane energy density is obtained
even without the tension.
Newton's constant is proportional to the brane thickness.
Other than the energy density equation, one also obtains the pressure equation
which depends on the bulk pressure in a nontrivial way.
These two cosmological equations together with the energy conservation law
determine the system completely.
The time-variation of the brane thickness was discussed in Ref. \cite{cosmo11}.
It was shown that
in the absence of the pressure
along the extra dimension in the brane energy-momentum tensor the thickness of the brane
decreases with time,
while at {a} late time {the}
negative transverse pressure can lead to an increase in
the brane thickness.
According to Eq. \eqref{id},
this suggests the time variation of the constants $G_4$ and $\Lambda_4$.

\vspace{0.3cm}

\paragraph{Quasi-static approximation}
The authors of Ref. \cite{navarro1} considered a thick codimension-one brane model
in which
the brane energy-momentum
tensor
includes
the pressure component along the extra dimension.
It is assumed that there is no matter outside the brane.
By integrating {the} four-dimensional time and spatial components of
the five-dimensional Einstein equations along the fifth dimension,
the first-order
transverse {derivatives}
of the metric functions at the brane boundaries
are related to the {components} of the brane energy-momentum tensor
integrated over the brane.
Here,
the assumption that
the derivatives parallel to the brane are negligible inside the thick brane
in comparison with the transverse ones, is made.
These {\it matching conditions}
are plugged into the remaining extra-dimensional component
of the Einstein equations
and the cosmological acceleration equations become
\beq
&&
3\Big(
\frac{\ddot{a}}{a}
+\frac{\dot{a}^2}{a^2}
\Big)
=\frac{1}{12 M^6}
\Big[2\Big(\sigma+\sigma_r\Big)^2
    +\Big(\sigma+\sigma_r\Big)
     \Big(1-3w-4w_r\Big)\rho_m
\Big]
+\frac{\Lambda}{M^3},
\nonumber\\
&&
\dot{\rho}{}_m
+3\frac{\dot{a}}{a}
\big(1+w\big)
-w_r \dot{\rho}_m,
\eeq
where $\rho=\sigma+\rho_m$,
$p=-\sigma+w\rho_m $ and $p_r=-\sigma_r +w_r \rho_{m}$
are energy density, pressure and bulk pressure in the brane, which
are defined by the integration of the bulk stress tensor
over the thick brane, respectively.
$\sigma$ and $\sigma_r$ are constant parts of the four-dimensional
matter, i.e., tension
and the bulk pressure, and $(w,w_r)$ specify the equations of state
of the time-dependent parts.
The presence of the bulk pressure {violates}
the energy conservation
on the brane.
The effective equation of state of the brane matter is given by
\beq
w_{\rm eff}=\frac{w+w_r}{1-w_r}.
\eeq
{An accelerating}
universe can be obtained for $w_{\rm eff}<-1/3$.
It implies that for example even for cosmic dust $w=0$,
{the}
cosmic acceleration can be {realized}
as long as $w_{r}<-1/2$
\cite{navarro1}.

This method was originally developed in the codimension-two case and
also generalized to the cases of higher codimensions,
which will be reviewed in the next subsection.

\subsection{Higher-codimensional cases}

\subsubsection{UV behavior of higher-codimensional branes}

In general,
a 3-brane with more than two codimensions is a singular object
because the gravitational potential sourced by the brane
is divergent in approaching it.
This is understood as follows.
Imagine a codimension-$n$ ($n=1,2,3,\cdots$)
self-gravitating object.
Gauss's law shows that
the gravitational potential becomes
$V(r)\propto 1/r^{n-2}$,
where $r$ is the radial distance from the object.
In the case of $n=1$, which is a singular domain wall case,
the jump across the object is finite
and
the matching conditions are available.
The case of $n=2$, where the object is a string-like defect,
is marginal
and only the tensional matter
can be put on the object and induces the deficit angle of the bulk.
In the case of $n>3$, for any kind of the localized matter,
the potential becomes singular and a naked singularity is usually formed.
In such a case, one needs some prescription, such as UV regularization,
to put matter there.

\vspace{0.3cm}

Several prescriptions for such divergence 
{were}
suggested
in the field theory approaches.
In Refs. \cite{carter, carter2, carter3}, an UV regularization
of the self-forces of various fields (scalar, graviton and form fields),
for extended objects with codimensions more than two, 
{were}
discussed
in the linearized approximation.
The regularization was performed by
replacing {a} singular delta functional source
with a regular profile function
(An IR cut-off is also required for a codimension two brane
to cure the logarithmic divergence).
For the gravitational force alone, the force is proportional to the extrinsic curvature vector, i.e., the trace of the second fundamental tensor.
For the special
case of a codimension-two brane, the force is actually zero and
so the self-interactions cancel, which
is a well-known result for cosmic strings in four dimensions, i.e.,
the tension determines the conical deficit angle and does not affect the local
geometry both inside and outside the string.
When a dilaton and a form field are included,
certain combinations of couplings result in {the}
vanishing total self-force.
It was also shown that these regularized self-interactions can be expressed
as renormalizations of the action.
In Ref. \cite{renorm1,renorm2},
the classical renormalization schemes for {a} local divergence
at a codimension-two (conical) brane {were} discussed,
mainly for {a} scalar theory.
It {was}
shown that the logarithmic (classical) divergence of a bulk field
in the thin brane limit
can be renormalized into the brane localized mass and couplings,
at the tree and one-loop levels,
and {the} renormalization group equations for them were obtained.
In Ref. \cite{renorm1}, {the} brane dynamics was not taken into account and
in Ref. \cite{renorm2} it was done.
The philosophy behind the procedure {is}
that
the low-energy physics on the brane
should be independent of the high-energy regularization mechanism:
1) bulk fields evaluated away from the brane should not depend on the regularization mechanism and thus be finite in the thin-brane limit,
2) bulk fields evaluated on the brane itself would be sensitive to the
regularization procedure and thus it would not be required that
bulk fields evaluated on the brane are finite in the thin-brane limit,
3) brane fields should have a well defined low-energy theory independent of
the brane regularization.

\subsubsection{Cosmology on regularized higher-codimensional branes}

To discuss nonlinear gravity and cosmology on a codimension-two brane,
various ways of regularization
{were} derived,
in e.g., Refs. \cite{vc1,vc2,codim2_a,codim2_b,codim2_c,codim2_d,codim2_e,navarro2}.
In the works  Ref. \cite{vc1,vc2}, the authors considered
a way of regularizations to smooth
out the matter over the brane with a finite thickness
and investigated {the}
effective cosmology in a
class of six-dimensional models with stable and compact extra dimensions.
The {time-dependence} of {the} brane matter
was taken into consideration
perturbatively around a static solution.
By integrating over the brane, the effective cosmological
equations are obtained and coincide with
the conventional one at the low energy scales.

Another way of regularization of a codimension-two brane has been discussed
in Refs. \cite{codim2_a,codim2_b,codim2_c,codim2_d,codim2_e}, such that
a codimension-two brane was replaced with a regular bulk geometry, i.e.,
a capped region.
Between the original bulk and the capped region,
a ring-like 4-brane with the compact fifth dimension, wrapping
around the axis of symmetry of the bulk, appears.
This 4-brane is a codimension-one object and
the Israel junction conditions can be applied.
A motion of the brane realizes cosmology, but
as long as the bulk is static, the effective cosmology is not
the realistic one.
An appropriate inclusion of the time-dependence of the bulk helps
the recovery of it on the brane.

\vspace{0.3cm}

The work \cite{navarro2} also discussed cosmology on a thick
codimension-two brane.
Apart from the previous approaches,
the discussion of \cite{navarro2} was {given}
in the general background
and may be able to be applied
to various codimension-two brane models, as long as
several assumptions are satisfied.
The authors of Ref. \cite{apple} extended the method
employed in \cite{navarro2}
to the {case}
of an arbitrary number of codimensions.
The discussion starts from the general metric ansatz that
\beq
ds^2=-N(t,r)^2dt^2
    +A(t,r)^2 g_{ij}dx^{i}dx^j
    +dr^2
    +\alpha(t,r)^2\gamma_{ab}dy^a dy^b,
\eeq
where $\gamma_{ab}$ is the metric of $(m-1)-$sphere.
Following Ref. \cite{carter, carter2,carter3}, regularization of the brane
is performed by replacing the delta function with a regular profile
function.
As the profile function, here, it is assumed that
all the brane matter is uniformly distributed
over the brane region $0<r<\epsilon$.

Here several assumptions are going to be made:
The important one is that
the derivatives tangential to the brane are small enough
in comparison with those normal to the brane.
It is also assumed
that $A(t,0)\approx A(t,\epsilon)=a(t)$
and the energy momentum tensor averaged over the brane takes the form
{of}
$\tilde T_A{}^B
={\rm diag}\big(-\rho, p,p,p,p_r,p_b,\cdots\big)
$, where $\rho$, $p$, $p_r$ and $p_{b}$ are the energy density
and
pressures along the three-dimensional space, bulk radial direction
and bulk angular directions, respectively.

In the case of a codimension-two brane {of} $m=2$,
in Ref. \cite{navarro2},
the pressure equation is obtained as
\beq
 3\Big(\frac{\ddot{a}}{a}+\frac{\dot{a}^2}{a^2}\Big)
&=&-\frac{\Lambda}{M^4}
+\frac{1}{2\pi \alpha^2 M^4}
   \big(p+p_r\big)\frac{\sqrt{-g}|_{0}}{\sqrt{-g}|_{\epsilon}}
\nonumber\\
&-&\frac{1}{32\pi^2\alpha^2M^8}
\Big[
3\big(\rho+p\big)^2+3p_b^2
+2\big(p_b+p_r\big)\big(\rho-3p\big)
-2pp_r-5p_r^2
\Big].
\eeq
Finally, in the codimension $m>3$ cases,
\beq
 3\Big(\frac{\ddot{a}}{a}+\frac{\dot{a}^2}{a^2}\Big)
&=&\omega_1
+\omega_2\big(\rho+p\big)^2
+\omega_3p(p-\rho)
+\omega_4 (\rho-3p_r),
\eeq
where
\beq
\omega_1
&=&-\frac{\Lambda}{M^{m+2}}
-\frac{A_m^2}{8(m+2)}
\Big[3(m-1)p_b^2
    -2\big(m-1\big)p_r p_b
    -(m+3)p_r^2
\Big]
\nonumber\\
&+&\frac{(m-1)(m-2)}{\epsilon^2}
   -\frac{(m-1)A_m}{2\epsilon} \big(p_r+p_b\big),
\nonumber\\
\omega_2&=&
-\frac{A_m^2(1+m)}{8(m+1)},
\nonumber\\
\omega_3&=&-\frac{A_m^2(m-2)}{4(m+2)},
\nonumber\\
\omega_4&=&-\frac{A_m^2}{4(m+2)}
\big[(m-1)p_b+p_r\big],\label{fried_high}
\eeq
and, $A_m=2/\big(\Omega_{m-1}\alpha^{m-1}M^{m+2}\big)$
and $B_m=-2-m$.
As in the five-dimensional case,
the brane matter is decomposed into the constant (tension) part
and time-dependent part
as $\rho=\sigma+\rho_m$ and $p=-\sigma+w\rho_m$.
In the low energy regime,
expanding the equation of motion up to the linear order $\rho_m$,
\beq
3\Big(\frac{\ddot{a}}{a}+\frac{\dot{a}^2}{a^2}\Big)
&=&\big(\omega_1+2 \sigma^2\omega_3+4\omega_4\sigma\big)
+\big(\omega_4+\sigma \omega_3\big)\big(1-3w\big)\rho_m
+O(\rho_m^2),
\eeq
which has the form in the conventional cosmology.

In the case of higher-codimensions that $m>2$,
under several assumptions such that the time evolution
of the brane thickness is negligible, 
$p_r$ is constant,
and
the energy flow into the bulk is negligible $T_{\mu r}\big|_{\epsilon}=0$,
the standard energy conservation law
is satisfied.
The form is similar to the Randall-Sundrum one.
But the effective gravitational constant is proportional
to $\omega_4$ which explicitly depends on the bulk
pressures $p_b$ and $p_r$.
However, in the higher codimensional
case, there is the subtlety on the presence of
{a localized} zero mode.

In the case of a codimension-two brane $m=2$,
the difference from the cases of $m>2$ comes from the term proportional
to $\sqrt{-g}\big|_{r=\epsilon}/\sqrt{-g}\big|_{r=0}$.
But by using the matching condition,
as long as $|p_r|$ and $|p_{\theta}|$ are small enough
in comparison with $|\rho|$ and $|p|$,
$\sqrt{-g}\big|_{r=\epsilon}/\sqrt{-g}\big|_{r=0}$
can be approximated to be unity.
Under such an approximation,
the cosmological behavior has
no essential difference from the cases of $m>2$,
but
there is no problem on {the}
localization of {a} graviton zero mode.

\section{S-branes}
\label{section-5}

Usually, a brane is considered as {a} time-like
manifold. However, it is also possible  to consider time-dependent solutions in multidimensional
gravitational theories and to interpret them as spacelike branes (S-branes).
Such an approach was suggested in the paper~\cite{Gut}.
S-branes which are thick and regular everywhere
{are interesting cosmological solutions} for solving the cosmological singularity problem.

The authors of the paper \cite{Gut} support their interest for investigations in this direction by the following arguments:
 SD-branes (S-branes with Dirichlet boundary conditions) may holographically reconstruct a time dimension just as in the AdS/CFT correspondence the D-brane field theory holographically reconstructs a spatial dimension.

In {\cite{Gut}},
the physical situations giving
{the solutions of
an S2-brane domain wall,
an S1-brane vortex,
and a charged S-brane}
are considered.
The arguments in favor of
existence of S-branes in string theory are presented.

Several solutions of the supergravity equations corresponding to S-branes with odd codimensions are obtained. The following S0-brane solution in
{the}
D=4 Einstein-Maxwell gravity is obtained.
The starting action is
\begin{equation}
	S = \int d^4 x \sqrt{-g} \left( R - F^2 \right).
\label{sbrane-15}
\end{equation}
The corresponding Einstein-Maxwell equations have
{the} following solution
\begin{equation}
	ds^2 = - \frac{Q^2}{\tau_0^2} \frac{\tau^2}{\tau^2 - \tau_0^2} d \tau^2 +
	\frac{\tau_0^2}{Q^2} \frac{\tau^2 - \tau_0^2}{\tau^2} d z^2 +
	\frac{Q^2 \tau_0^2}{\tau_0^2} d H_2^2,
\label{sbrane-25}
\end{equation}
where $Q$ and $\tau_0$ are constants and $d H_2^2$ is the unit metric on a 2-dimensional space with the constant negative curvature,
{respectively}.
The solution describes a charged S0-brane.

An S5-brane in {the} D=11 supergravity is obtained as well.
The starting action is
\begin{equation}
	S = \int d^{11} x \sqrt{-g} \left( R - \frac{1}{2 \cdot 4!} F^2 \right).
\label{sbrane-30}
\end{equation}
The solution of corresponding filed equations is
\begin{equation}
	ds^2 = - \left( \frac{q^2}{24} \right)^{1/3}
	\frac{\left[ \cosh \sqrt{24} (t-t_0) \right]^{2/3}}
	{\left( \sinh 3t \right)^{8/3}}
	\left[
		-dt^2 + \left( \sinh 3t \right)^2 d H_4^2
	\right] +
	\left( \frac{q^2}{24} \right)^{-1/6}
	\frac{1}{ \left[ \cosh \sqrt{24} (t-t_0) \right]^{1/3} } dx_6^2,
\label{sbrane-40}
\end{equation}
where $q,t_0$ are constants and $dH_4^2$
is the unit metric on a 4-dimensional space with the constant negative
curvature, {respectively}.
It was shown that there exists a singularity of this metric
near the brane for a large $t$.

In Ref.
\cite{Chen:2002yq} S-branes containing a graviton, $q-$form field strength, $F_{[q]}$, and a dilaton scalar, $\phi$, coupled to the form field with the coupling constant $a$, are discussed. In the Einstein frame, the action is given by
\begin{equation}
	S = \int d^d x \sqrt{-g} \left(
		R - \frac{1}{2} \partial_\mu \phi \partial^\mu \phi -
		\frac{1}{2 q!} e^{a \phi} F^2_{[q]}
	\right).
\label{sbrane-50}
\end{equation}
The corresponding equations of motion are
\begin{eqnarray}
	R_{\mu\nu} - \frac12 \partial_\mu \phi \partial_\nu \phi -
	\frac{{\rm e}^{a\phi}}{2(q-1)!} \left[
	F_{\mu\alpha_2\cdots\alpha_q} F_\nu{}^{\alpha_2\cdots\alpha_q}-
	\frac{q-1}{q(d-2)} F_{[q]}^2 \, g_{\mu\nu} \right] &=& 0,
\label{sbrane-60} \\
	\partial_\mu \left( \sqrt{-g} \, {\rm e}^{a\phi} \,
	F^{\mu\nu_2\cdots\nu_q} \right) &=& 0,
\label{sbrane-70} \\
	\frac1{\sqrt{-g}}\, \partial_\mu \left( \sqrt{-g} \partial^\mu
	\phi \right) - \frac{a}{2\, q!} {\rm e}^{a\phi} F_{[q]}^2 &=& 0.
\label{sbrane-80}
\end{eqnarray}
A solution describing {an}
S-brane is sought in the form
\begin{equation}
	ds^2 = - {\rm e}^{2A} dt^2 + {\rm e}^{2B} (dx_1^2 + \cdots +
	dx_p^2) + {\rm e}^{2C} \, d\Sigma_{k,\sigma}^2 + {\rm e}^{2D}
	(dy_1^2 + \cdots + dy_{q-k}^2).
\label{sbrane-90}
\end{equation}
The hyperspace $\Sigma_{k,\sigma}$ is a maximally symmetric space
with a constant scalar curvature
\begin{equation}
	d\Sigma_{k,\sigma}^2 = \bar g_{ab} dz^a dz^b = \left\{
 \begin{array}{ll}
 d \psi^2 + \sinh^2\psi \, d\Omega_{k-1}^2, \qquad & \sigma=-1,\\
 d \psi^2 + \psi^2 \, d\Omega_{k-1}^2, 			\qquad & \sigma=0,\\
 d \psi^2 + \sin^2\psi \, d\Omega_{k-1}^2, 	\qquad & \sigma=+1.
 \end{array} \right.
\label{sbrane-100}
\end{equation}
The corresponding field equations can be written in the following form
\begin{eqnarray}
	\dot h^2 + \frac{(q-1) b^2}{(d-2)\chi} {\rm e}^{\chi h} &=& \alpha^2,
\label{sbrane-110}\\
	\dot g^2 + \sigma {\rm e}^{2(k-1) g} &=& \beta^2,
\label{sbrane-120}
\end{eqnarray}
where the following notations are introduced
\begin{eqnarray}
	f(t) &=& h(t) - {a c_1 \over \chi} t - {a c_2\over \chi},
\label{sbrane-130}\\
	\chi &=& 2 p + \frac{a^2(d-2)}{q-1},
\label{sbrane-140}\\
	A(t) &=& k g(t) - \frac{p}{q-1} f(t), \quad B(t) = f(t),
	\quad C(t) = g(t) - \frac{p}{q-1} f(t), \quad D(t) = -\frac{p}{q-1} f(t).
\label{sbrane-150}
\end{eqnarray}
The constants $\alpha$ and $\beta$ satisfy
\begin{equation}
	{p\, c_1^2 \over \chi} + {(d-2) \chi \alpha^2\over 2(q-1)} -
	k(k-1) \beta^2 = 0.
\label{sbrane-160}
\end{equation}
The equations \eqref{sbrane-110} \eqref{sbrane-120} have the following solution
\begin{eqnarray}
	f(t) &=&  \frac2{\chi} \ln \left\{ \frac{\alpha}{\cosh \left[ {\chi
	\alpha \over 2}(t-t_0) \right]} \right\} +
	\frac1{\chi} \ln \left[	{ (d-2) \chi \over (q-1) b^2 } \right] -
	{a c_1\over	\chi}t - {a c_2\over \chi},
\label{sbrane-170}\\
	g(t) &=& \left\{ \begin{array}{ll}
 \frac1{k-1}\ln\left\{ \frac{\beta}{\sinh[(k-1) \beta (t-t_1)
   ]} \right\}, \qquad & \sigma=-1. \\
 \pm \beta (t-t_1), & \sigma=0. \\
 \frac1{k-1}\ln\left\{ \frac{\beta}{\cosh[(k-1) \beta (t-t_1)
   ]} \right\}& \sigma=+1. \end{array} \right\},
\label{sbrane-180}
\end{eqnarray}
which describes S-branes.

The S-brane solution obtained in Ref.\cite{Dzhunushaliev:2006xh} is based on the use of two strongly interacting scalar fields giving regular solutions over all spacetime.
The solution is based on the fact that the potential of the scalar fields has two global and two local minima.
As $|t| \rightarrow \infty$ the scalar fields go to one of the local minima.
This is the necessary condition for existence of such type of a solution.
The numerical analysis showed that, apparently, there is no 
other solution
which tends toward one of the global minima. From the mathematical point of view, the obtained S-brane solution in Ref.\cite{Dzhunushaliev:2006xh}
is very similar to the usual (time-like) brane solution considered
in the section \ref{two_scalar_field}.
\par
\begin{figure}[h]
\begin{minipage}[t]{.49\linewidth}
  \begin{center}
  \includegraphics[width=8cm]{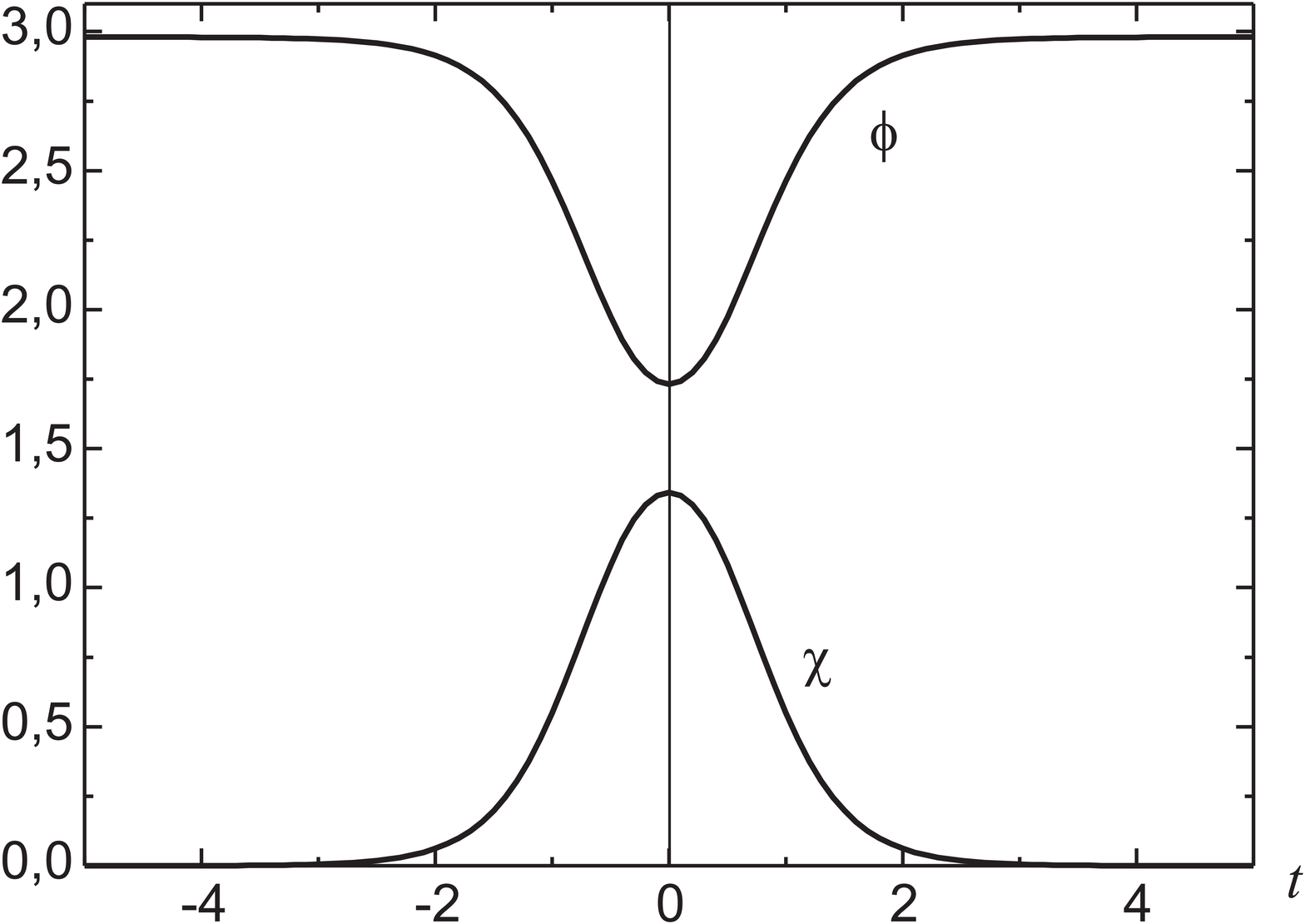}
\caption{The graphs for the scalar fields $\phi^*(t), \chi^*(t)$. Both functions tend asymptotically to the local minimum at $\phi^*(t)=m_1^*, \chi^*(t)=0$.  }
    \label{sbrane-10}
  \end{center}
\end{minipage}\hfill
\begin{minipage}[t]{.49\linewidth}
  \begin{center}
  \includegraphics[width=8cm]{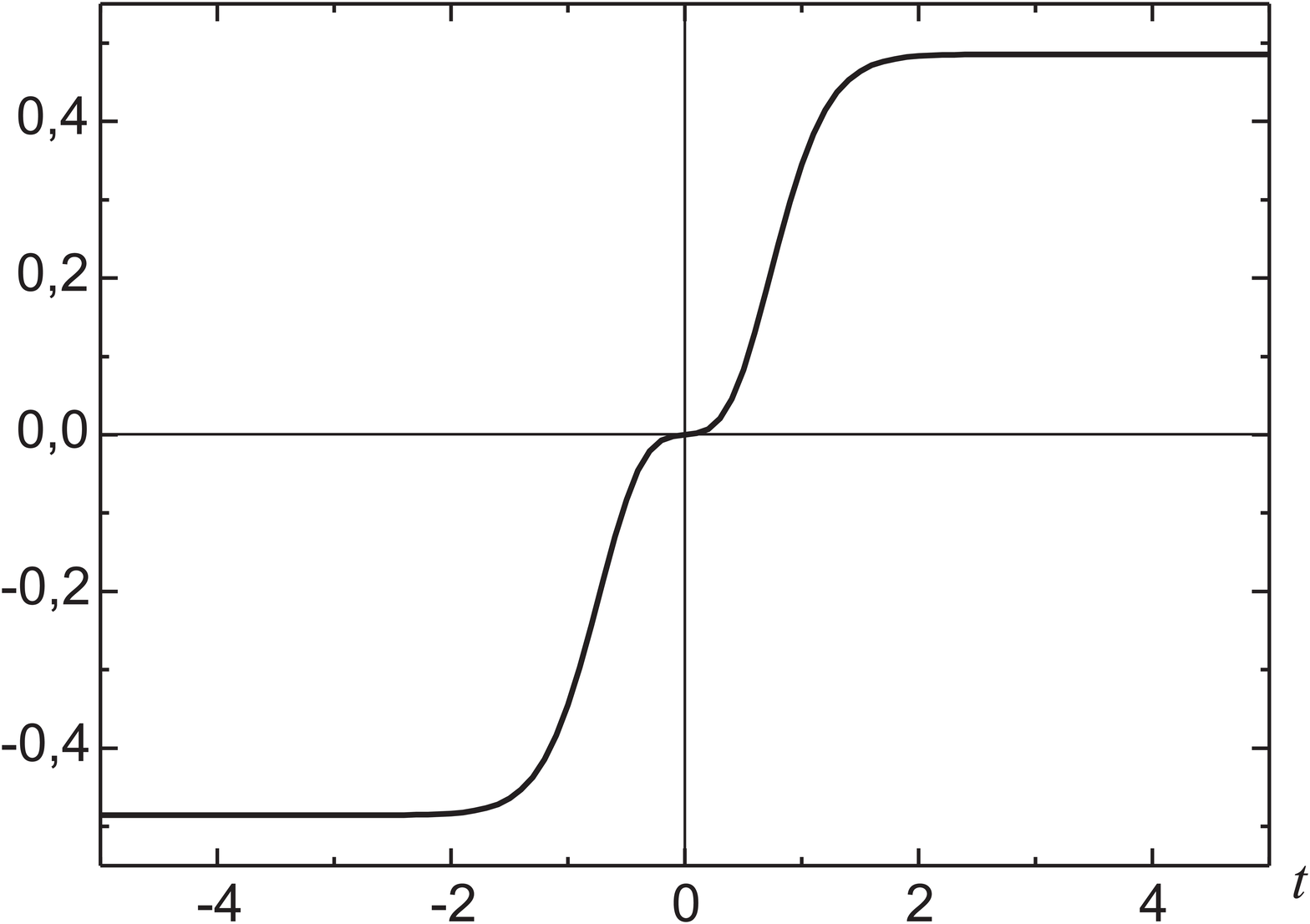}
  \caption{The Hubble parameter $H=\dot a/a$ tends asymptotically to
  the de Sitter stages: if $H<0$ - the stage of contraction, if $H>0$ - the
  stage of expansion.}
    \label{sbrane-20}
  \end{center}
\end{minipage}\hfill
\end{figure}

The Lagrangian for gravitating phantom scalar fields $\phi, \chi$ is:
\begin{equation}
	L=-\frac{R}{16 \pi G}-\frac{1}{2}\partial_\mu \phi \partial^\mu
	\phi-\frac{1}{2}\partial_\mu \chi \partial^\mu
	\chi-V(\phi,\chi),
\label{sbrane-190}
\end{equation}
where the potential $V(\phi,\chi)$ describing strong interaction between scalar fields is given by
\begin{equation}
	V(\phi,\chi)=\frac{\lambda_1}{4}(\phi^2-m_1^2)^2+
	\frac{\lambda_2}{4}(\chi^2-m_2^2)^2+\phi^2 \chi^2-V_0
\label{sbrane-200}
\end{equation}
where $V_0$ is defined
by initial conditions. The metric describing
{an}
S-brane is
\begin{equation}
	ds^2=dt^2-a(t)^2 (dx^2+dy^2+dz^2).
\label{sbrane-210}
\end{equation}
After finding the corresponding equations for the gravitating scalar fields $\phi$ and $\chi$,
one can see, from the physical point of view, that the metric describes
the bounce of the universe at the moment $t=0$. Corresponding equations (in the Planck units, i.e. at $c=\hbar=G=1$) are
\begin{eqnarray}
	\left( \frac{\dot a}{a}\right)^2 &=&
	\frac{1}{6}\left[-\dot \phi^2-\dot \chi^2+V(\phi,\chi)\right]
\label{sbrane-220} \\
	\frac{\ddot a}{a}-\left (\frac{\dot a}{a}\right)^2 &=&
	\frac{1}{4}\left ( \dot \phi^2+\dot \chi^2 \right),
\label{sbrane-230} \\
	\ddot \phi +3\frac{\dot a}{a}\dot \phi &=&
	\phi \left[\chi^2+\lambda_1\left(\phi^2-m_1^2\right)\right],
\label{sbrane-240} \\
	\ddot \chi +3\frac{\dot a}{a}\dot \chi &=&
	\chi \left[\phi^2+\lambda_2\left(\chi^2-m_2^2\right)\right],
\label{sbrane-250}
\end{eqnarray}
and
\begin{equation}
	V_0=\frac{\lambda_1}{2}(\phi_0^2-m_1^2)^2+
	\frac{\lambda_2}{2}(\chi_0^2-m_2^2)^2+\phi_0^2 \chi_0^2.
\label{sbrane-260}
\end{equation}
The numerical solution with the boundary conditions
\begin{eqnarray}
	a(0)=a_0, \quad \dot a(0)=0,
\label{sbrane-265} \\
	\phi(0)=\phi_0, \quad \dot \phi(0)=0,
\label{sbrane-270}\\
	\chi(0)=\chi_0, \quad \dot \chi(0)=0
\label{sbrane-280}
\end{eqnarray}
gives the result presented in Figs. \ref{sbrane-10} and \ref{sbrane-20}.
It is necessary to note that, as well as for time-like branes described in the section \ref{two_scalar_field}, regular solutions (presented in Figs. \ref{sbrane-10} and \ref{sbrane-20}) exist only for definite values of the parameters  $m_1 = m_1^*, m_2 = m_2^*$. In this sense, the solution of
{the}
equations \eqref{sbrane-220}-\eqref{sbrane-250} is a solution of a problem for two eigenvalues $m_1$ and $m_2$ and two eigenfunctions $\phi^*(t)$ and $\chi^*(t)$.


\section{Conclusion}

In this paper, we gave
a review of works
devoted to study the thick brane solutions.
Our main objective is
to introduce the recent progress.
As mentioned in the introduction (section I),
there are
the important motivations to consider the thick brane models:
The first motivation comes from the interesting feature of
a multidimensional topological defect
that can trap the various fields inside it.
The second one comes from the fact
that
along the developments of the (thin) brane cosmology,
in particular in considering models
{of} a spacetime with more than six
dimensions, it
turned out
that even at the classical level
a thin brane exhibits a singular behavior,
due to the stronger self-gravity of the brane.
To cure it, it would be crucial to take
the effects of brane thickness into consideration.

We showed
the previously known solutions
and reviewed their basic properties.
But as mentioned in the introduction, {in this article},
little was
referred to about
{the applications of these solutions to
the high energy physics and}
the localizability of the fields with
various spins, because
they concern both the thin and thick brane models and
should be reviewed in separate publications.
We tried to introduce classifications of the thick branes
on the basis of their division into some classes by combining
these models on topological and physical properties .
They are summarized below (see also tables I and II).

\vspace{0.5cm}

\paragraph*{\it Topological feature:}
The first important classification concerns
whether a solution is topologically trivial
or nontrivial.
This is the division of the models in terms of
topological properties.
The detailed explanation about the concept of the topological triviality
was given in the subsection II A.

The presence of topologically nontrivial vacuum states is typical for the topologically nontrivial solutions (section II).
Here there are two cases:

\begin{enumerate}

{\item codimension is equal to one}

{\item codimensions are more than two.}

\end{enumerate}

In the first case, there are kink-like solutions.
Such solutions
were
found in the theory
with scalar field(s) with {a} non-linear potential.
In the second case, there are hedgehog-like solutions.
Such a topological structure provides the stability of the solutions (at least at the classical level) that, of course, is a great advantage of such models.

\vspace{0.2cm}

In the case of the topologically trivial solutions, there is only one vacuum state (see section III):
a solution starts from a vacuum
and returns to the same vacuum.
The solutions were
constructed by
employing two strongly coupled scalar fields.
A question concerning the stability still remains.
The stability analysis should be done in case by case.

\vspace{0.3cm}

\paragraph*{\it Brane geometry:}

The geometry of the four-dimensional section on the
brane is also an important point.
The most works were
devoted to
search for a solution
with {a} maximally symmetric
four-dimensional section,
i.e.,
Minkowski, de Sitter or anti-de Sitter(AdS) spacetime.
This is because
the system can be described by a set of
the coupled ordinary differential equations
with respect to the (radial) bulk coordinate.
In particular, in the five-dimensional problem,
the solution generation techniques as the superpotential method,
discussed in the subsection II B,
were developed.
In {a} Minkowski
{or a} de Sitter thick brane background,
the trapping
gravitons are possible,
but in the AdS one, it may be impossible.
The (stable) thick de Sitter brane solutions could be used
for modeling inflation or dark energy.

\vspace{0.2cm}

In terms of obtaining a
realistic cosmological model,
{a}
solution with
the cosmological four-dimensional section
{is}
more desired (see section V).
They would be used
for describing the evolution of a brane configuration changing with time.
In this case, generically,
the problem depends both on extra space-like coordinates and time.
It is obvious that
both {the search
and interpretation of solutions}
would be much more complicated.
Therefore, the research in this field
{was}
performed
by introducing the brane thickness by hand.
Although
there is always ambiguity
{on how to define and handle the thickness},
several procedures {were}
proposed,
{such as the averaging approach,
the thin-shell approximation and
the quasi-static approximation.}
In such models, generically, new effects due to the thickness of the brane
are induced onto the effective Friedmann-like equations,
which for instance
could accelerate the Universe
even though there is only the ordinary matter inside the brane.
{But the nature of these effects depends on the regularization procedure.}

\vspace{0.3cm}

\paragraph*{\it Bulk geometry:}

In known solutions in five-dimensional models,
the asymptotic bulk geometry can be either flat,
de Sitter or AdS.
In the model with a canonical scalar field,
to obtain an asymptotically AdS thick brane solution,
the fine-tuning condition Eq. (\ref{cond}) must be satisfied.
In {a multidimensional spacetime},
the asymptotic geometry and
regularity of the spacetime
crucially depend on
the boundary conditions
and the choice of parameters.
It is not only the case that the bulk spacetime is regular,
but in some cases the bulk spacetime contains a naked singularity
{at a finite distance from the brane}.

\vspace{0.3cm}

\paragraph*{\it {Special} models :}

The model with an unusual source function put in by hand provides
another {special kind} of solutions (see section IV). In this case, the aim is to simplify a model as much as possible
{to obtain analytic solutions}.
It allows analyzing properties of the model in the most evident form. However, this advantage of the model is smoothed over by the fact that the source is put in by hand.

The next {special kind}
of solutions is the so-called S-brane model,
in which it is assumed that the warp factor is some function of time (see section VI). It allows one to describe a cosmological evolution of multidimensional models which are both regular in a whole multidimensional space and do not have the initial cosmological singularity.

\vspace{0.2cm}

Within the framework of multidimensional theories
including string and Kaluza-Klein theories,
the possibility of 
extra time-like dimensions may not be excluded.
This may provide the disappearance of the cosmological constant
in {the}
four-dimensional spacetime \cite{Arefeva,Arefeva1}.
However, using the extra time-like dimensions leads to considerable
difficulties in fundamental properties,
like appearance of tachyons and ghosts \cite{Scherk,Arefeva2},
and violation of causality \cite{Yndurain:1990fq}.
A model of a thick brane with extra time-like coordinate can be obtained, for example, with the use of two interacting scalar fields from section
\ref{two_scalar_field}. In {the} case of a
 5-dimensional model
from \ref{5-dim_problem}, one can choose the metric in the form
\begin{equation}
ds^2= \phi ^2(r) \eta_{\alpha \beta }(x^\nu)dx^\alpha dx^\beta -
\delta \,dr^2~,
\end{equation}
where $\delta=+1$ corresponds to the problem with one time-like dimension considered in \ref{two_scalar_field},
and $\delta=-1$ refers to the problem with {the}
additional time-like coordinate. In the latter case, the same thick brane solutions
like in section \ref{two_scalar_field} will exist if one takes the reverse sign {for}
the potential \eqref{pot2}.

\begin{table}[ht]
 \caption{ \small Classification of the thick brane models {with scalar fields}}
\label{tab1}
\begin{center}
\begin{tabular}{|p{4cm}|p{4cm}|p{7cm}|}
\hline
{\bf Class of brane models} \vspace{.3cm}&{\bf Matter sources} &
{\bf Basic feature} \\
\hline { Topologically non-trivial thick branes}  &
{Scalar field(s) with non-linear potentials}&
{The presence of topologically nontrivial vacuum states provide stability of the solutions at the classical level.
A
question about the quantum stability requires further
{studies}.
}\\
\hline
{ Topologically trivial thick branes} &
{Two strongly interacting usual or phantom scalar fields }&
{
A
question about the classical and quantum stability requires further
{studies}.}\\
\hline
\end{tabular}
\end{center}

\end{table}

\begin{table}[ht]
 \caption{ \small Special thick brane models}
\label{tab2}
\begin{center}
\begin{tabular}{|p{4cm}|p{4cm}|p{7cm}|}
\hline
{\bf 
} \vspace{.3cm}&{\bf Matter sources} &
{\bf Basic feature} \\

\hline
{Branes with unusual source functions} &
{Sources put in by hand via the special ansatz of functions}&
{The advantage is that
it is possible to make use of analytical {studies}.
The main drawback is that the source is put in by hand.}\\
\hline
{S-branes} &
{Scalar fields, including fields with spontaneous symmetry breaking}&
{{The}
multidimensional time-dependent solutions can be both topologically trivial and nontrivial ones.}\\
\hline
{Thick branes 
with additional time coordinates} &
{Two strongly interacting scalar fields}&
{{They}
are not excluded but possess the considerable difficulties such as appearance
of tachyons, ghosts and violation of the causality.}\\
\hline
\end{tabular}
\end{center}

\end{table}


\vspace{0.5cm}

We shall close this review article after
suggesting the possible perspective directions of investigations:

\begin{enumerate}

\item
{
To search new thick brane solutions,
with the maximally symmetric four-dimensional section,
in more generalized models of the field theory and the modified gravity.}

\item
{
To search
self-consistent solutions describing objects localized on the brane (monopoles, black holes and so on) depending both on
extra space-like coordinates and 4-dimensional coordinates.}

\item
{
To search
the cosmological thick brane solutions which depend both on time and
extra space-like coordinates.
Having explicit solutions of such cosmological thick branes,
the ambiguity {mentioned above}
 would be somewhat resolved}.

\end{enumerate}

\section*{Acknowledgements}
We thank C. Adam,
I. Antoniadis, N.  Barbosa-Cendejas, M. Cvetic, A. Flachi, A. Herrera-Aguilar,
Y. Liu, I. Neupane, M. Pavsic and A. Wereszczynski
for comments and suggestions.
{The authors also would like to express the special gratitude
to W. Naylor and D. Singleton
for their careful reading of the manuscript and fruitful suggestions.}
V. D. and V. F. are grateful to the Research Group Linkage Programme of the
Alexander von Humboldt Foundation for the support of this research.
V. F. would also like to thank the ICTP for their hospitality during his visit.
The work of M.~M. was supported 	
by the National Research Foundation of Korea (NRF) 	
grant funded by the Korea government(MEST) (No. 	
20090063070).

\end{document}